\documentclass{emulateapj} 
\usepackage{subfigure} \usepackage{url} \usepackage{multirow} \usepackage{color}
\usepackage{graphicx} \usepackage{xspace} \usepackage{natbib} \usepackage{amsmath}
\usepackage[figuresright]{rotating} 
\usepackage{subfigure}
\usepackage{float}
\newcommand{\water}{$\rm H_2O$} \newcommand{\methane}{$\rm CH_4$}
 \newcommand{\cotwo}{$\rm CO_2$}
\newcommand{\ozone}{$\rm O_3$} \newcommand{\nitrogen}{$\rm N_2$}
\newcommand{\oxygen}{$\rm O_2$}

\shorttitle{Retrievals of Earth-like Atmospheres} \shortauthors{Feng et al.}

\begin{document}

\title{\uppercase{Characterizing Earth Analogs in Reflected Light: Atmospheric Retrieval Studies for Future Space Telescopes}}

\author{\textsc{Y. Katherina Feng\altaffilmark{1,2,3}, Tyler D.
Robinson\altaffilmark{1,4,5}}, Jonathan J. Fortney\altaffilmark{1,3}, Roxana E. Lupu\altaffilmark{6,7}, Mark S. Marley\altaffilmark{6}, Nikole K. Lewis\altaffilmark{8,9}, Bruce Macintosh\altaffilmark{10},  Michael R. Line\altaffilmark{11}}

\altaffiltext{1}{Department of Astronomy \& Astrophysics, University of California, Santa Cruz, CA 95064, USA} 
\altaffiltext{2}{NSF Graduate Research Fellow}
\altaffiltext{3}{University of California, Santa Cruz, Other Worlds Laboratory}
\altaffiltext{4}{now at Department of Physics \& Astronomy, Northern Arizona University, Flagstaff, AZ 86011, USA}
\altaffiltext{5}{NASA Astrobiology Institute's Virtual Planetary Laboratory}
\altaffiltext{6}{NASA Ames Research Center, Moffett Field, CA 94035, USA}
\altaffiltext{7}{Bay Area Environmental Research Institute, Petaluma, CA 94952, USA}
\altaffiltext{8}{Space Telescope Science Institute, Baltimore, Maryland 21218, USA}
\altaffiltext{9}{Department of Earth and Planetary Sciences, Johns Hopkins University, Baltimore, MD 21218, USA}
\altaffiltext{10}{Department of Physics, Kavli Institute for Particle Astrophysics and Cosmology, Stanford University, Stanford, CA 94305, USA}
\altaffiltext{11}{School of Earth and Space Exploration, Arizona State University, Tempe, AZ 85287, USA}

%
\begin{abstract} 

Space-based high contrast imaging mission concepts for studying rocky exoplanets in reflected light are currently under community study.  We develop an inverse modeling framework to estimate the science return of such missions given different instrument design considerations. By combining an exoplanet albedo model, an instrument noise model, and an ensemble Markov chain Monte Carlo sampler, we explore retrievals of atmospheric and planetary properties for Earth twins as a function of signal-to-noise ratio (SNR) and resolution ($R$). Our forward model includes Rayleigh scattering, single-layer water clouds with patchy coverage, and pressure-dependent absorption due to water vapor, oxygen, and ozone. We simulate data at $R = 70$ and $R = 140$ from 0.4--1.0~$\mu$m with SNR $ = 5, 10, 15, 20$ at 550~nm (i.e., for HabEx/LUVOIR-type instruments). At these same SNR, we simulate data for {\it WFIRST} paired with a starshade, which includes two photometric points between 0.48--0.6~$\mu$m and $R = 50$ spectroscopy from 0.6--0.97~$\mu$m. Given our noise model for {\it WFIRST}-type detectors, we find that weak detections of water vapor, ozone, and oxygen can be achieved with observations with at least $R = 70$ / SNR$\ = 15$, or $R = 140$ / SNR$\ = 10$ for improved detections. Meaningful constraints are only achieved with $R = 140$ / SNR$\ = 20$ data.   The {\it WFIRST} data offer limited diagnostic information, needing at least SNR = 20 to weakly detect gases. Most scenarios place limits on planetary radius, but cannot constrain surface gravity and, thus, planetary mass.
\end{abstract}

%
\section{Introduction} 
The scientific field of exoplanets has been rapidly advancing since the hallmark discovery of the first planet orbiting a Sun-like star \citep{mayor1995}. Following the launch of NASA's {\it Kepler} mission  \citep{borucki2003, Borucki11}, the field has seen the discovery of thousands of transiting exoplanets and the exciting result that planets with radii between $0.75$--$2.5\ R_{\oplus}$ are common around solar-type stars \citep{burke2015}. Only within the last decade have observational studies for exoplanet atmospheric characterization seen substantial development, starting with the first detection of an exoplanet's atmosphere by \citet{Charb02}. 

To date, the majority of exoplanet atmospheric characterization investigations have focused on transiting worlds. Hot Jupiters, owing to their large sizes and short orbital periods, are typically emphasized as targets for these studies.  Characterization of small, potentially rocky exoplanets is limited to worlds with cool stellar hosts (K and M dwarfs), which offer favorable planet-to-star size ratios.  Recently, \citet{dewit2016} studied the combined transmission spectra of two transiting Earth-sized planets orbiting the ultracool dwarf TRAPPIST-1 using the \textit{Hubble Space Telescope}.  While no gas absorption features were detected by \citet{dewit2016}, this work highlights the improvements in signal size when terrestrial-sized transiting planets are studied around low-mass stars.  Additionally, since the Habitable Zone \citep{kasting1993} around a low-mass star is relatively close-in, characterization studies of potentially habitable exoplanets around cool stars can benefit from the frequency of transit events.  However, for Sun-like hosts, the planet-to-star size ratio is much less favorable and the Habitable Zone is located far from the star, thus severely limiting the potential for atmospheric characterization. 

Direct, high-contrast imaging has now emerged as an essential technique for studying the atmospheres of planets at larger orbital separations from their host star (i.e., at orbital distances $\gtrsim 1$~au).  Thus far, high-contrast imaging has been proven successful in studying atmospheres of young, self-luminous gas giants in the near-infrared and mid-infrared \citep[e.g.,][]{barman2011,Skemer14,macintosh2015}. These worlds, owing to their intrinsic brightness, have typical contrast ratios of $10^{-4}$ with respect to their hosts. A true Jupiter analog at visible wavelengths, by comparison, would have a contrast ratio of $10^{-9}$, while an Earth analog would have a contrast ratio of order $10^{-10}$. Reflected light in the visible probes to atmospheric depths of up to $\sim 10$ bar for giant planets \citep{marley2014}, which is complimentary to the relatively low pressures probed in transit observations (typically less than 10--100~mbar). Additionally, the wavelength range of 0.4--1.0 \micron\ holds rich information about a planet's atmosphere, including signatures of methane, water vapor, and haze \citep{marley2014,burrows2014}.

In spite of the incredible technological challenges, there are multiple planned or in-development space-based missions that would be capable of high-contrast imaging of exoplanets in reflected light. First among these will be NASA's Wide-Field InfraRed Survey Telescope \citep[{\it WFIRST},][]{spergel2013}, which was identified as the top priority space mission in the 2010 National Academy of Sciences Decadal Survey of Astronomy and Astrophysics\footnote{\url{http://sites.nationalacademies.org/bpa/bpa\textunderscore049810}}.  The {\it WFIRST} mission will carry a Coronagraphic Instrument (CGI) with imaging capability and a visible-light integral field spectrograph of wavelength resolution $\sim 50$  \citep{noecker2016,trauger2016, seo2016,cady2016,balasubramanian2016,groff2018}. Although envisioned primarily as a technology demonstrator, it may study the atmospheres of relatively cool gas giant exoplanets that have been previously detected using the radial velocity technique \citep{Traub2016}. 

While \textit{WFIRST} could also have some capability to survey stars in the solar neighborhood for lower-mass planetary companions \citep{burrows2014, greco2015, spergel2015,savransky&garrett2016,robinson2016}, it is anticipated that the core optical throughput of the {\it WFIRST} CGI will be low for planetary signals. This stems primarily from the complexities of accommodating for {\it WFIRST}'s on-axis secondary mirror and support structures within the high-contrast instruments \citep{Traub2016,krist2016}.  Low throughput drives long requisite integration times, thereby likely making spectroscopic observations of smaller, less-bright worlds (such as super-Earth exoplanets) unfeasible except around the very closest stars \citep{robinson2016}.  However, if the {\it WFIRST} spacecraft were to be paired with an external starshade \citep{cash2006,Kasdin2012}, the CGI can be operated in a direct mode without coronagraphic masks, substantially increasing throughput. High-contrast imaging of sub-Neptune and terrestrial-sized exoplanets may then become possible. The feasibility of a starshade ``rendezvous'' with the \textit{WFIRST} spacecraft is under active investigation \citep{seager2015,crill2017}. 

In advance of the 2020 astronomy and astrophysics decadal survey, several large-scale space-based mission concepts are being studied\footnote{\url{https://science.nasa.gov/astrophysics/2020-decadal-survey-planning}}.  Of these, two have a strong focus on the characterization of rocky exoplanets with direct imaging: the Habitable Exoplanet Imaging Mission \citep[HabEx;][]{habexRef} and the Large Ultra-Violet/Optical/InfraRed Surveyor \citep[LUVOIR;][]{luvoirRef}. HabEx and LUVOIR are incorporating aspects of design that would allow the detection of water vapor and biosignatures on planets in the Habitable Zones of nearby Sun-like stars. It is therefore timely and critical that we explore observational approaches that maximize science yield during the development of these large-scale mission concepts as well as the \textit{WFIRST} rendezvous concept. To accomplish this, we must perform atmospheric and instrument modeling to simulate the types of spectra we can expect to measure, and we must develop tools to infer planetary properties from these simulated observations.

Traditionally, the comparison to a limited range of forward models has been used to infer atmospheric properties (such as temperature structure and gas abundances) from spectral observations. This involves iterating to a radiative-convective solution for a given set of planetary parameters (e.g., gravity, metallicity, equilibrium abundances, incident flux), and can include detailed treatment of aerosols, chemistry, and dynamics within the model atmosphere \citep{marley2015}. The goal is to generate a spectrum that matches available data and, thus, offers one potential explanation for the world's atmospheric state \citep[e.g.,][]{Konopacky13,macintosh2015,barman2015}.  A more data-driven interpretation of atmospheric observations is accomplished through inverse modeling, or retrievals. Developed for Solar System studies and remote sensing \citep[e.g.,][]{rodgers76,irwin2008}, retrievals have become a valuable tool in constraining our understanding of the atmospheres of transiting exoplanets.  Early exoplanet retrieval work invoked grid-based optimization schemes \citep{madhu2009}, while subsequent works have taken advantage of Bayesian inference with methods such as optimal estimation and Markov chain Monte Carlo (MCMC) \citep[e.g.,][]{lee2012,Benneke12,Line2013}.

Several studies have examined the hypothetical yield from characterizing giant exoplanets observed with a space-based coronagraph (such as \textit{WFIRST}) with retrieval techniques. \citet{marley2014}, for example, modeled spectra we could expect from known radial velocity gas giants if observed by the \textit{WFIRST} CGI. Given the diversity of cool giant planets, the model spectra have a variety of input assumptions for clouds, surface gravity, and atmospheric metallicity. \citet{marley2014} then applied retrieval methods to these synthetic spectra, enabling the exploration of how well atmospheric parameters are constrained under varying quality of data. \citet{lupu2016} further investigated the feasibility of characterizing cool giant planet atmospheres through retrieval, focusing on the ability to constrain the \methane\ abundance and cloud properties. The systematic study of the impact of conditions like signal-to-noise ratios or wavelength resolution is essential to quantifying the scientific return of these reflected-light observations. \citet{nayak2017} considered the impact of an unknown phase angle on the inference of properties such as planet radius and gravity. In all of these studies, the signal-to-noise ratio (SNR) of the data has a significant influence on the constraints of atmospheric properties.

Previous work on smaller planets in the context of possible future space missions includes \citet{vonparis13}, who synthesized infrared emission observations of a cloud-free, directly-imaged Earth-twin, and employed a least-squares approach and $\chi^2$ maps to perform retrievals and explore parameter space (considering the effects of instrument resolution and SNRs). A collection of recent studies \citep{wang2017a,mawet2017,wang2017b} examined atmospheric species detection using ``High Dispersion Coronagraphy'', which couples starlight suppression technologies with high resolution spectroscopy. In these studies, simulated observations (typically at spectral resolutions, $R=\lambda/\Delta\lambda$, of many hundreds to tens of thousands) are cross-correlated with template molecular opacity spectra to explore the feasibility of species detection. While this novel approach can yield detections of key atmospheric constituents, the abundance of these these atmospheric species cannot be robustly constrained.

To date, there still does not exist a systematic study of atmospheric characterization of small exoplanets using retrieval techniques on reflected light observations at spectral resolutions relevant to {\it WFIRST} rendezvous, HabEx, and LUVOIR.  Motivated by this need, we present here our extension of Bayesian retrieval techniques into the terrestrial regime. We construct a forward model suitable for simulating reflectance spectra of Earth-like planets in the visible wavelength range of 0.4 \micron\ to 1.0 \micron.  We explore retrievals of planetary and atmospheric properties from simulated data sets at varying spectral resolutions and SNRs. A retrieval framework such as this allows us to quantify uncertainties we expect for key planetary parameters given certain observing scenarios.  Thus, our approach enables us to search for the minimal observing conditions that achieve the scientific goal of identifying traits associated with habitability and life.  In particular, we are interested in our ability to detect and constrain abundances of molecules such as water, oxygen, and ozone, characterize basic properties of a cloud layer, and measure bulk parameters such as radius.

In section \ref{methods}, we describe our forward model and construction of simulated data. In section \ref{buildup}, we validate our forward model by building up retrieval complexity (i.e., number of retrieved parameters). We perform a study of retrieval performance with respect to spectral resolution and SNR in section \ref{snr_study}, with implications for HabEx/LUVOIR. We also study the retrieval performance for data sets expected from a {\it WFIRST} rendezvous scenario, where the CGI would provide modest-resolution spectroscopy in the red (600--970~nm) and photometry in the blue (480--600~nm). We present our discussion and conclusions in sections \ref{discussion} and \ref{conclusion}, respectively. 

\section{Methods} \label{methods}

The observed quantity for a directly imaged exoplanet in reflected light at a given phase (i.e., planet-star-observer) angle, $\alpha$, is the wavelength dependent planet-to-star flux ratio, 
\begin{equation} 
\frac{F_p}{F_s} = A_g \Phi (\alpha)
\Big(\frac{R_p}{r}\Big)^2, \label{fpfs} 
\end{equation} 
where $A_g$ is the geometric albedo, $\Phi(\alpha)$ is the phase function, $R_p$ is the radius of the planet, and $r$ is the orbital separation. The phase function (which depends on wavelength) translates the planetary brightness at full phase (i.e., where  $\alpha=0^{\circ}$) to its brightness at different phase angles. The wavelength dependent geometric albedo is the ratio between the measured flux from the planet at full phase to that from a perfectly reflecting Lambert (i.e., isotropically-reflecting) disk with the size of the planet.  We denote the product of the geometric albedo and the phase function as the phase dependent ``reflectance'' of the planet. In general, the geometric albedo encodes information about the composition and structure (i.e., ``state'') of an atmosphere, while the phase function is strongly related to the scattering properties of an atmosphere \citep[e.g.,][]{Marley99,burrows2014}. 

To understand the information contained in direct imaging observations of exoplanets in reflected light, we employ a retrieval (or inverse analysis) framework that consists of several linked simulation tools and models.  Of central importance is a well-tested three-dimensional albedo model---described in greater depth below---that computes a reflectance spectrum at high resolution for a planet given a description of its atmospheric state \citep{Mckay89,Marley99,cahoy2010,lupu2016,nayak2017}.  When coupled with a simulator for degrading a high resolution spectrum to match the resolution of an instrument, we refer to these two tools as the ``forward model.''  By adding simulated noise to forward model spectra, we generate faux ``observations'' of worlds as would be studied by future high-contrast imaging missions.  To create ``observed'' spectra, we adopt a widely-used direct imaging instrument simulator \citep{robinson2016} that generates synthetic observations given an input, noise-free spectrum.

Given an ``observed'' planet-to-star flux ratio spectrum, our inverse analyses use a Bayesian inference tool that compares the observation to forward model outputs to sample the posterior probability distributions for a collection of atmospheric state parameters.  In other words, our inverse analyses indicate what range of atmospheric state parameters (e.g., gas abundances) adequately describe a direct imaging observation.  Our  Bayesian parameter estimations use an open-source affine invariant Markov-Chain Monte Carlo (MCMC) ensemble sampler---{\tt emcee} \citep{goodman2010,foreman2013}.

In this work, retrieval analyses generally proceed by first simulating a noise-free spectrum of a world with a known atmospheric state (e.g., Earth).  We then add simulated observational noise to this spectrum to create a synthetic observation.  Following Bayesian parameter estimation on this synthetic observation, we can compare a retrieved atmospheric state to the original, known atmospheric state, thereby allowing us to understand how observational noise affects our ability to deduce the true nature of an exoplanetary atmosphere.

\subsection{Albedo Model} \label{fm}
Our three-dimensional albedo model \citep[see also][]{cahoy2010} divides a world into a number of plane-parallel facets with coordinates of longitude ($\zeta$) and co-latitude ($\eta$), with the former referenced from the sub-observer location and the latter ranging from 0 at the northern pole to $\pi$ at the southern pole. A single facet has downwelling incident stellar radiation from a zenith angle $\mu_{\rm s} = \cos{\theta_{\rm s}} = \sin{\eta}\cos{(\zeta - \alpha)}$, where, as earlier, $\alpha$ is the phase angle. The facet reflects to the observer in a direction whose zenith angle is given by $\mu_{\rm o} = \cos{\theta_{\rm o}} = \sin{\eta} \cos{\zeta}$.  Note that, at full phase (where the geometric albedo is defined) the observer and the source are colinear such that $\mu_{\rm o} = \mu_{\rm s}$ for all facets.

The atmosphere above each facet is divided into a set of pressure levels, and we perform a radiative transfer calculation to determine the emergent intensity. With the intensities calculated for an entire visible hemisphere, we follow the methods outlined by \citet{horak1950} and \citet{horak1965} to perform integration using Chebychev-Gauss quadrature, thus producing the reflectance value at a given wavelength. We repeat this procedure at each of the wavelength points within a specified range to build up a reflectance spectrum. 

Taking $I(\tau,\mu,\phi)$ to be the wavelength-dependent intensity at optical depth $\tau$ in a direction determined by the zenith and azimuth angles $\mu$ and $\phi$, we ultimately need to determine the emergent intensity from each facet in the direction of the observer, $I(\tau=0,\mu_{\rm o},\phi_{\rm o})$.  Thus, for each facet we must solve the one-dimensional, plane-parallel radiative transfer equation,
\begin{equation}
    \mu\frac{dI}{d\tau} = I(\tau,\mu,\phi) - S(\tau,\mu,\phi),
\end{equation}
where $S$ is the wavelength dependent source function.  Following \citet{meador1980}, \citet{Toon89}, and \citet{marley2015}, the source function is 
\begin{equation}
\begin{aligned}
    S(\tau,\mu,\phi) =  \frac{\bar{\omega}}{4\pi}F_{\rm s} \cdot p(\tau,\mu,\phi,-\mu_{\rm s},\phi_{\rm s}) \cdot \mathrm{e}^{-\tau/\mu_{\odot}} \\
     +\, \bar{\omega}\int_0^{2\pi} d\phi' \int_{-1}^{1}\frac{d\mu'}{4\pi} \cdot I(\tau,\mu',\phi') \cdot p(\tau,\mu,\phi,\mu',\phi'),
\end{aligned}
\label{eq:sourcefunction}
\end{equation}
where $\bar{\omega}$ is the single scattering albedo, $F_{\rm s}$ is the incoming stellar flux at the top of the atmosphere (which we normalize to unity so that emergent intensities correspond to reflectivities), $\phi_{\rm s}$ is the stellar azimuth angle, and $p$ is the scattering phase function. Note that our source function does not include an emission term since we are not computing thermal spectra.  Recall that the first term in Equation~\ref{eq:sourcefunction} describes directly scattered radiation from the direct solar beam while the final term describes diffusely scattered radiation from the ($\mu',\phi'$) direction scattering into the ($\mu,\phi$) direction.

Like most standard tools for solving the radiative transfer equation, we separate treatments of directly-scattered radiation from diffusely-scattered radiation, and, for both, it is convenient to express the scattering phase function in terms of a unique scattering angle, $\Theta$.  As single scattered radiation typically has more distinct forward and backward scattering features, we choose to represent the scattering phase function for the direct beam with a two-term Henyey-Greenstein (TTHG) phase function \citep{kattawar1975},
\begin{equation}
    p_{\rm TTHG}(\Theta) = fp_{\rm HG}(g_{\rm f},\Theta) + (1-f)p_{\rm HG}(g_{\rm b},\Theta) \,
\end{equation}
where $p_{\rm HG}$ is the Henyey-Greenstein (HG) phase function with,
\begin{equation}
    p_{\rm HG} = \frac{1}{4\pi}\frac{1 - \bar{g}^2}{(1 + \bar{g}^2 - 2\bar{g}\cos{\Theta})^{3/2}} \ .
\end{equation}
Recall that a TTHG phase function can represent both forward and backward scattering peaks, while the (one term) Henyey-Greenstein phase function only has one peak (typically in the forward direction).  In the previous expressions, $\bar{g}$ is the asymmetry parameter, $f$ is the forward/backward scatter fraction, $g_{\rm f}$ is the asymmetry parameter for the forward-scattered portion of the TTHG, and $g_{\rm b}$ is the asymmetry parameter for the backward-scattered portion of the TTHG.  For the forward and backward scattering portions of the TTHG phase function, we use $g_{\rm f} = \bar{g}$, $g_{\rm b} = -\bar{g}/2$, and $f = 1 - g_{\rm b}^2$. Substituting these into the TTHG phase function expression yields, 
\begin{equation}
    p_{\rm TTHG} = \frac{\bar{g}^2}{4}p_{\rm HG}(-\bar{g}/2,\Theta) + (1-\frac{\bar{g}^2}{4})p_{\rm HG}(\bar{g},\Theta).
\end{equation}
For radiation that is single-scattered from the solar beam to the observer, the scattering geometry is fixed by the planetary phase angle such that $\Theta = \pi-\alpha$. Our choice of parameters, and their relation to $\bar{g}$, in the TTHG was designed by \citet{cahoy2010} to roughly reproduce the phase function of liquid water clouds.  This parameterization, however, is different from that proposed by \citet{kattawar1975}.  We do not expect our results to be sensitive to the details of a particular phase function treatment as \citet{lupu2016} showed that scattered-light retrievals struggle to constrain phase function parameters.

We adopt a standard two-stream approach to solving the radiative transfer equation \citep{meador1980}.  In this case, the diffusely-scattered component of the source function is azimuthally averaged. Combined with our representation of the directly-scattered component, we have,
\begin{equation}
\begin{aligned}
S(\tau,\mu, \mu_{\rm s}, \alpha) = {}& \frac{\bar{\omega}}{4\pi}F_{\rm s} \cdot p_{\rm TTHG}(\mu,-\mu_{\rm s}) \cdot \mathrm{e}^{-\tau/\mu_{\rm s}} \\
 & +  \frac{\bar{\omega}}{4\pi}\int_{-1}^{1}  I(\tau,\mu')p(\mu,\mu') d\mu',
\end{aligned}
\label{eq:directdiffuse}
\end{equation}
where the azimuth-averaged phase functions are given by,
\begin{equation}
    p(\mu,\mu') = \frac{1}{2\pi} \int_{0}^{2\pi} p(\mu,\phi,\mu',\phi) d\phi \ .
\end{equation}
We represent the azimuth-averaged scattering phase functions as a series of Legendre polynomials, $P_l(\mu)$, expanded to order $M$ with,
\begin{equation}
    p(\mu,\mu') = \sum_{l = 0}^{M}g_l P_l(\mu) P_l(\mu') \ ,
\end{equation}
where the phase function moments, $g_l$, are defined according to,
\begin{equation}
    g_l = \frac{2l+1}{2}\int_{-1}^1 p(\cos{\Theta}) P_l(\cos{\Theta}) d\cos{\Theta} \ .
\end{equation}
The first moment of the phase function is related to the asymmetry parameter, with $\bar{g} = g_1/3$. We use a second order expansion of the phase function, giving, 
\begin{equation}
    p(\mu,\mu') = 1 + 3\bar{g}\mu\mu' + \frac{g_2}{2}(3\mu^2 - 1)(3\mu'^2 - 1) \ . 
\end{equation}

In a given atmospheric layer of our albedo model, the optical depth is the sum of the scattering optical depth and the absorption optical depth, $\tau = \tau_{\rm scat} + \tau_{\rm abs}$. The scattering optical depth has contributions from Rayleigh scattering and clouds, so that $\tau_{\rm scat} = \tau_{\rm Ray} + \bar{\omega}_{\rm cld}\tau_{\rm cld}$, where $\bar{\omega}_{\rm cld}$ is the cloud single scattering albedo. The single scattering albedo for a layer is then $\bar{\omega}=\tau_{\rm scat}/\tau$.  We determine the asymmetry parameter, $\bar{g}$, with an optical depth weighting on the Rayleigh scattering asymmetry parameter (which is zero) and the cloud scattering asymmetry parameter, yielding $\bar{g} = \bar{g}_{\rm cld} (\tau_{\rm cld}/\tau_{\rm scat})$.  When representing the second moment of the phase function, we use $g_2 = \frac{1}{2}(\tau_{\rm Ray}/\tau_{\rm scat})$ so that $g_2$ tends towards the appropriate value for Rayleigh scattering \citep[i.e., 1/2;][]{hansen1974} when the Rayleigh scattering optical depth dominates the scattering optical depth.

\subsection{Model Upgrades} \label{subsec:upgrades}
As compared to prior investigations that have used the \citet{cahoy2010} albedo tool \citep[e.g.,][]{lupu2016,nayak2017}, we have updated the model to include an optional isotropically-reflecting (Lambertian) lower boundary (mimicking a planetary surface), and have added pressure-dependent absorption due to \water, \ozone, \oxygen, and \cotwo. \methane\ remains a radiatively active species in the model, as in previous studies.  We also include Rayleigh scattering from \water, \oxygen, \cotwo, and N$_2$ (in addition to H$_2$ and He from previous studies).  As in \citet{lupu2016}, we allow for an extended gray-scattering cloud in our atmospheres.

In \citet{lupu2016}, their two-layer cloud model atmosphere includes a deeper, optically thick cloud deck that essentially acts as a reflective surface. Unlike the gas giants within that study, though, terrestrial planets have a solid surface we can probe. 
We characterize our isotropically-reflecting lower boundary using a spherical albedo for the planetary surface, $A_{\rm s}$, which represents the specific power in scattered, outgoing radiation compared to that in incident radiation.  For this study, we simply adopt gray surface albedo values, which reduces complexity and computation time.  For the inhomogeneous surface of a realistic Earth, featuring oceans and continents, the surface albedo is wavelength-dependent, and we hope to investigate the significance of such surfaces in future work.

We undertook a test to check our reflective lower boundary condition in the limit of a transparent atmosphere. Without atmospheric absorption or scattering, our assumption of a Lambertian surface would imply that the reflectivity (or phase function) determined by our
albedo code should follow the analytic Lambert phase function, \begin{equation}
\Phi_{\rm{L}}(\alpha) = \frac{\sin{\alpha} + (\pi - \alpha)\cos{\alpha}}{\pi}.
\label{lamb_phase} \end{equation} 

\begin{figure}[ht!] \begin{center}
\includegraphics[width=0.47\textwidth]{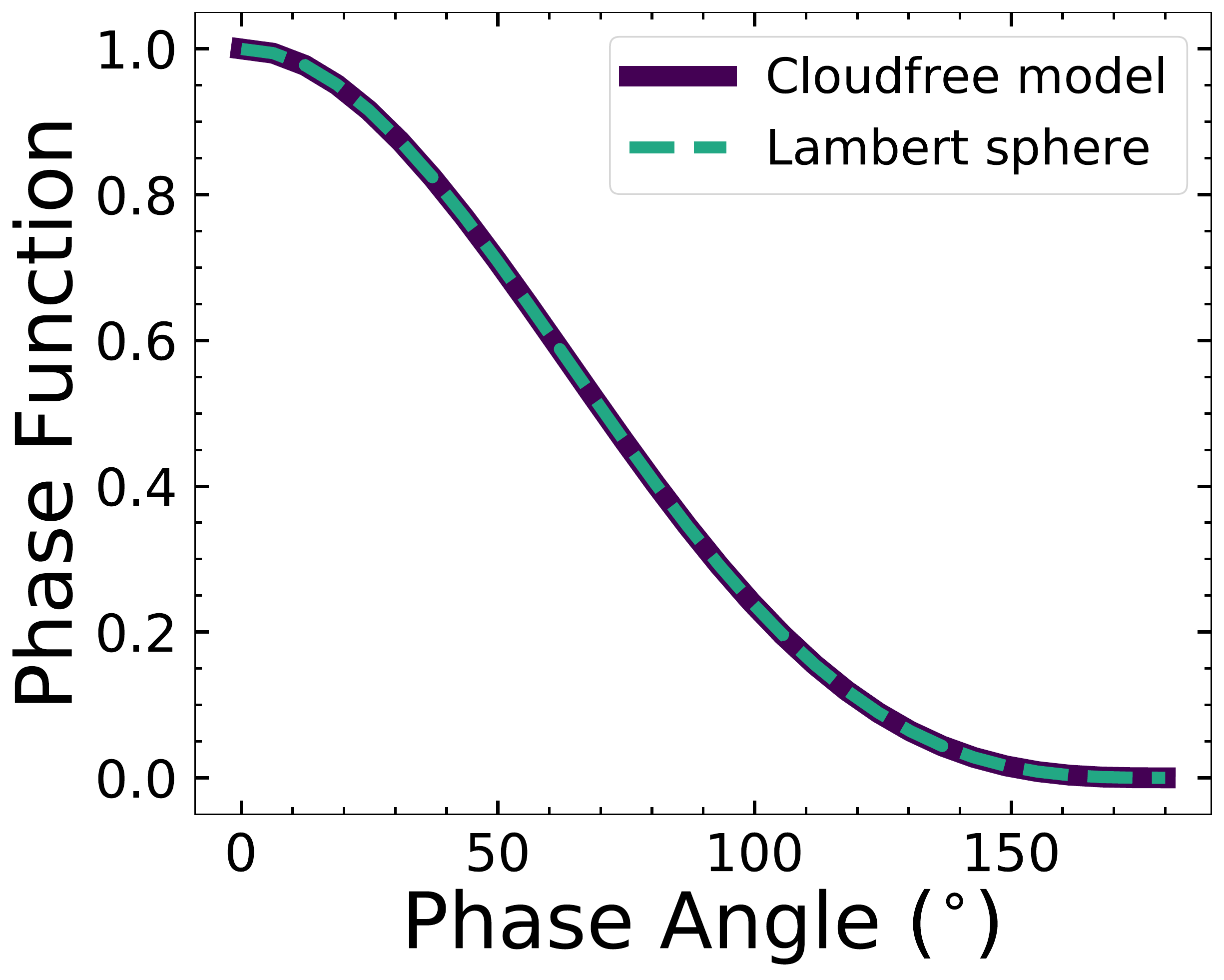} \end{center}
\caption{Comparing our model phase function to the analytic Lambertian phase function
(Equation \ref{lamb_phase}). No atmospheric absorption or scattering is present in
the forward model.  
} \label{lambert_test} \end{figure}

Figure \ref{lambert_test}
compares the model phase function with the analytic phase function and shows complete
agreement, confirming that our treatment of the surface is correct.

Previous work featuring the albedo model adopted here used a pre-defined atmospheric pressure grid. To accommodate the finite surface pressures of rocky planets as well as the various combinations of cloud parameters our retrievals will explore, we instead establish an adaptive method of determining the pressure grid.  Here, we divide the atmosphere into a pressure grid of $N_{\rm level}$, bounded by
$P = P_{\rm top}$ at the top of the atmosphere and $P = P_0$ at the
surface. In a cloud-free scenario, we simply divide the atmosphere evenly in log-$P$ space.  

For our simulations that include a single cloud deck, we adaptively determine the pressure value at each level depending on the location, thickness, and optical depth of the cloud.  The quantities that define the cloud deck
are $p_{\rm t}$, the cloud-top pressure, $dp$, the atmospheric pressure across the cloud, and $\tau$, the cloud optical depth.  We begin by assigning a number of layers to the cloud, imposing two conditions: (1) there should be at least three model pressure layers to each atmospheric pressure scale height (${\tt perH} =3$), and (2) the cloud optical depth in a layer must remain below at most 5 (${\tt maxtau} = 5$). This allows us to avoid any one layer from spanning a large extent within the atmosphere, and also avoids cloud layers that have extremely large scattering optical depths.

\begin{figure}[ht!] \begin{center}
\includegraphics[width=0.41\textwidth]{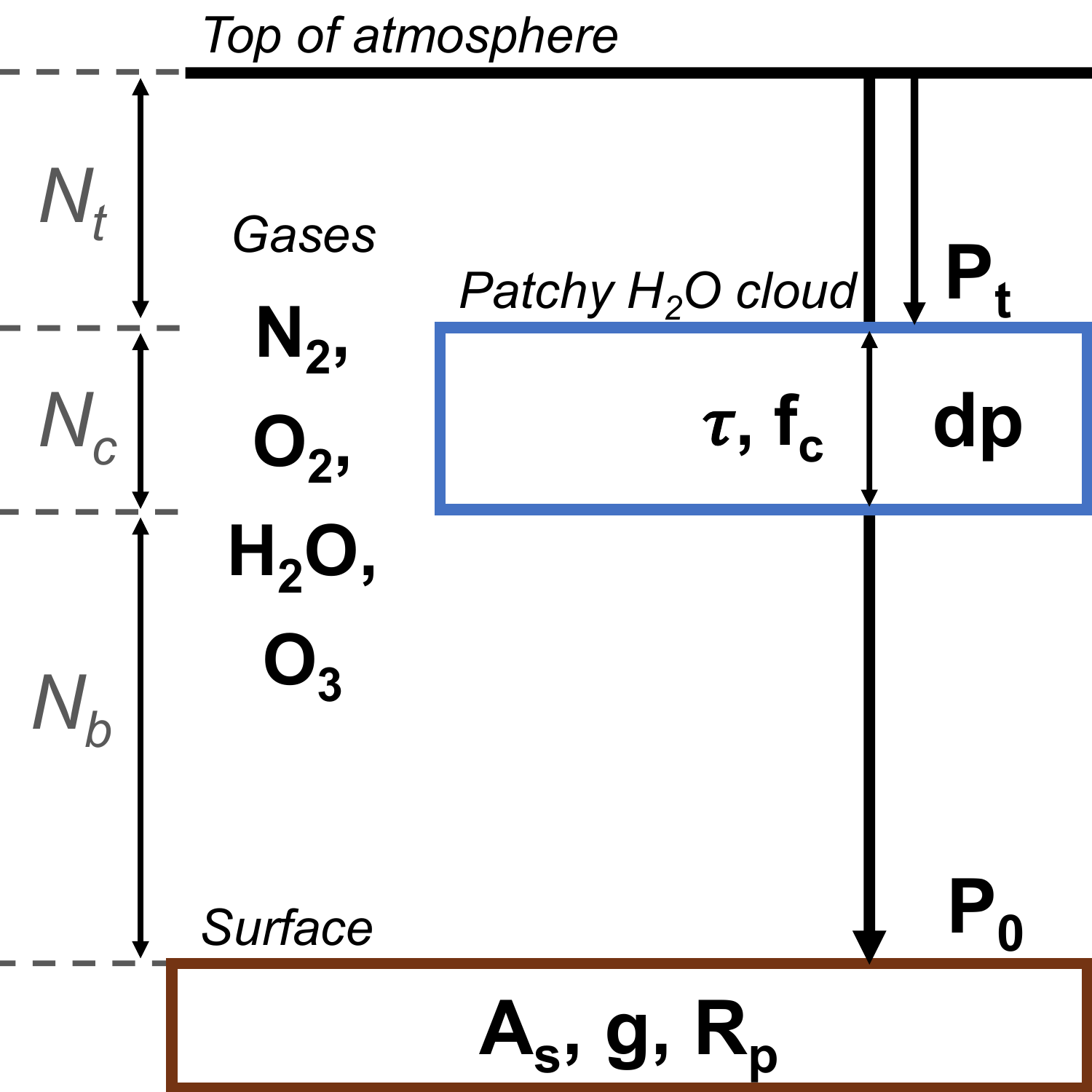} \end{center}
\caption{Illustrative schematic of our model atmosphere's structure. The atmosphere has $N_t + N_c + N_b$ layers.
Table \ref{tab:params} lists the definitions, fiducial values, and priors of the presented parameters.} \label{fig:atmscheme} \end{figure}

When beginning our gridding process, we propose an initial number of cloud layers, $N_{\rm{c}} = {\tt
  perH} \times {\tt numH}$, where $ {\tt numH} =
\ln{\frac{p_{\rm t}+dp}{p_{\rm t}}}$ is the number of e-folding distances through
the cloud (serving as a proxy for scale height). The aerosol optical depth for each pressure layer within the cloud would then simply be $\Delta\tau = \frac{\tau}{N_{\rm c}}$. However, if $\Delta \tau > {\tt maxtau}$, we adjust 
the cloud resolution by increasing $N_{\rm{c}}$ by a factor of $\frac{\Delta\tau}{{\tt maxtau}}$ and then round up to the nearest integer. In other words, we increase the resolution of the pressure grid through the cloud until the layer optical depth is under {\tt maxtau}. We determine successive pressure level values through the cloud with $p[i] = p[i-1] + \Delta\ln{p}$, where $\Delta\ln{p}
= \frac{\ln{(p_{\rm t} + dp)} - \ln{p_{\rm t}}}{N_{\rm c}}$, starting from the top of the cloud. We divide the remaining $N_{\rm level} - N_{\rm c}$ levels in uniform $\ln{p}$ space on either side of the cloud, weighted by the number of pressure scale heights above ($N_{\rm t}$) and below ($N_{\rm b}$) the cloud.  
Figure \ref{fig:atmscheme} visualizes the three portions of the atmosphere. 

\begin{figure*}[ht!]
\centering
\includegraphics[width = 0.49\textwidth]{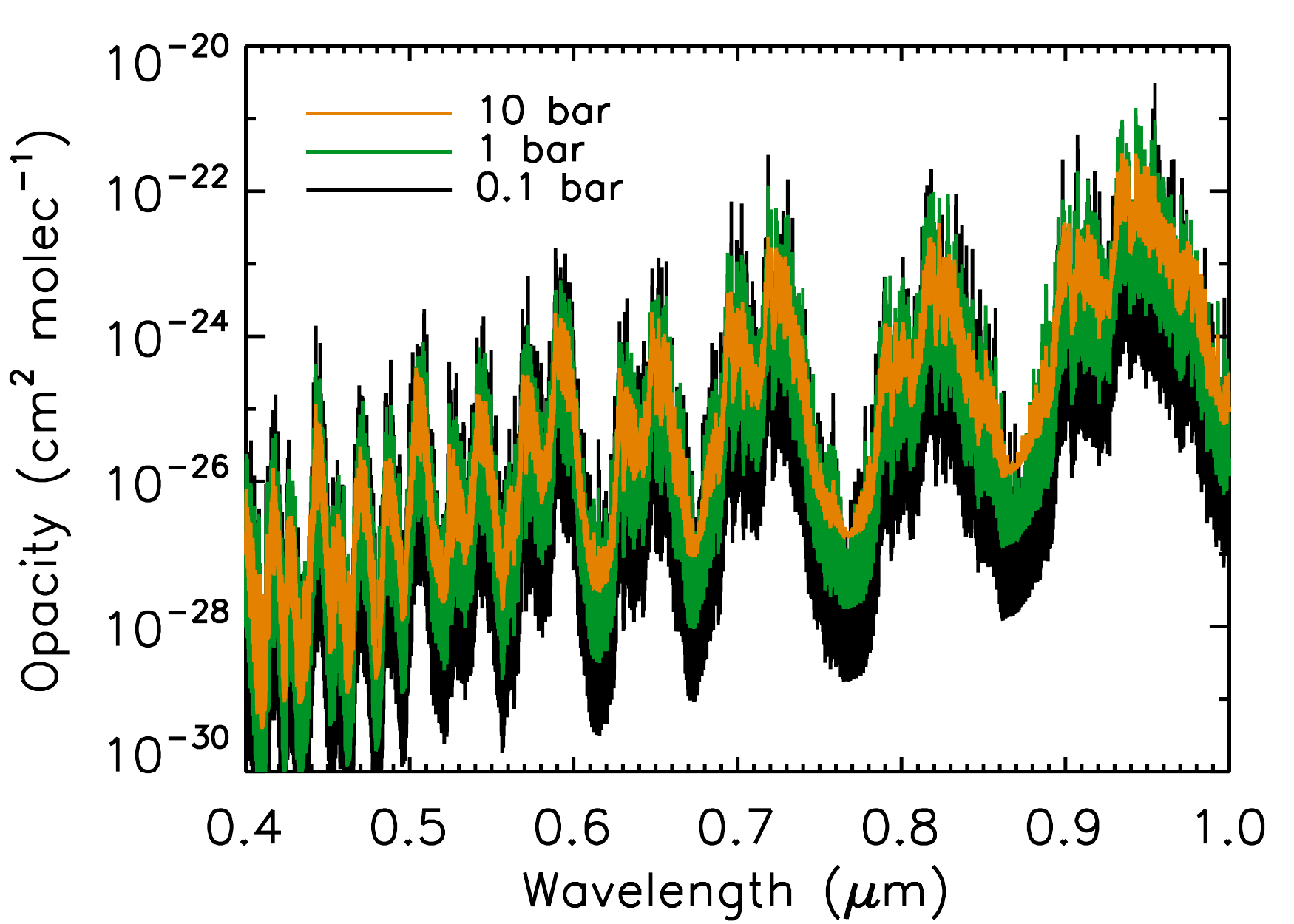}
\includegraphics[width = 0.49\textwidth]{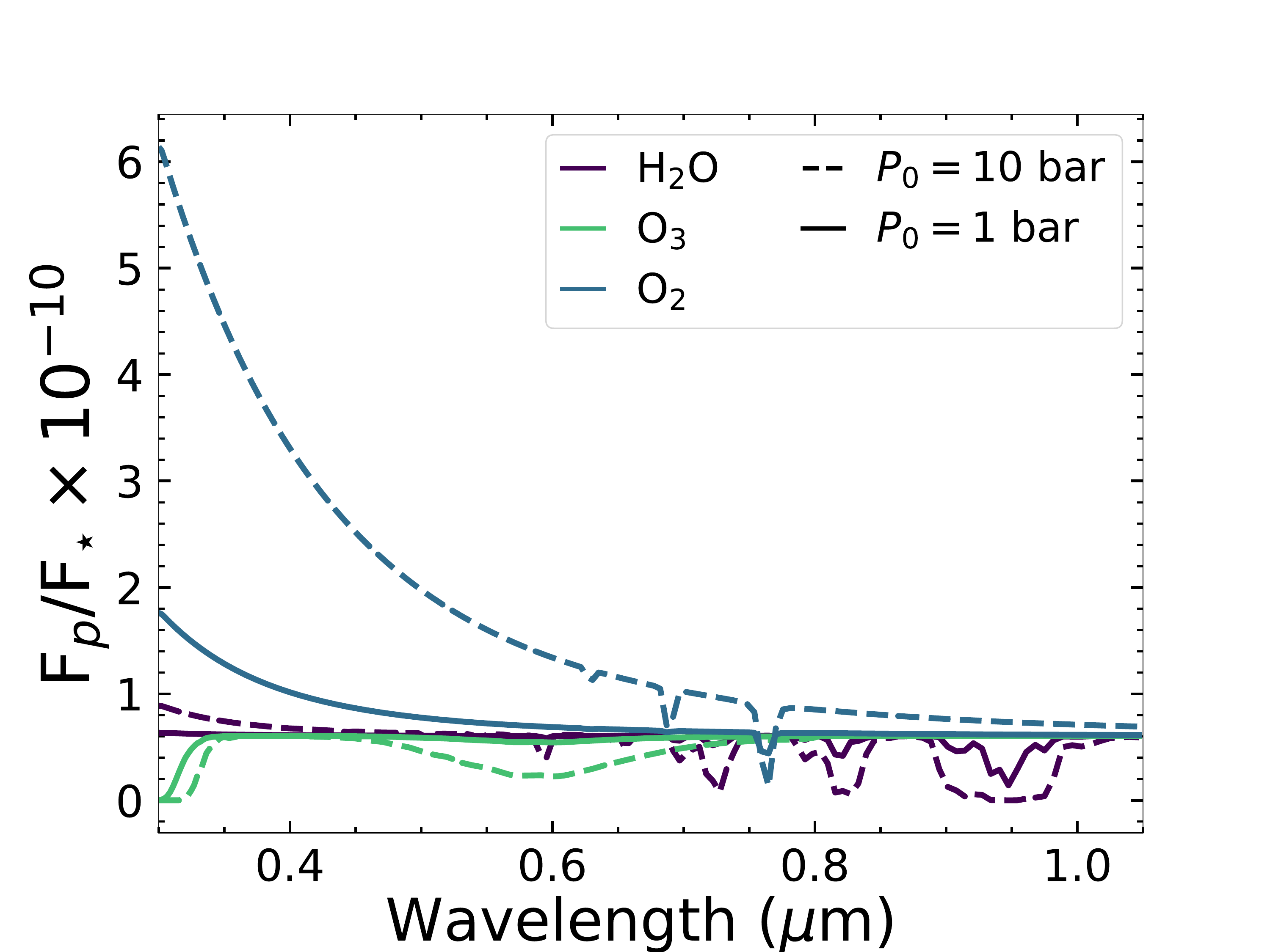}

\caption{{\it Left:} High resolution (1 cm$^{-1}$) \water\ opacities from 0.4-1.0 $\mu$m at three different pressures: 0.1 bar, 1 bar, and 10 bars. {\it Right:} Absorption features in a $R = 140$ spectrum from 0.3 - 1.05 $\mu$m of \water, \oxygen, and \ozone\ at fiducial mixing ratios listed in Table \ref{tab:params} at $P = 1$ bar and $P = 10$ bar.  For each spectrum here, the atmosphere only contains the stated molecule and a radiatively inactive filler gas to match the pressure. }
\label{fig:pdep}
\end{figure*}

For simplicity, we assume an isothermal atmosphere (at $T=250$~K), as temperature has little effect on the reflected-light spectrum \citep{robinson2017}.
Pressure, however, has a strong impact on molecular opacities, as
seen in Figure \ref{fig:pdep}. We incorporated high-resolution pressure-dependent opacities for all molecules in our atmosphere. The absorption opacities are generated line-by-line from the HITRAN2012 line list \citep{rothman2013} for seven orders of magnitude in pressure ($10^{-5} - 10^{2}$ bar) at $T$,  spanning our entire wavelength range at $<1$ cm$^{-1}$ resolution. Figure \ref{fig:pdep} also illustrates how absorption features of \water, \oxygen, and \ozone~change when in an atmosphere of 1~bar versus one of 10~bar. 

We interpolate our high-resolution opacity tables to the slightly lower resolution of the forward model in order to maintain short model runtimes while not affecting the accuracy of the output spectra. For each model layer, we interpolate over the opacities from our table given the pressure. The chemical abundances in our forward model atmosphere are constant as a function of pressure, and we also adopt a uniform acceleration due to gravity. 

We have also added an option to include partial cloudiness across a planetary disk, whose fractional coverage is described by $f_{c}$. To mimic partial cloudiness as we see on Earth, we call the forward model twice. We use the same set of atmospheric and planetary parameters for both calls, except for the cloud optical depth. ``Cloudy'' is the call that has a non-zero cloud optical depth, while ``cloud-free'' is the call where we set cloud optical depth to zero. Each call returns a geometric albedo spectrum, and we combine the two sets with the fractional cloudiness parameter such that the combined spectrum follows $f_c \times {\rm cloudy} + (1 - f_c) \times$~{cloud-free}.  

\subsection{Albedo Model Fiducial Values and Validation} \label{subsec:fiducialvals}
The generalized three-dimensional albedo model described above can simulate reflected-light spectra of a large diversity of planet types, spanning solid-surfaced worlds to gas giants with a variety of prescribed atmospheric compositions.  For the present study, however, we choose to focus on Earth-like worlds, which are described in detail below.  Thus, we define a set of fiducial model input parameters that are designed to mimic Earth and thereby enable us to generate simulated observational datasets for an Earth twin.

Table \ref{tab:params} summarizes the fiducial model parameter values adopted for our Earth twin.  Also shown are the priors for these parameters, which we use when performing retrieval analyses.  For an Earth-like setup, the surface atmospheric pressure is $P_0 = 1$ bar and we adopt a surface albedo of $A_{\rm s} = 0.05$, which is representative of mostly ocean-covered surface.  We adopt a uniform acceleration due to gravity of $g = 9.8$ m~s$^{-2}$ and set the planetary radius to $R_{\oplus}$.  For convenience, we sometime refer to these four variables ($P_0$, $A_{\rm s}$, $g$, and $R_{\rm p}$) as the bulk planetary and atmospheric parameters.

\begin{figure*}[ht]
\centering
    \includegraphics[width=0.49\linewidth]{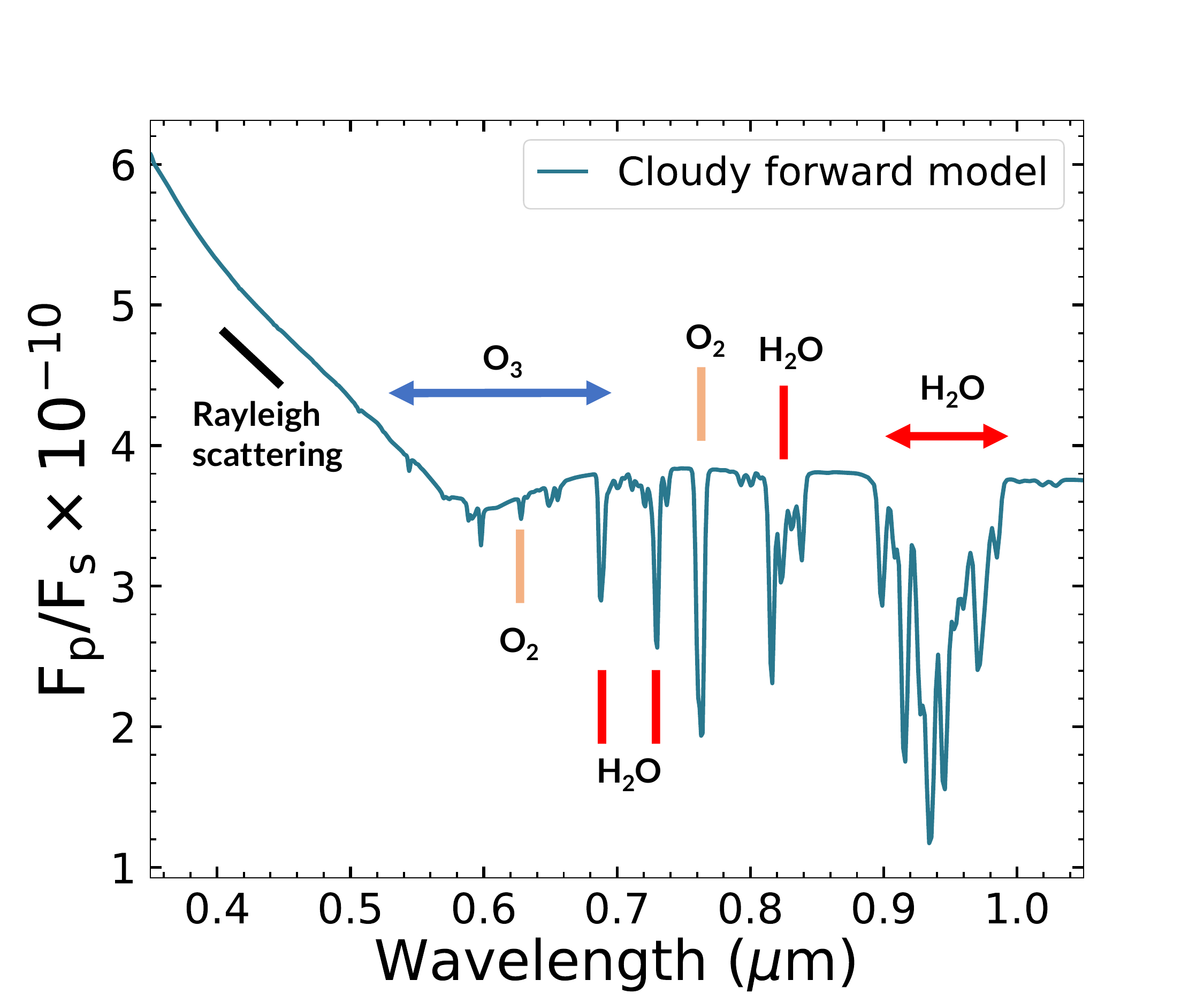}\hfil
    \includegraphics[width=0.49\linewidth]{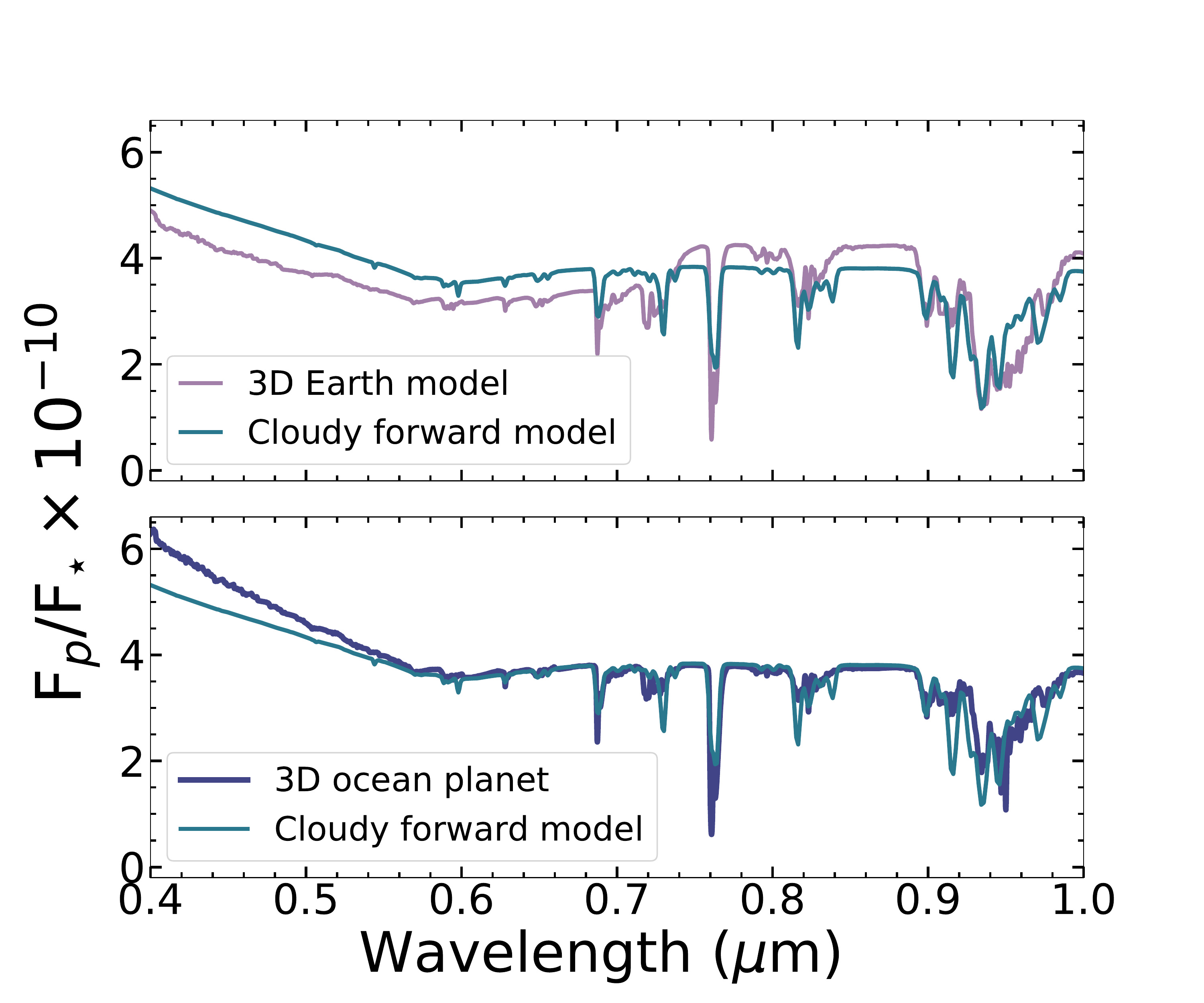}
\caption{\textit{Left:} The spectrum generated with the forward model in this study using fiducial values from Table \ref{tab:params}. Key spectral features from the atmospheric species in our model are labeled. \textit{Right, top:} Comparison of the cloudy forward model in this study using fiducial values from Table \ref{tab:params} to a spectrum from a more computationally complex three-dimensional (3D) forward model of Earth at full phase described in \citet{robinson2011}.  \textit{Right, bottom:} Comparison of the cloudy forward model to a spectrum of a planet generated using the 3D model from \citet{robinson2011} that is like Earth except it only has ocean coverage.}
\label{fig:3dcompare}
\end{figure*}

We focus on molecular absorption due to \water, \ozone, and \oxygen. While our albedo model includes opacities from \methane\ and \cotwo\ as well, we omit these two species as the reflected-light spectrum of Earth in the visible contains no strong features for these molecules. The input values for the molecular abundances (or volume mixing ratios) are \water\ = $3\times10^{-3}$, \ozone\ = $7\times10^{-7}$, and \oxygen = 0.21. These abundance values are based on column weighted averages from a standard Earth model atmosphere with vertically-varying gas mixing ratios \citep{mcclatchey1972}. The primary Rayleigh scatterer and background gas in our fiducial model is \nitrogen, whose abundance makes up the remainder of the atmosphere after all other gases are accounted for (i.e., roughly 0.79). Rayleigh scattering is treated according to \citet{hansen1974} with constants to describe the scattering properties of \nitrogen, \oxygen, and \water\ from \citet{allen2000}.  We do not include polarization or Raman scattering effects.

Our cloud model was designed to be minimally parametric while still enabling us to sufficiently reproduce realistic spectra of Earth.  Our single-layer gray \water\ cloud has $\bar{\omega} = 1$ and $\bar{g} = 0.85$, which are characteristic of water clouds across the visible range.  These two parameters were fixed to minimize retrieval model complexity, as we believe that water is the most likely condensate for worlds in the Habitable Zone.  Nevertheless, future studies may not wish to assume values of $\bar{\omega}$ and $\bar{g}$ {\it a priori}.  Cloud top pressure ($p_{\rm t}$) and fractional coverage ($f_{c}$) are set at 0.6 bar and 50\%, respectively, which are roughly consistent with observations of optically thick cloud coverage on Earth \citep{stubenrauch13}.  Cloud thickness ($dp$) and optical depth ($\tau$) were set to 0.1~bar and 10, respectively, based on results from the MODIS instrument (http://modis-atmos.gsfc.nasa.gov) used in \citet{robinson2011}.  

With fiducial values chosen, we validate our forward model against a simulated high-resolution disk-integrated spectrum of Earth at full phase, as shown in Figure~\ref{fig:3dcompare}. The comparison spectrum is produced by the NASA Astrobiology Institute's Virtual Planetary Laboratory (VPL) sophisticated 3D line-by-line, multiple scattering spectral Earth model \citep{robinson2011}.  The \citet{robinson2011} tool can simulate images and disk-integrated spectra of Earth from the ultraviolet to the infrared. It has been validated against observations at visible wavelengths taken by NASA's {\it EPOXI} mission \citep{robinson2011} and NASA's {\it LCROSS} mission \citep{robinsonetal2014}.

Features of the \citet{robinson2011} model include Rayleigh scattering due to air molecules, realistic patchy clouds, and gas absorption from a variety of molecules, including \water, \cotwo, \oxygen, \ozone, and \methane. Surface coverage of different land types (e.g., forest, desert) is informed by satellite data, and water surfaces incorporate specular reflectance of sunlight. A grid of thousands of surface pixels are nested beneath a grid of 48 independent atmospheric pixels, all of equal area.  For each surface pixel, properties from the overlying atmospheric pixels are used as inputs to a full-physics, plane-parallel radiative transfer solver--- the Spectral Mapping Atmospheric Radiative Transfer (SMART)  model \citep{meadows1996}.  Intensities from this solver are integrated over the pixels with respect to solid angle, thereby returning a disk-integrated spectrum.  

The sophistication of the \citet{robinson2011} model makes it unsuitable to retrieval studies, however, as model runtimes are measured in weeks for the highest-complexity scenarios.  This, in part, justifies our adoption of a minimally-parameteric albedo model (whose runtime is measured in seconds).  Furthermore, as in Figure~\ref{fig:3dcompare}, our efficient albedo model reproduces all of the key features of the \citet{robinson2011} model.  The most notable differences---that the efficient model, as compared to the \citet{robinson2011} model, is more reflective in the blue and less reflective in the red---are simply due to our adoption of a gray surface albedo.  Land and plants, which cover roughly 29\% of Earth's surface, are generally more reflective in the red than in the blue. Figure \ref{fig:3dcompare} also compares a spectrum from our forward model against a spectrum of a partially clouded ocean planet generated with the \citet{robinson2011} model. This ocean world is identical to Earth except for the fact that its surface is covered entirely by an ocean, with no land present. The surface albedo in the ocean model is gray beyond 500~nm; shortward of this the reflectivity increases, likely leading to the discrepancy in our comparison at the bluest wavelengths. Still, with a more accurate match to a planet that has a nearly gray albedo through the visible, we consider our assumption of gray surface albedo to be the main reason for the discrepancies when compared to the \citet{robinson2011} realistic model. 

Finally, in our albedo model we set 100 facets for the visible hemisphere and calculate a high-resolution geometric albedo spectrum at 1000 wavelength points between 0.35$\mu$m and 1.05$\mu$m. Like \citet{lupu2016}, we only consider a planet at full phase ($\alpha = 0^{\circ}$). While direct imaging missions will not obtain observations of exoplanets at full phase, this assumption makes little difference for our results as we are not computing integration times and only work in SNR space.  Additionally, as \citet{nayak2017} followed up \citet{lupu2016} by retrieving phase information from giant planets in reflected light, we anticipate performing a similar expansion in the future.  Our forward model has 61 pressure levels in an isothermal atmosphere of 250 K, bounded below by a reflective surface. The top of the atmosphere is set at $P_{\rm top} = 10^{-4}$ bar.

\begin{deluxetable*}{llll}
\tablecaption{List of the 11 retrieved parameters in the complete cloudy model, their descriptions, fiducial input values, and corresponding priors. \label{tab:params}}
\tablewidth{0pt}
\tabletypesize{\scriptsize}
\tablehead{Parameter	&  Description	& Input & Prior}
\startdata
$\log{P_0}$ (bar)	    & Surface pressure              & $\log{(1)}$               & [-2,2]        \\
$\log{\rm H_2 O}$       & Water vapor mixing ratio      & $\log{(3\times10^{-3})}$  & [-8,-1]       \\
$\log{\rm O_3}$         & Ozone mixing ratio            & $\log{(7\times10^{-7})}$  & [-10,-1]      \\
$\log{\rm O_2}$         & Molecular oxygen mixing ratio & $\log{(0.21)}$            & [-10,0]       \\
$R_{p}$ (R$_{\oplus}$) & Planet radius             & $1$                       & [0.5, 12]     \\
$\log{g}$ (m s$^{-2}$)  & Surface gravity               & $\log{(9.8)}$             & [0,2]         \\
$\log{A_s}$             & Surface albedo                & $\log{(0.05)}$            & [-2, 0]       \\
$\log{p_t}$ (bar)       & Cloud top pressure            & $\log{(0.6)}$             & [-2,2]        \\
$\log{dp}$ (bar)        & Cloud thickness               & $\log{(0.1)}$             & [-3,2]        \\
$\log{\tau}$            & Cloud optical depth           & $\log{(10)}$              & [-2,2]        \\
$\log{f_c}$             & Cloudiness fraction           & $\log{(0.5)}$             & [-3,0]        \\
\enddata
\end{deluxetable*}

\begin{deluxetable*}{l|l|l}
\tablecaption{Simulated data sets. \label{tab:runs}}
\tablewidth{0pt}
\tabletypesize{\scriptsize}
\tablehead{	{}&  $R = 70,\ R = 140$	& \textit{WFIRST} Rendezvous \tablenotemark{a}}
\startdata
Wavelength ($\mu$m) 		& 0.4 -- 1.0	    & 	0.506, 0.575\tablenotemark{b}, $R = 50$: 0.6 -- 0.97\tablenotemark{c}   \\
Data quality            	& SNR$_{\rm 550nm} = 5,\ 10,\ 15,\ 20$         	&  SNR$_{\rm 600nm} = 5,\ 10,\ 15,\ 20$            \\
\enddata
\tablecomments{We do not randomize the noise for any of the data sets.}
\tablenotetext{a}{Using \textit{WFIRST} Design Cycle 7 values from \\ \url{https://wfirst.ipac.caltech.edu/sims/Param\textunderscore db.html}}
\tablenotetext{b}{The first photometric band is centered on 0.506 $\mu$m and covers 0.48--0.532 $\mu$m. The second photometric band is centered on 0.575 $\mu$m and covers 0.546--0.6 $\mu$m. We assume $100\%$ transmission.}
\tablenotetext{c}{We combine three integral field spectrograph bands into one at $R = 50$ from 0.6 $\mu$m to 0.97 $\mu$m. Separated, they are 0.6--0.72 $\mu$m, 0.7--0.84 $\mu$m, and 0.81-0.97 $\mu$m.}
\end{deluxetable*}

\begin{deluxetable*}{llll}
\tablecaption{Four cumulative models for retrieval validation, as described in Section \ref{buildup}. \label{tab:buildup}}
\tablewidth{0pt}
\tabletypesize{\scriptsize}
\tablehead{\multicolumn{2}{l}{Model Variant}	&  Retrieved Parameters	& N$_{\rm param}$}
\startdata
I 		& Surface conditions	& $P_0$, $A_{\rm s}$				& 2	\\
II 		& $+$ Bulk properties	& $P_0$, $A_{\rm s}$, $g$, $R_{\rm p}$	& 4	\\
III		& $+$ Gas mixing ratios	& $P_0$, $A_{\rm s}$, $g$, $R_{\rm p}$	& 7	\\
    	&						& \water, \oxygen, \ozone	&	\\
IV		& $+$ Cloud properties	& $P_0$, $A_{\rm s}$, $g$, $R_{\rm p}$	& 11\\
    	&						& \water, \oxygen, \ozone	&	\\
    	&						& $p_{\rm t}$, $dp$, $\tau$, $f_{\rm c}$&	\\
\enddata
\tablecomments{See Table \ref{tab:params} for the corresponding definition and prior of each parameter. Model IV represents the full suite of parameters and can serve as a reference for the fixed parameters in Models I through III.}
\end{deluxetable*}

\subsection{Retrieval Setup and Noise Model} \label{subsec:retrieval}
We convert a high resolution geometric albedo spectrum to a synthetic planet-to-star flux ratio spectrum given the resolution of an instrument and a noise model. We then apply a Bayesian inference tool on the synthetic data set to sample the posterior probability distributions of the forward model input parameters. To perform Bayesian parameter estimation, we utilize the open-source affine invariant Markov-Chain Monte Carlo (MCMC) ensemble sampler {\tt emcee} \citep{goodman2010,foreman2013}. Ensemble refers to the use of many chains, or walkers, to traverse parameter space; as a massively parallelized algorithm, it is computationally efficient.  Affine-invariance refers to the invariant performance under linear transformations of parameter space, enabling the algorithm to be insensitive to parameter covariances \citep{foreman2013}. With a cloudy retrieval, we can expect complex correlations that a sampler should be able to reveal. As it is more agnostic to the shape of the posterior, we choose {\tt emcee} following \citet{nayak2017} over {\tt Multinest}, another sampler \citet{lupu2016} considered that yielded consistent results.  The albedo model is coded in Fortran; we convert it into a Python-callable library with the F2PY package. Each call to the forward model takes approximately 10 seconds of clock time on an 8-core processor. To visualize the MCMC results, we utilize the {\tt corner} plotting package developed by \citet{corner}.

Table \ref{tab:params} lists the priors for our parameters. We offer a generous range on the molecular abundances; we allow \oxygen\ in particular to be able to dominate the atmospheric composition. Our choice of radius range (0.5--12 $\rm R_\oplus$) reflects the range of of planetary sizes from Mars to Jupiter. Also, when performing retrievals, we impose two limiting conditions to maintain physical scenarios. First, we limit the mixing ratio of \nitrogen, $f_{\rm N_2} =  1 - \sum (\rm gas\ abundances)$, to be between 0 and 1. Second, for the cloud pressure terms, we reject any drawn value that does not satisfy $10^{p_{\rm t}} + 10^{dp} < 10^{P_0}$ (i.e., that the cloud base cannot extend below the bottom of the atmosphere). Note that for the purposes of the retrieval, we consider pressures in log space. 

We simulate noise in our observations following the expressions given in \citet{robinson2016}.  For simplicity, we include only read noise and dark current, as \citep{robinson2016} showed that detector noise will be the dominant noise source in \textit{WFIRST}-type spectral observations of exoplanets.  Detector and instrument parameters for the HabEx and LUVOIR concepts are only loosely defined, and advances in detector technologies for these missions may move observations out of the detector-noise dominated regime.  In the detector-noise dominated regime the signal-to-noise ratio is simply,
\begin{equation}
{\rm SNR} = \frac{c_p \times t_{\rm int}}{\sqrt{(c_d + c_r) \times t_{\rm int}}} \,
\label{snr}
\end{equation}
where $t_{\rm int}$ is the integration time, $c_p$ is planet count rate, and $c_d$ is the dark noise count rate, and $c_r$ is the read noise count rate. More rigorously, it can be shown that, at constant spectral resolution, ${\rm SNR} \propto q \mathcal{T} A_g \Phi(\alpha) B_\lambda \lambda$, where $q$ is the wavelength-dependent detector quantum efficiency, $\mathcal{T}$ is throughput, and $B_\lambda$ is the host stellar specific intensity (taken here as a Planck function at the stellar effective temperature). We use a stellar temperature of 5780 K for the blackbody. When the SNR at one wavelength is specified, this scaling implies that the the calculation of the signal-to-noise ratio at other wavelengths is independent of the imaging raw contrast of the instrument. We can expect the noise at the redder end of our range to be large, as the detector quantum efficiency (taken to be appropriate for the {\it WFIRST}/CGI) rapidly decreases. Since we treat only SNR rather than modeling exposure times, the exact mix of noise sources is not relevant (so that, e.g., dark current and readnoise are indistinguishable). The key relevant properties of the noise model is that it is uncorrelated between spectral channels and its magnitude only depends on wavelength via a dependence on point spread function area \citep[][their Equation 26]{robinson2016}, which will be true for detector-limited cases but may not be true for large-aperture instruments limited by speckle noise.

\begin{figure}[ht!] \begin{center}
\includegraphics[width=0.45\textwidth]{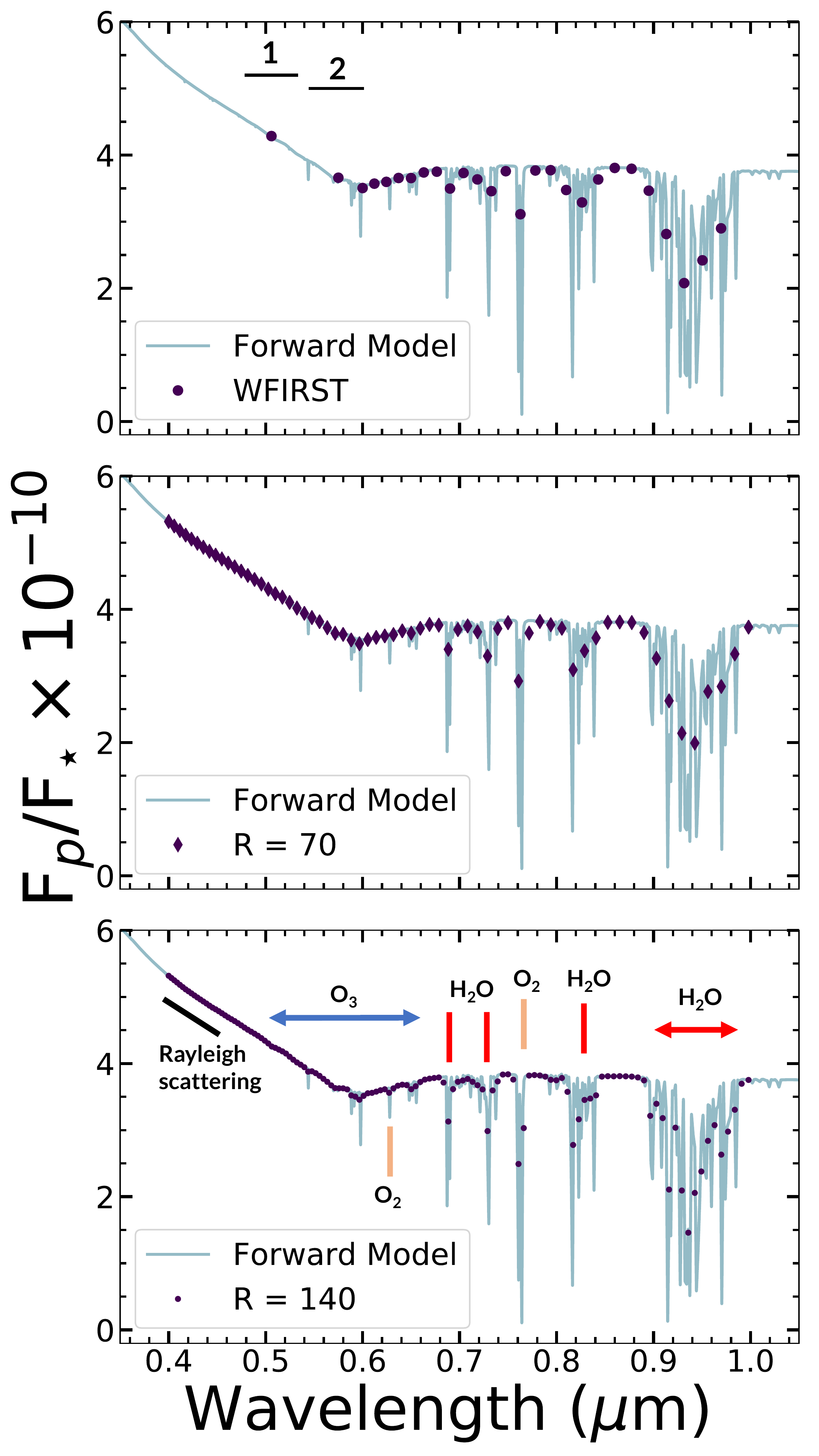} \end{center}
\caption{The high resolution (1000 wavelength points from $0.35 - 1.05\ \mu$m) forward model spectrum, overplotted with simulated \textit{WFIRST} rendezvous, $R = 70$, $R = 140$ data, from top panel to bottom. Key spectral features for atmospheric gases in our model are labeled. In the top panel, ``1'' and ``2'' mark the span of the \textit{WFIRST} Design Cycle 7 filters (see Table \ref{tab:runs}). } \label{fig:labeledSpec} \end{figure}

For our study, we will consider multiple wavelength resolutions, $R$, and SNRs. Working in SNRs (instead of integration times) makes our investigations independent of telescope diameter, target distance, and other system-specific or observing parameters. Because the SNR is dependent on wavelength, we reference our values to be at V-band (550 nm) for all resolutions for HabEx/LUVOIR.  Since the {\it WFIRST}/CGI spectrograph is currently planned to only extend to 600 nm at the blue end, we opt to reference our {\it WFIRST} SNRs to this wavelength.  Unlike previous studies \citep{lupu2016,nayak2017}, our simulated \textit{WFIRST} rendezvous data include two photometric points in the blue, which is consistent with current CGI designs. We set the SNR in the {\it WFIRST} filters to be equal to that at 600~nm.  

\begin{figure}[ht!] \begin{center}
\includegraphics[width=0.47\textwidth]{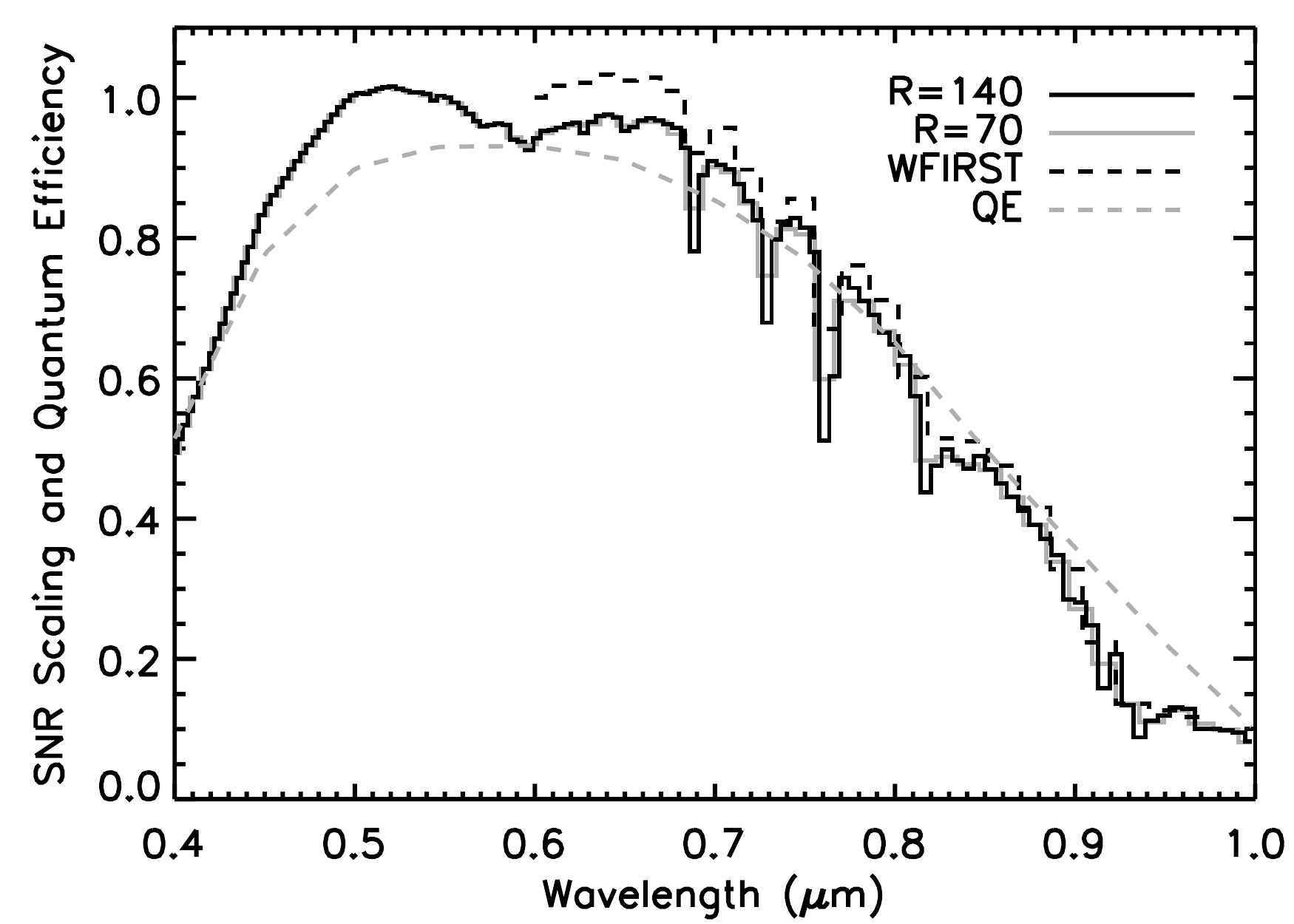} \end{center}
\caption{Scaling of SNR with wavelength for {\it WFIRST} rendezvous, $R = 70$, and $R = 140$ cases.  The {\it WFIRST} curve is normalized to unity at 600~nm while the $R=70$ and $R=140$ curves are normalized to unity at 550~nm, following our definite of simulation SNR at these respective wavelengths.  Also shown is the wavelength-dependent detector quantum efficiency (QE) that we adopt.} \label{fig:snrscaling} \end{figure}

Our simulation grid setup is shown in Table \ref{tab:runs}, where the spectral resolutions and SNRs assumed for different observing scenarios are indicated. Figure \ref{fig:labeledSpec} demonstrates the \textit{WFIRST} rendezvous scenario data along with $R = 70$ and $R = 140$ data points (for HabEx/LUVOIR) plotted over the forward model spectrum before noise is added. The scaling of SNR with wavelength for {\it WFIRST} rendezvous (normalized to unity at 600 nm) as well as our $R = 70$ and $R = 140$ cases (normalized to unity at 550~nm) is shown in Figure~\ref{fig:snrscaling}.  The impact of the host stellar SED sets the overall shape of the SNR scaling, with additional influence from atmospheric absorption bands detector as well as quantum efficiency effects (that have strong impacts at red wavelengths).  Thus, Figure~\ref{fig:snrscaling} can be used to translate our stated SNR to the SNR at any other wavelength (e.g., a SNR$=10$ simulation has a SNR in the continuum shortward of the 950~nm water vapor band of roughly $0.3\times10=3$).

When generating simulated data with a noise model, there are several options for handling the placement/sampling of the mock observational data points.  Previous studies \citep{lupu2016,nayak2017} have generated a single, randomized dataset for a given SNR.  The placement of a single spectral data point is determined by randomly sampling a Gaussian distribution whose width is determined by the wavelength-dependent SNR.  While this treatment can accurately simulate a single observational instance, it also runs the risk (especially at lower spectral resolution and SNR) of biasing retrieval results, as the random placement of only a small handful of spectral data points can significantly impact the outcome.  Given this, it is ideal to retrieve on a large number ($\gtrsim 10$) of simulated data sets at a given spectral resolution and SNR, where a comprehensive view of all the posteriors from the collection of instances will indicate expected telescope/instrument performance.  Unfortunately, given the large number of $R$/SNR pairs in our study (10) and the long runtime of an individual retrieval (of order 1~week on a cluster), running $\sim$10 noise instances for each of our $R$/SNR pairs is computationally unfeasible (requiring $\sim$100 weeks of cluster time).  Thus, we opt for an intermediate approach that maintains computational feasibility and avoids potential biases from individual noise instances.  Here, we run only a single noise instance at a given $R$/SNR pair, but we do not randomize the placement of the individual spectral points.  In other words, the individual simulated spectral points are placed on the ``true'' planet-to-star flux ratio point and are assigned error bars according to the SNR and noise model.  While this approach prevents having a small handful of randomized data points from biasing retrieval results, it does lead to likely optimistic results, especially at modest SNR (i.e., SNR$\sim$10), since data point randomization is, in effect, an additional ``noise'' source that we are omitting. This means that the posterior distributions will usually be centered on the true values in an unrealistic fashion. However, the width and shape of the posterior covariances will be representative of ‘real’ observations, so the fidelity of retrievals can be assessed. We keep this optimism in mind when discussing results in later sections; in particular, we compare the performance of retrievals on multiple noise instances of a subset of the cases we consider to the non-randomized case in Section \ref{subsec:multiple}.

\section{Retrieval Validation} \label{buildup}
Before using our framework on simulated data, we validate its accuracy and examine its performance. For this initial validation, we use non-randomized, wavelength-independent noise at a signal-to-noise ratio of 20 for a spectrum at a resolution of $R = 140$. Table \ref{tab:buildup} lists our four validation model variants, each increasing in complexity as we systematically explore how the addition of retrieved parameters influences the posterior distributions and correlations. In Model I, we fix all parameters except $P_0$ and $A_{\rm s}$.  In Model II, we add $g$ and $R_p$; in Model III, we then add in gases as retrieved parameters (\water, \ozone, \oxygen); and in Model IV, we add all cloud parameters. Incrementally increasing the number of free parameters (from 2 to 11) allows us to see the interconnections between them, and helps us understand how clouds can obscure our inferences. 

\begin{figure}[ht] \begin{center}
\includegraphics[width=0.45\textwidth]{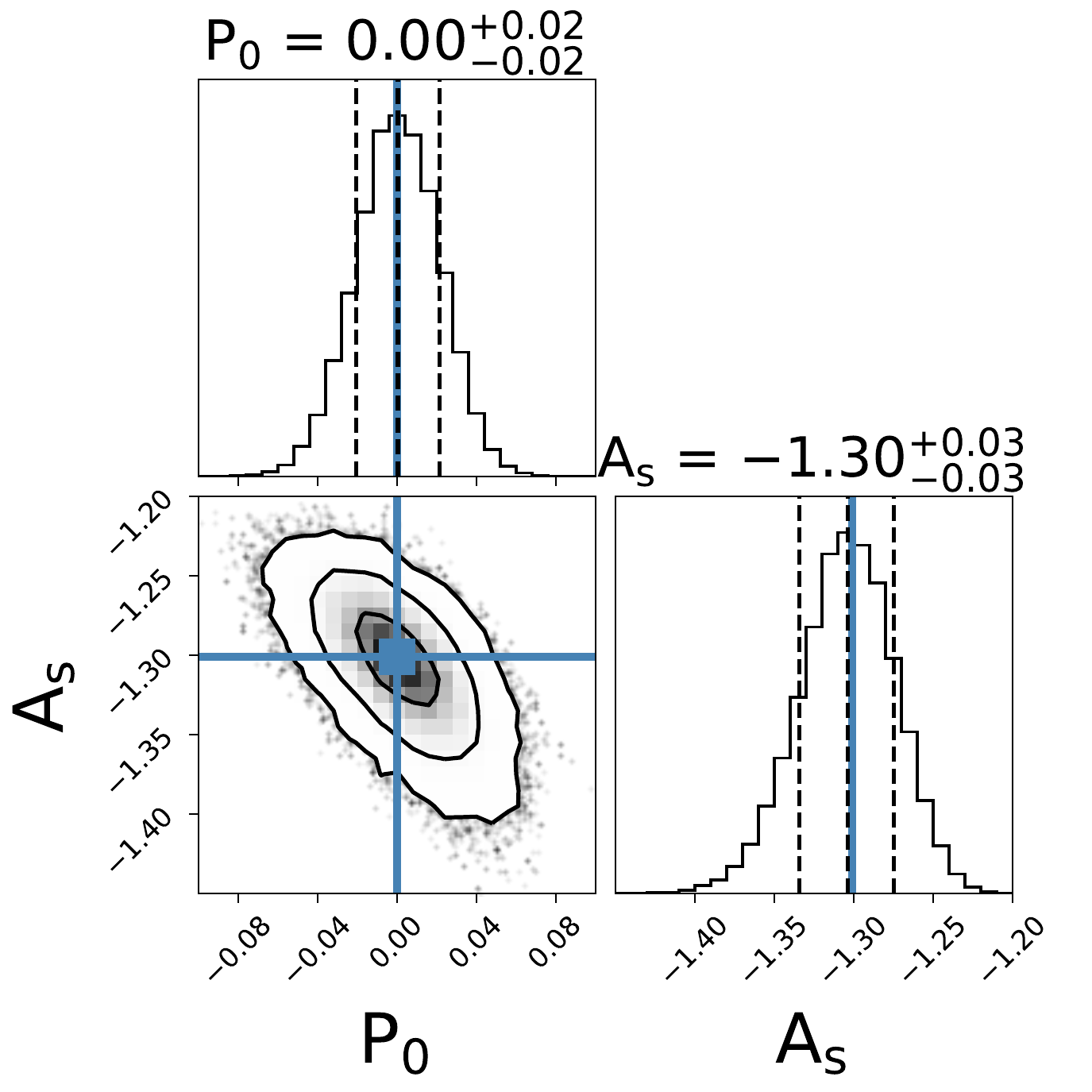} \end{center}
\caption{Posterior distributions of Model I from Table \ref{tab:buildup}, where we fix all parameters but $P_0$ and $A_{\rm s}$. We retrieve on $R = 140$, SNR$\ = 20$ data with wavelength-independent noise. Overplotted in solid light-blue color are the fiducial parameter values. The 2D marginalized posterior distribution, used in interpreting correlations, is overplotted with the 1-, 2-, and 3-$\sigma$ contours. Above the 1D marginalized posterior for each parameter, we list the median retrieved value with uncertainties that indicate the 68\% confidence interval. Dashed lines (left to right) mark the 16\%, 50\%, and 84\% quantiles.} \label{fig:b1stairs} 
\end{figure}

\begin{figure*}[ht] \begin{center}
\includegraphics[width=0.7\textwidth]{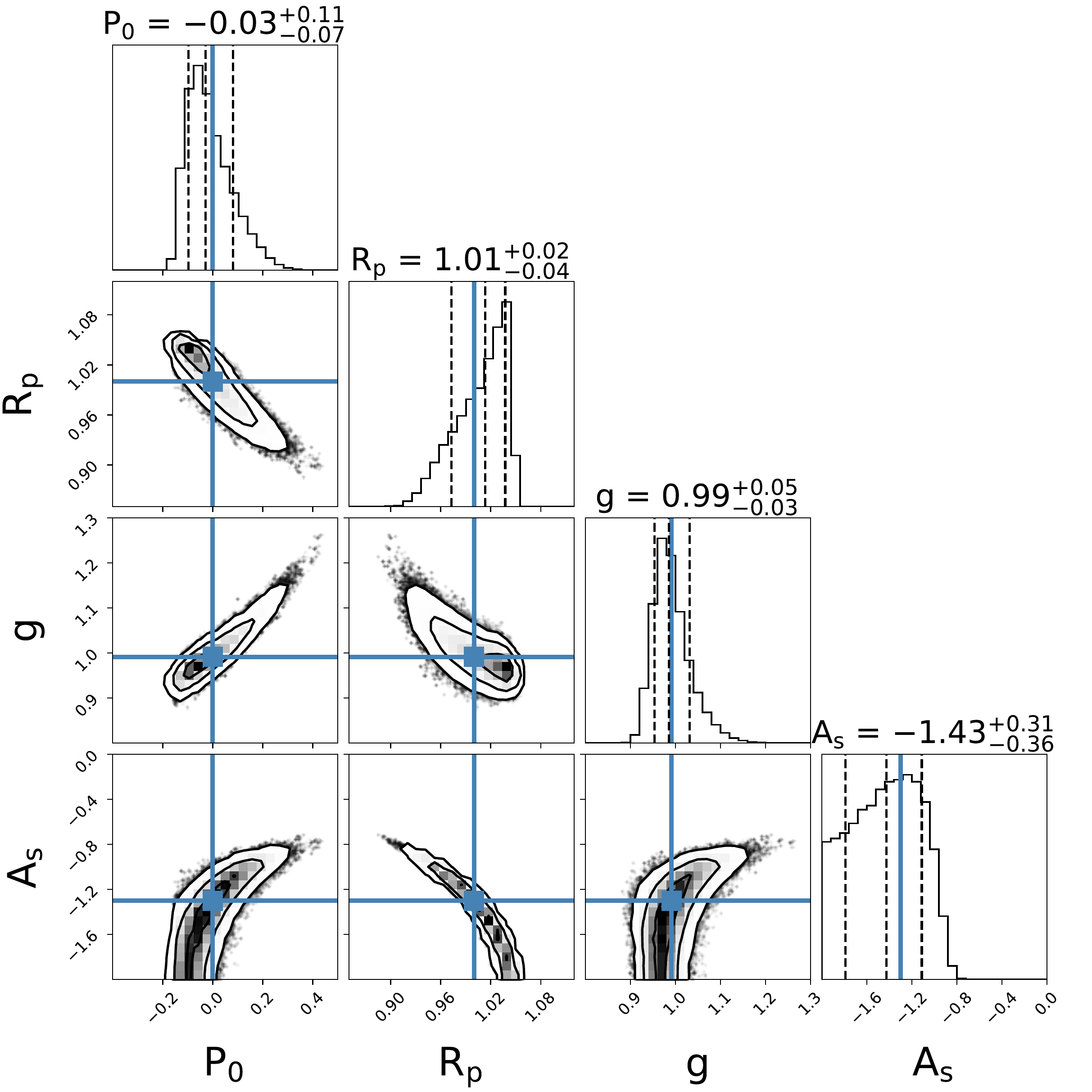} \end{center}
\caption{Posterior distributions of Model II from Table \ref{tab:buildup}, where we fix all parameters except for $P_0$, $A_{\rm s}$, $g$, and $R_{\rm p}$. We retrieve on $R = 140$, SNR$\ = 20$ data with wavelength-independent noise. Overplotted in solid light-blue color are the fiducial parameter values. The 2D marginalized posterior distribution, used in interpreting correlations, is overplotted with the 1-, 2-, and 3-$\sigma$ contours. Above the 1D marginalized posterior for each parameter, we list the median retrieved value with uncertainties that indicate the 68\% confidence interval. Dashed lines (left to right) mark the 16\%, 50\%, and 84\% quantiles. } \label{fig:b2stairs} 
\end{figure*}

\begin{figure*}[ht] \begin{center}
\includegraphics[width=\textwidth]{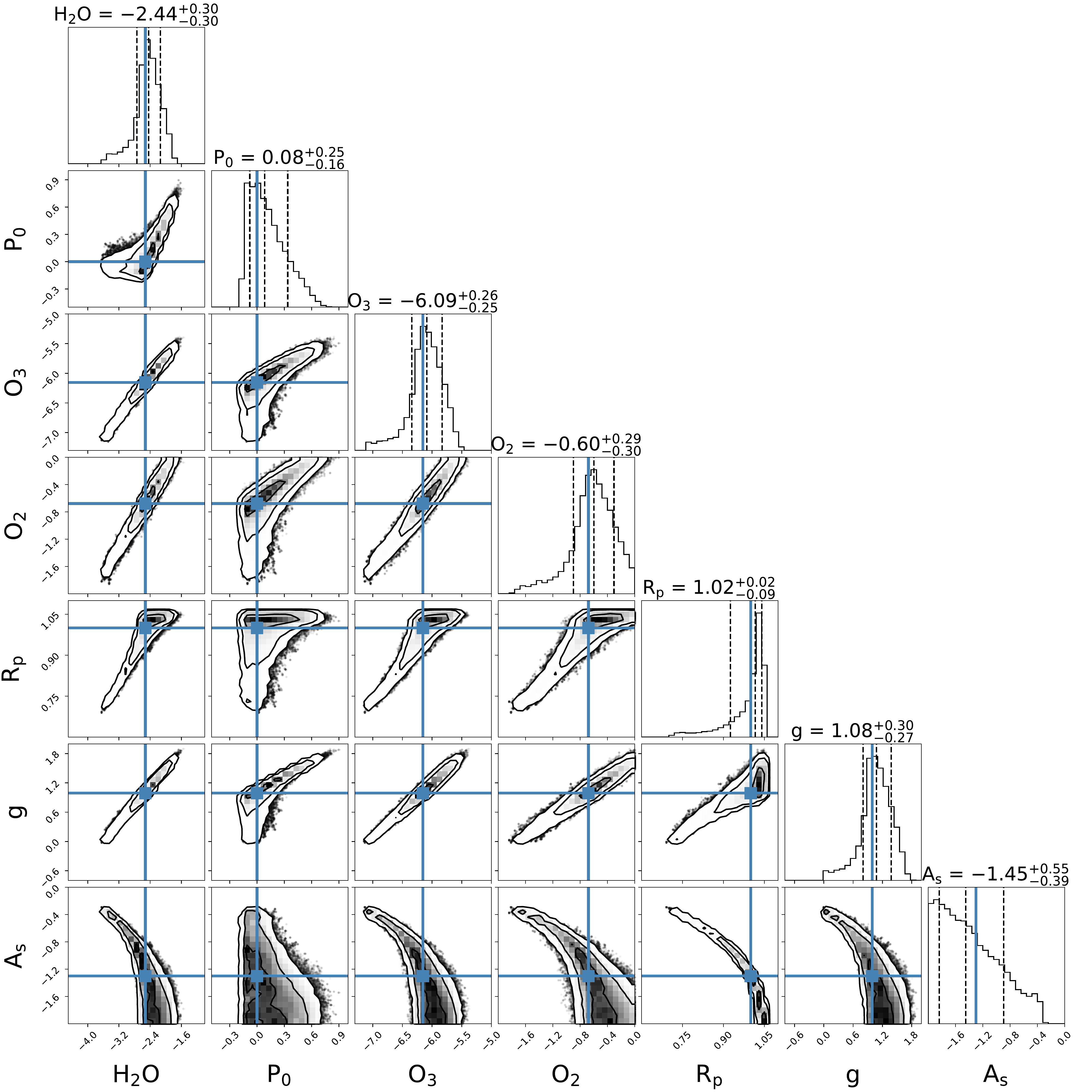} \end{center}
\caption{Posterior distributions of Model III from Table \ref{tab:buildup}, where we retrieve $P_0$, $A_{\rm s}$, $g$, $R_{\rm p}$, \water, \oxygen, and \ozone. We retrieve on $R = 140$, SNR$\ = 20$ data with wavelength-independent noise. Overplotted in solid light-blue color are the fiducial parameter values. The 2D marginalized posterior distribution, used in interpreting correlations, is overplotted with the 1-, 2-, and 3-$\sigma$ contours. Above the 1D marginalized posterior for each parameter, we list the median retrieved value with uncertainties that indicate the 68\% confidence interval. Dashed lines (left to right) mark the 16\%, 50\%, and 84\% quantiles.} \label{fig:b3stairs} 
\end{figure*}

\begin{figure*}[ht] \begin{center}
\includegraphics[width=\textwidth]{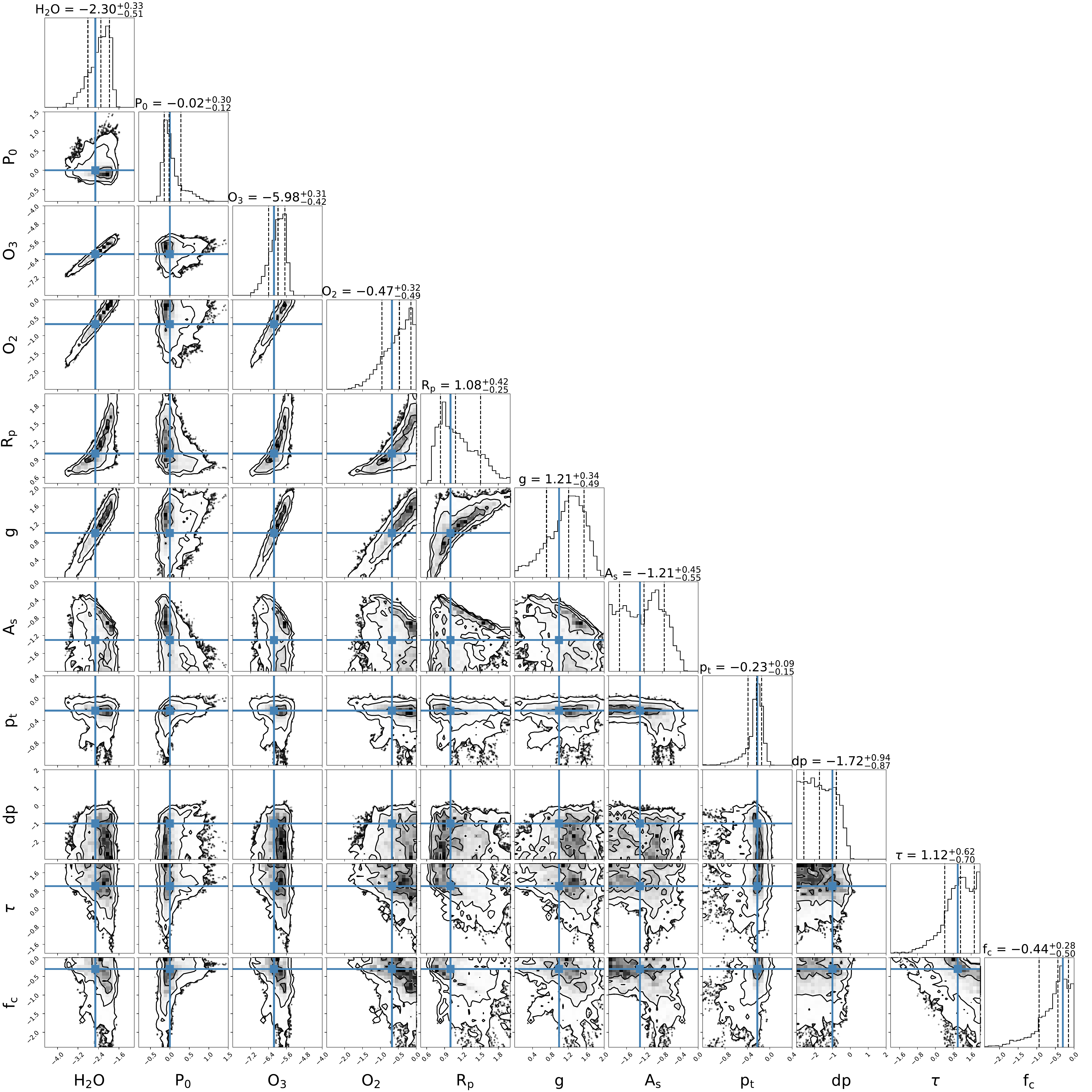} \end{center}
\caption{Posterior distributions of Model IV, or the complete model, from Table \ref{tab:buildup}. We retrieve for 11 parameters: $P_0$, $A_{\rm s}$, $g$, $R_{\rm p}$, \water, \oxygen, \ozone, $p_{\rm t}$, $dp$, $\tau$, and $f_{\rm c}$. We retrieve on $R = 140$, SNR$\ = 20$ data with wavelength-independent noise. Overplotted in solid light-blue color are the fiducial parameter values. The 2D marginalized posterior distribution, used in interpreting correlations, is overplotted with the 1-, 2-, and 3-$\sigma$ contours. Above the 1D marginalized posterior for each parameter, we list the median retrieved value with uncertainties that indicate the 68\% confidence interval. Dashed lines (left to right) mark the 16\%, 50\%, and 84\% quantiles.} \label{fig:b4stairs} 
\end{figure*}

In Figure \ref{fig:b1stairs}, we present the posterior distributions for Model I. In the two-dimensional correlation histogram, a higher probability corresponds to a darker shade.  With all else held constant, we see narrow posterior distributions and a slight correlation between $P_0$ and $A_s$. For lower values of surface pressure, which controls the turn off of the Rayleigh scattering slope, we need a brighter surface to maintain the measured brightness, especially in the red end of the spectrum, and vice versa. We mark the 16\%, 50\%, and 84\% quantiles in the marginalized one-dimensional posterior distributions. The posterior distributions for Model II are shown in Figure \ref{fig:b2stairs}, and are generally narrow (as only four parameters are being retrieved). There are two key correlations, one between $g$ and $P_0$, and one between $R_{\rm p}$ and $A_{\rm s}$. Both gravity and surface pressure influence the column mass, so that, when attempting to fit a spectrum, we can trade a larger gravity with a larger surface pressure and maintain a similar column mass (which controls, e.g., the Rayleigh scattering feature). Additionally, we can trade off a larger reflecting surface area (i.e., larger $R_{\rm p}$) with a darker surface (lower $A_{\rm s}$), which is a statement of the typical ``radius/albedo degeneracy'' problem. The posterior for surface albedo is now an upper limit instead of a constraint. As a result, the radius posterior distribution appears truncated at larger values given the tight correlation between these two parameters. The correlation seen originally in Model I, between $P_0$ and $A_{\rm s}$, then acts as a chain between the other two, more prominent, correlations to induce correlations between parameters such as $A_{\rm s}$ and $g$ or $R_{\rm p}$ and $P_0$. 

Once we allow gases to be free parameters in Model III (Figure \ref{fig:b3stairs}), the $P_0$ and $A_{\rm s}$ correlation becomes diminished as \water, due to its numerous bands across the spectral range, becomes a primary control of brightness. The significant impact of \water\ on the spectrum leads to a strong, positive correlation between \water\ abundance and planetary size, as additional water vapor absorption can be compensated by a larger planetary size to maintain fixed brightness. We now see gravity linked to the molecular abundances, which is expected as surface gravity directly influences the column abundance of a species. This key correlation also causes the individual gas abundances to be correlated with each other. The main correlations from Model II are still present. We note once more that we do not have constraint on the surface albedo, again leading to an asymmetric distribution for radius. Thus, from the strong correlation of \water\ with with $R_p$, and the fundamental correlation between $R_{\rm p}$ and $A_{\rm s}$, we see correlations between planetary radius, surface albedo, and all gas abundances. Weak correlations between surface pressure and the gas abundances are due to column abundance effects.

Finally, as shown by Figure \ref{fig:b4stairs}, we retrieve on the data with the full forward model, adding in the cloud parameters $p_{\rm t}$, $dp$, $\tau$, and $f_{\rm c}$. This version of the model is what we apply when simulating direct-imaging data in the upcoming sections, and represents our most realistic (i.e., true to the actual Earth) scenario. The optical depth is shown to only have a lower limit constraint.  Thus, the retrieval detects a cloud but cannot constrain the optical depth beyond showing that the cloud is optically thick.  There is an expected correlation between $\tau$ and $f_{\rm c}$; a higher cloudiness fraction can complement a less optically thick cloud, and vice versa. There is only an upper limit to $dp$, which is a result of the lack of vertical sensitivity given the constant-with-pressure abundance distributions. The posterior distribution for \oxygen\ becomes a lower limit instead of a constraint as in Model III. Surface gravity is less precisely and less accurately constrained compared to the previous, less complex renditions of the model.

For optically thin clouds, we expect to better constrain surface albedo; however, we do not consider this scenario in our study. We examined instead the performance of a completely cloud-free model on data generated with our cloudy model. We find that while the model can fit the data and return accurate estimates of e.g., the mixing ratios, we get inaccurate estimates of the surface albedo and the surface pressure. These two parameters are biased, with lower surface pressure paired with higher surface albedo as the preferred configuration in the cloud-free case. As a result, we move forward with utilizing our cloudy forward model on our simulated data. However, we note that in realistic cases where we do not know the true state of a planet's atmosphere, we could obtain complementary information relating to the presence of clouds (e.g., variability) such that we may choose the most appropriate forward model.

\section{Results} \label{snr_study}
We generate data sets for HabEx and LUVOIR-like missions ($0.4 - 1.0\ \mu$m at $R=70$, $R = 140$) at SNR$= 5,\ 10,\ 15,\ 20$, and for the {\it WFIRST} rendezvous scenario (two photometric points within $0.4 - 0.6\ \mu$m plus a spectrum of $R = 50$ for $0.6 - 0.96\ \mu$m) also at SNR$=5,\ 10,\ 15,\ 20$. In all cases, we used the noise model to generate uncertainties expected for high-contrast imaging instead of the wavelength-independent noise for the validations in the previous section. As Section \ref{subsec:retrieval} described, the SNR refers to the value at 0.55 $\mu$m for $R = 70,\ 140$, and at 0.6 $\mu$m for {\it WFIRST}. We record the specific runs in Table \ref{tab:runs}. In place of showing the correlations for all parameters for all cases, we refer to Figure \ref{fig:b4stairs}, which represents the ideal case correlations among the parameter posteriors. We only show the posterior probability distributions themselves to better highlight any trends with respect to SNR and/or $R$. We grouped the posteriors in terms of bulk atmospheric and planetary parameters ($P_0$, $R_{\rm p}$, $g$, $A_{\rm s}$), then cloud parameters ($p_{\rm t}$, $dp$, $\tau$, $f_{\rm c}$), and finally gases (\water, \ozone, \oxygen). For each case, {\tt emcee} was run with 16 MCMC chains (walkers) per parameter for at least 12000 steps, the last 5000 of which are  used to determine the posterior distributions. From those 5000 steps, we randomly selected 1000 sets of parameters to calculate their corresponding high resolution spectra. These spectra are plotted with the data to show the 1-$\sigma$, 2-$\sigma$, and median fits.

\subsection{Results for $R = 70$, $R = 140$ simulated data}
\begin{figure*}[h]
\centering
\subfigure[$R = 70$ bulk parameters]{
    \includegraphics[width = 0.45\textwidth]{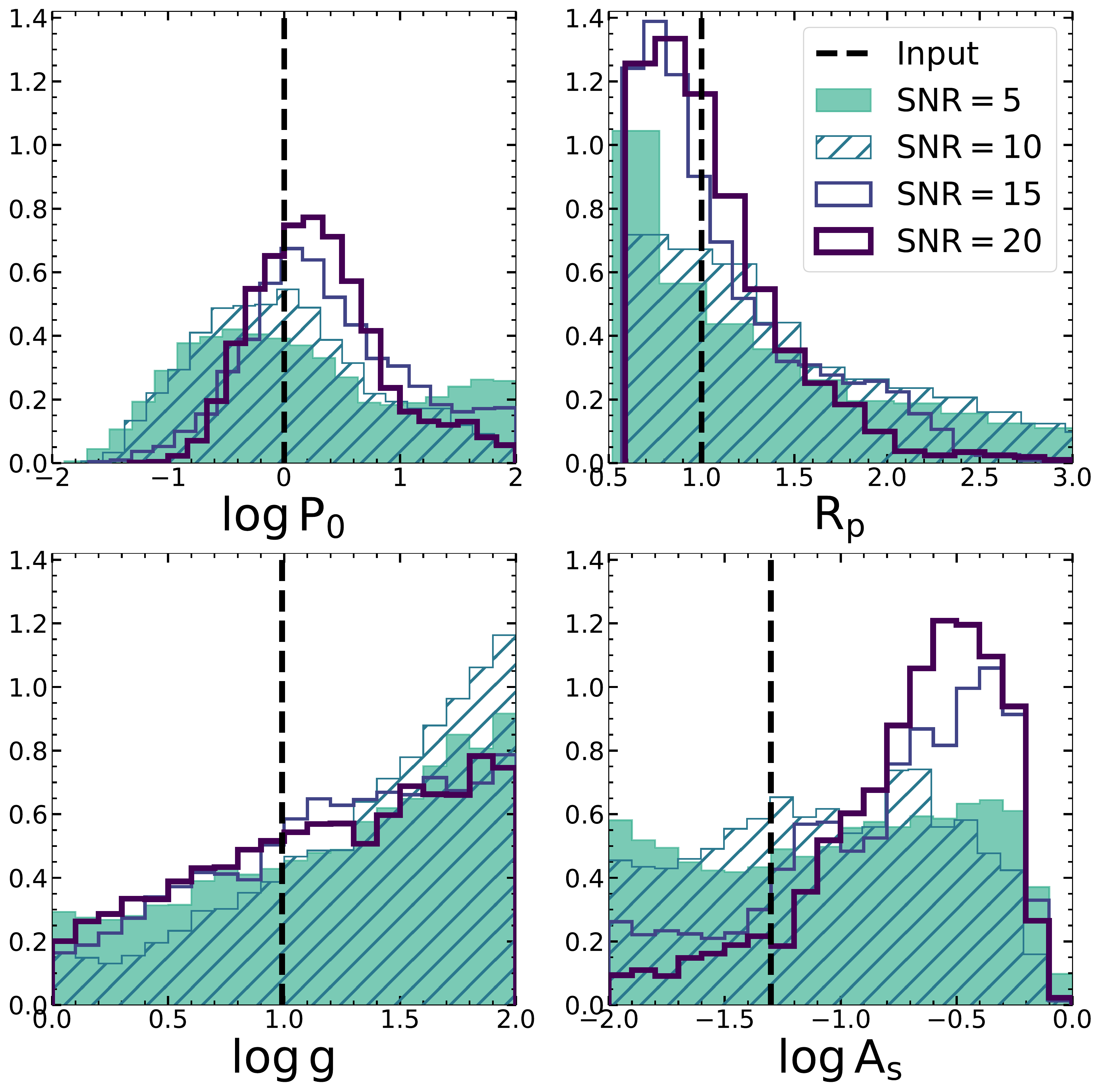}
    \label{r70bulk}}\qquad
\subfigure[$R = 70$ cloud parameters]{
    \includegraphics[width = 0.45\textwidth]{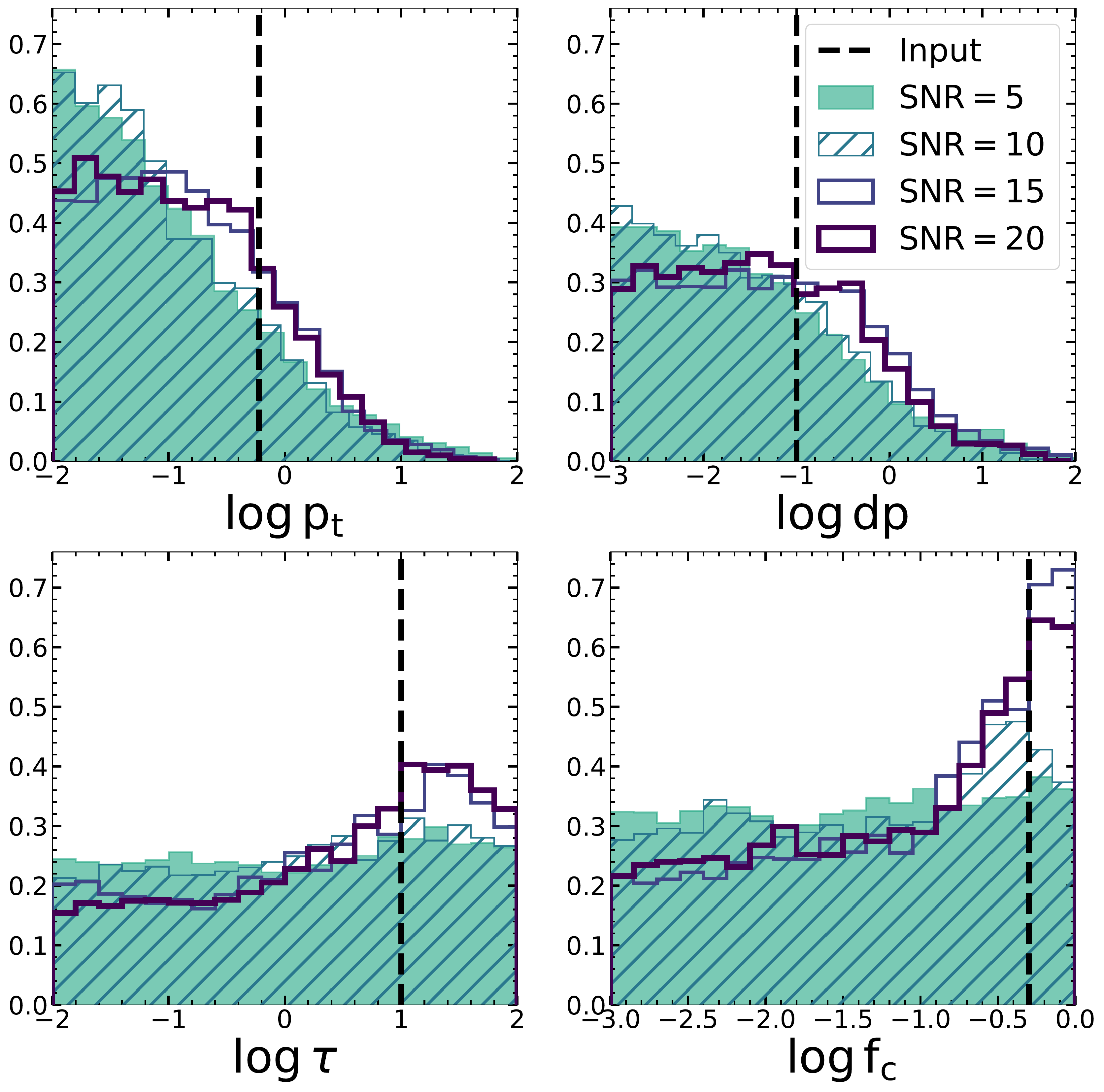}
    \label{r70clouds}}
\subfigure[$R = 70$ gas mixing ratios]{
    \includegraphics[width = 0.8\textwidth]{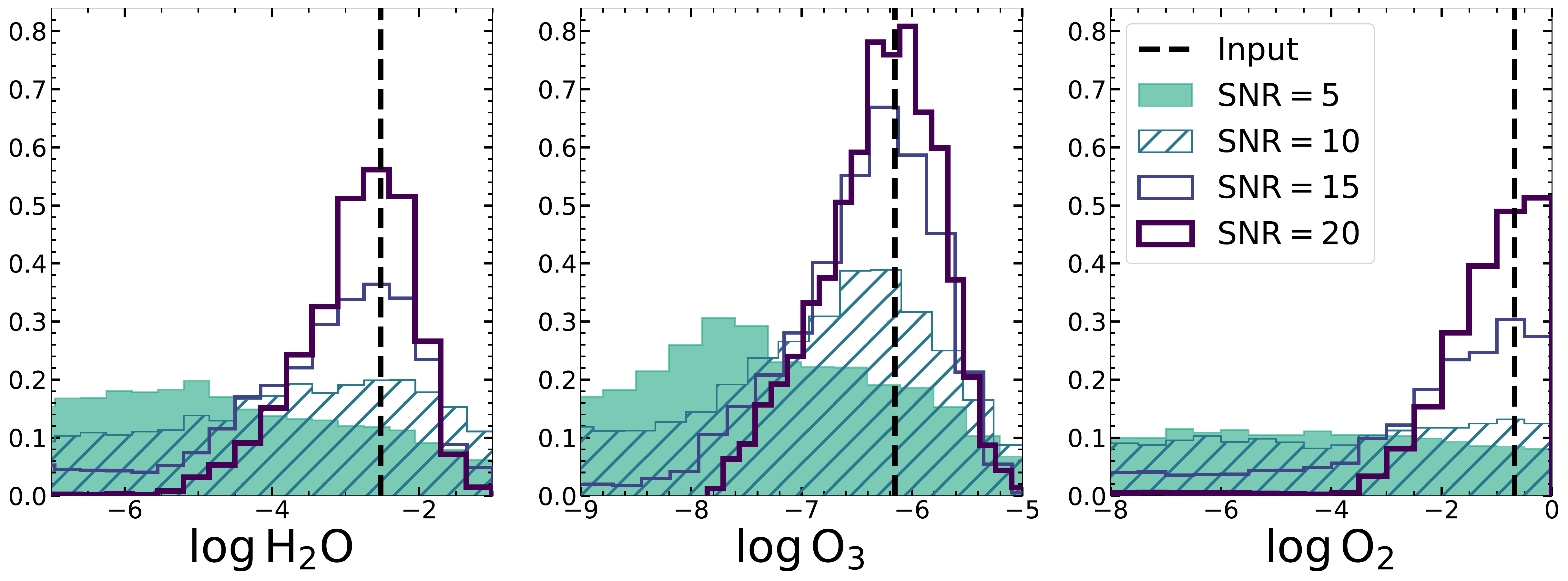}
    \label{r70gases} \qquad
    }
\caption{Comparing 1D marginalized posterior distributions for all parameters for all SNR cases of $R = 70$. See Table \ref{tab:r70results} for corresponding median retrieved value with uncertainties that indicate the 68\% confidence interval. Overplotted dashed line represents the fiducial values from Table \ref{tab:params}.}
\label{fig:compareR70}
\end{figure*}

\begin{figure*}[ht!]
\centering
\subfigure[$R = 140$ bulk parameters]{
    \includegraphics[width = 0.45\textwidth]{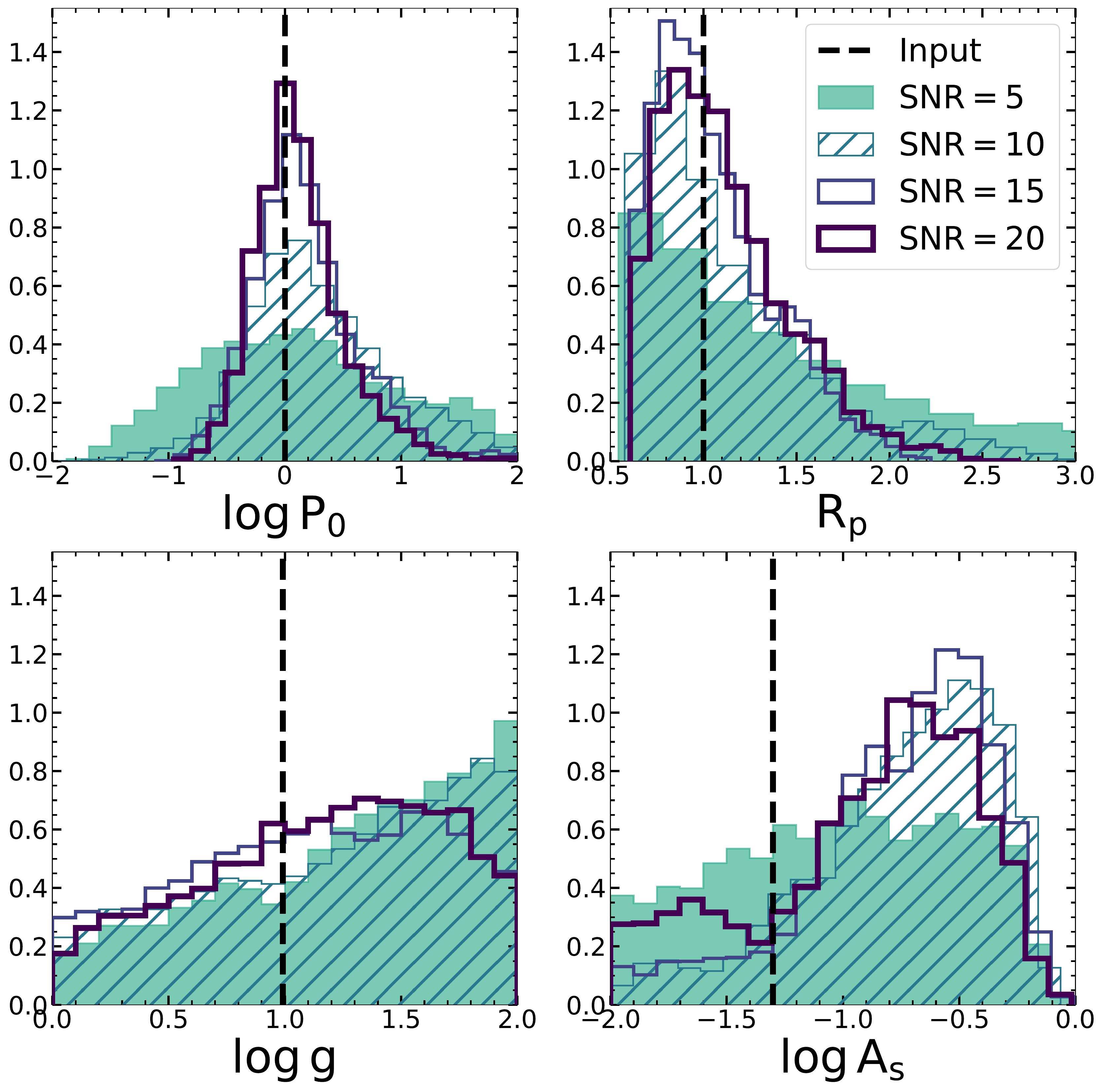}
    \label{r140bulk}}\qquad
\subfigure[$R = 140$ cloud parameters]{
    \includegraphics[width = 0.45\textwidth]{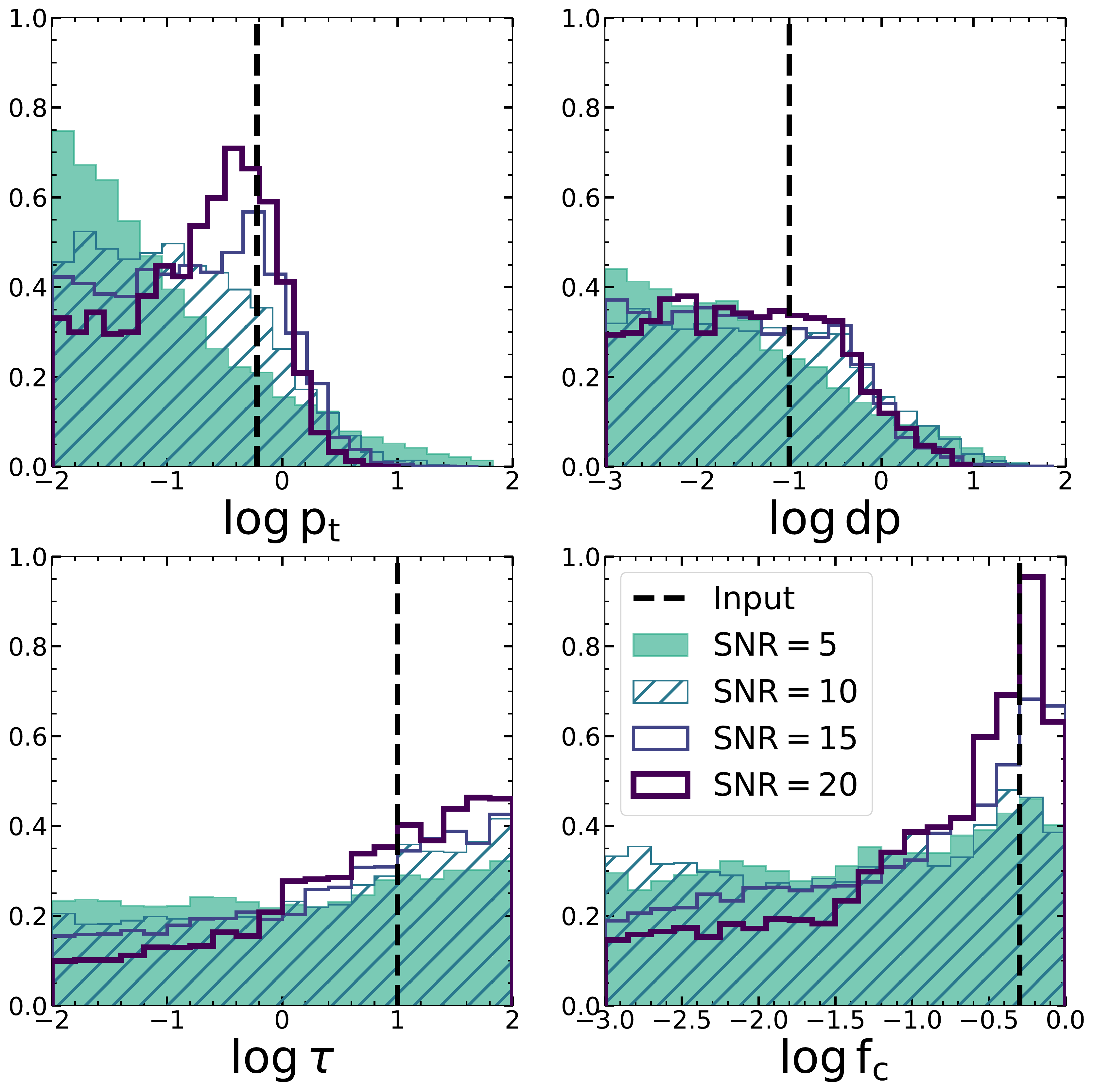}
    \label{r140clouds}}
\subfigure[$R = 140$ gas mixing ratios]{
    \includegraphics[width = 0.8\textwidth]{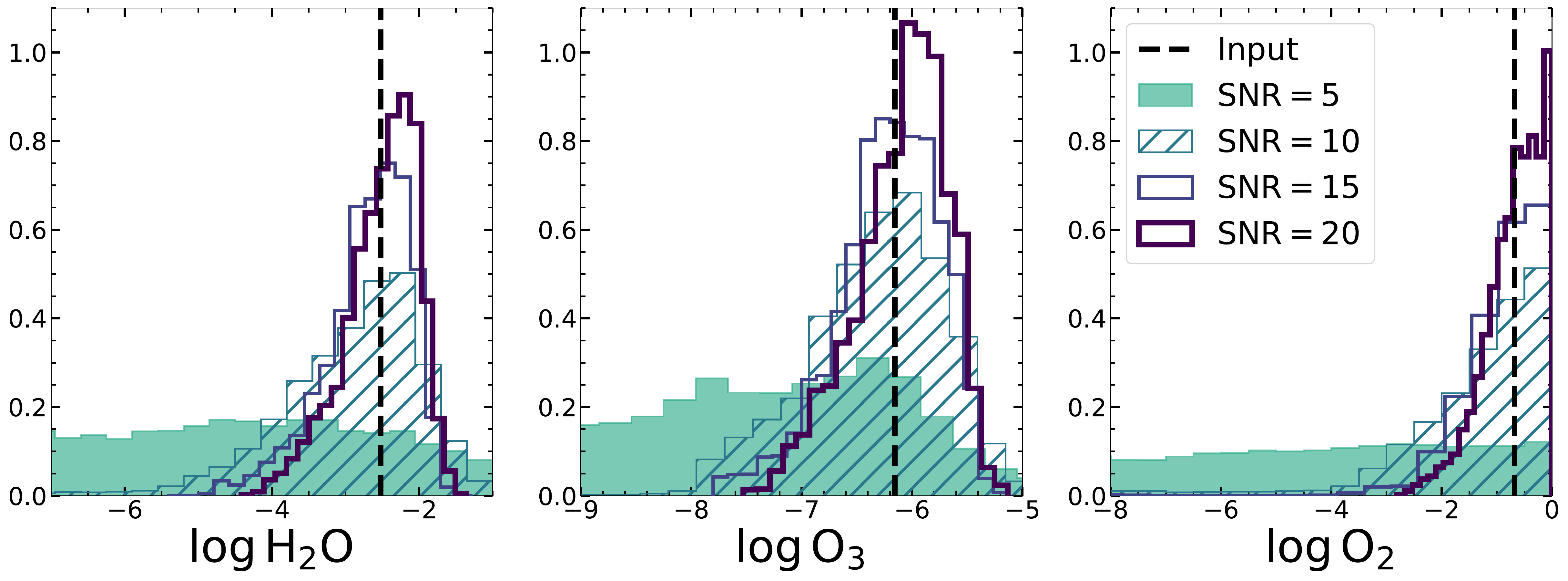}
    \label{r140gases} \qquad
    }
\caption{Comparing 1D marginalized posterior distributions for all parameters for all SNR cases of $R = 140$. See Table \ref{tab:r140results} for corresponding median retrieved value with uncertainties that indicate the 68\% confidence interval. Overplotted dashed line represents the fiducial values from Table \ref{tab:params}.}
\label{fig:compareR140}
\end{figure*}

For both $R = 70$ and $R = 140$, we simulated data sets at SNR = 5, 10, 15, 20. Table \ref{tab:r70results} lists the median and 1-$\sigma$ values of all retrieved parameters for each SNR at $R = 70$. Figure \ref{fig:compareR70} shows the marginalized posterior distributions for the model parameters for all SNR cases for $R = 70$, plotted with the fiducial or ``truth'' values. Table \ref{tab:r140results} lists the median and 1-$\sigma$ values of all retrieved parameters for each SNR at $R = 140$. Figure \ref{fig:compareR140} shows the posterior distributions for $R = 140$ for the model parameters for all SNR cases compared against their input values. Figure \ref{fig:yarr} shows the corresponding spread in fits and the median fit to the data for each SNR for both resolutions. 

\subsection{Results for \textit{WFIRST} rendezvous simulated data}
\begin{figure*}[ht!]
\centering
\subfigure[\textit{WFIRST} bulk parameters]{
    \includegraphics[width = 0.45\textwidth]{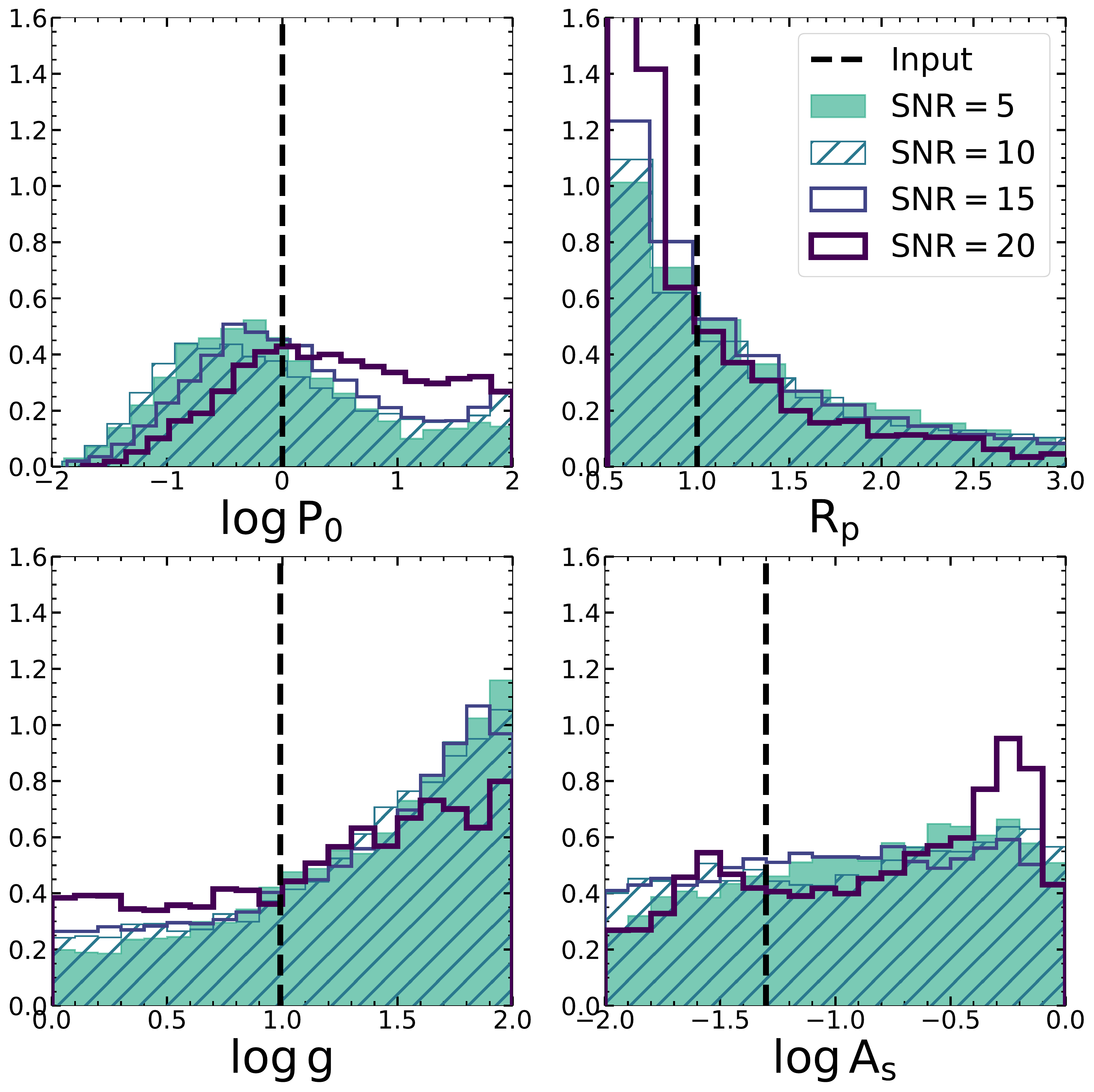}
    \label{wfirstbulk}}\qquad
\subfigure[\textit{WFIRST} cloud parameters]{
    \includegraphics[width = 0.45\textwidth]{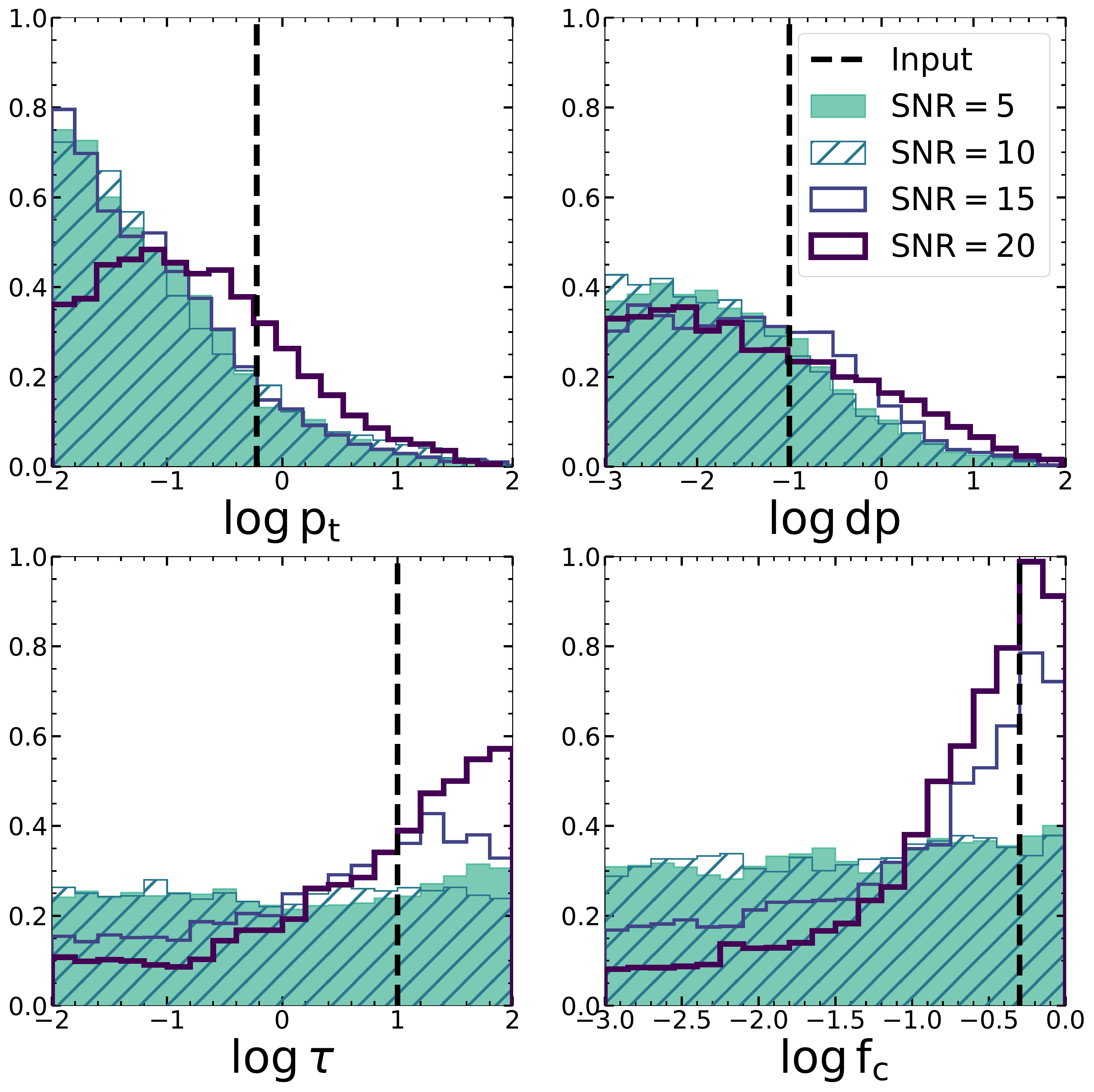}
    \label{wfirstclouds}}
\subfigure[\textit{WFIRST} gas mixing ratios]{
    \includegraphics[width = 0.8\textwidth]{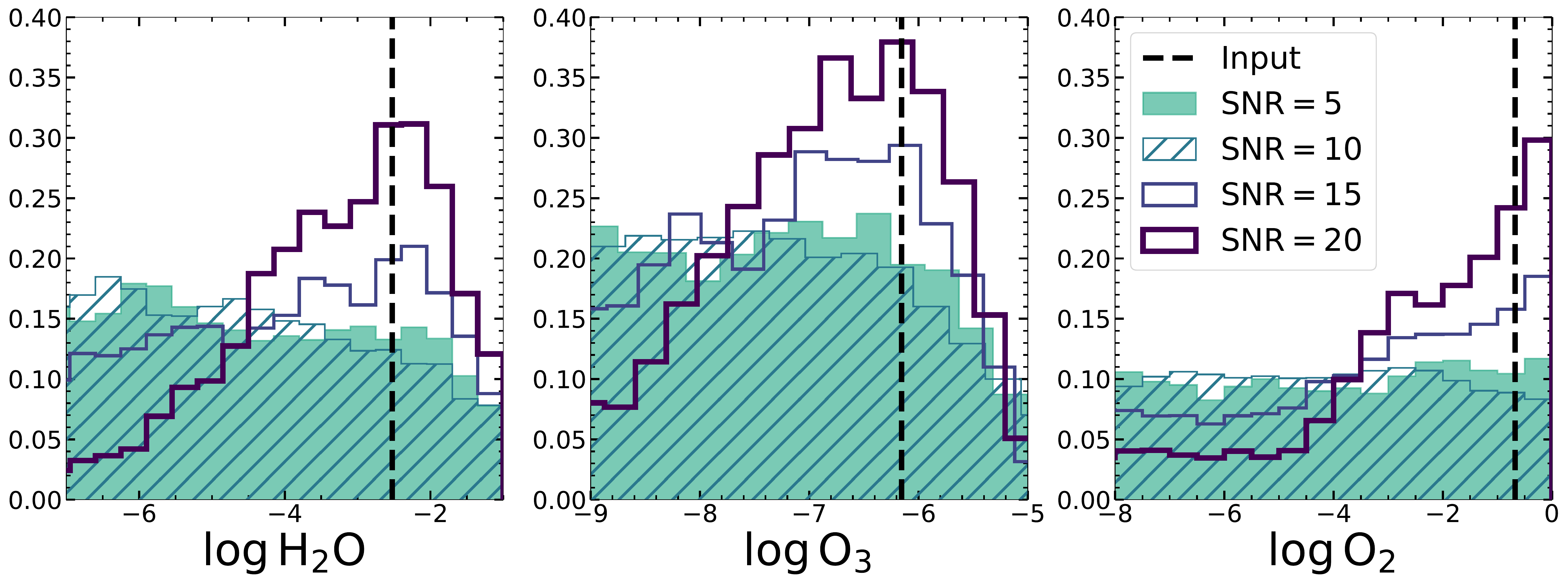}
    \label{wfirstgases} \qquad
    }
\caption{Comparing 1D marginalized posterior distributions for all parameters for all SNR cases of a \textit{WFIRST} rendezvous scenario. See Table \ref{tab:wfirstresults} for corresponding median retrieved value with uncertainties that indicate the 68\% confidence interval. Overplotted dashed line represents the fiducial values from Table \ref{tab:params}.}
\label{fig:compareWfirst}
\end{figure*}

\begin{figure*}[ht!]
\centering
    \includegraphics[scale = 0.3]{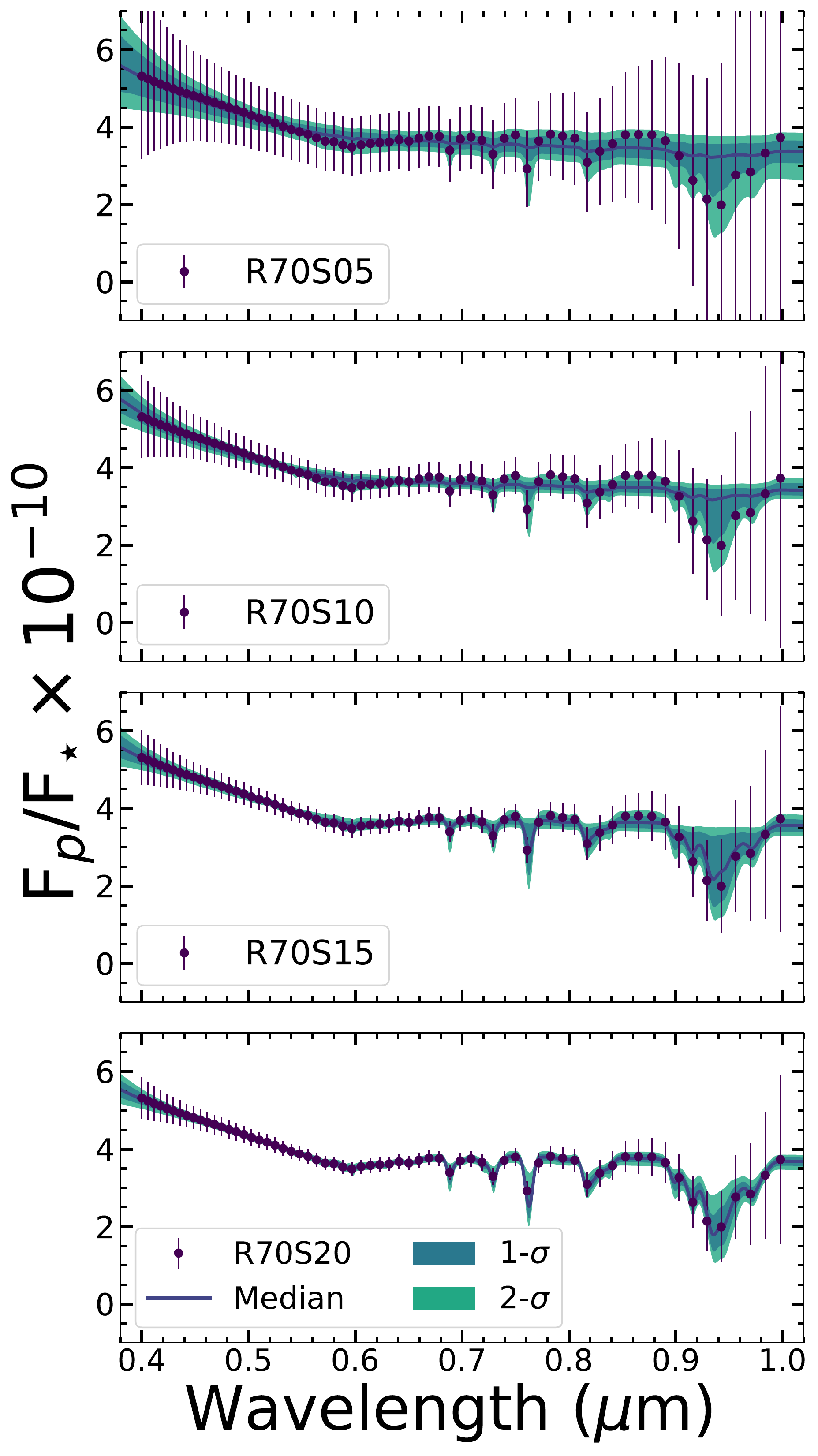}
    \includegraphics[scale = 0.3]{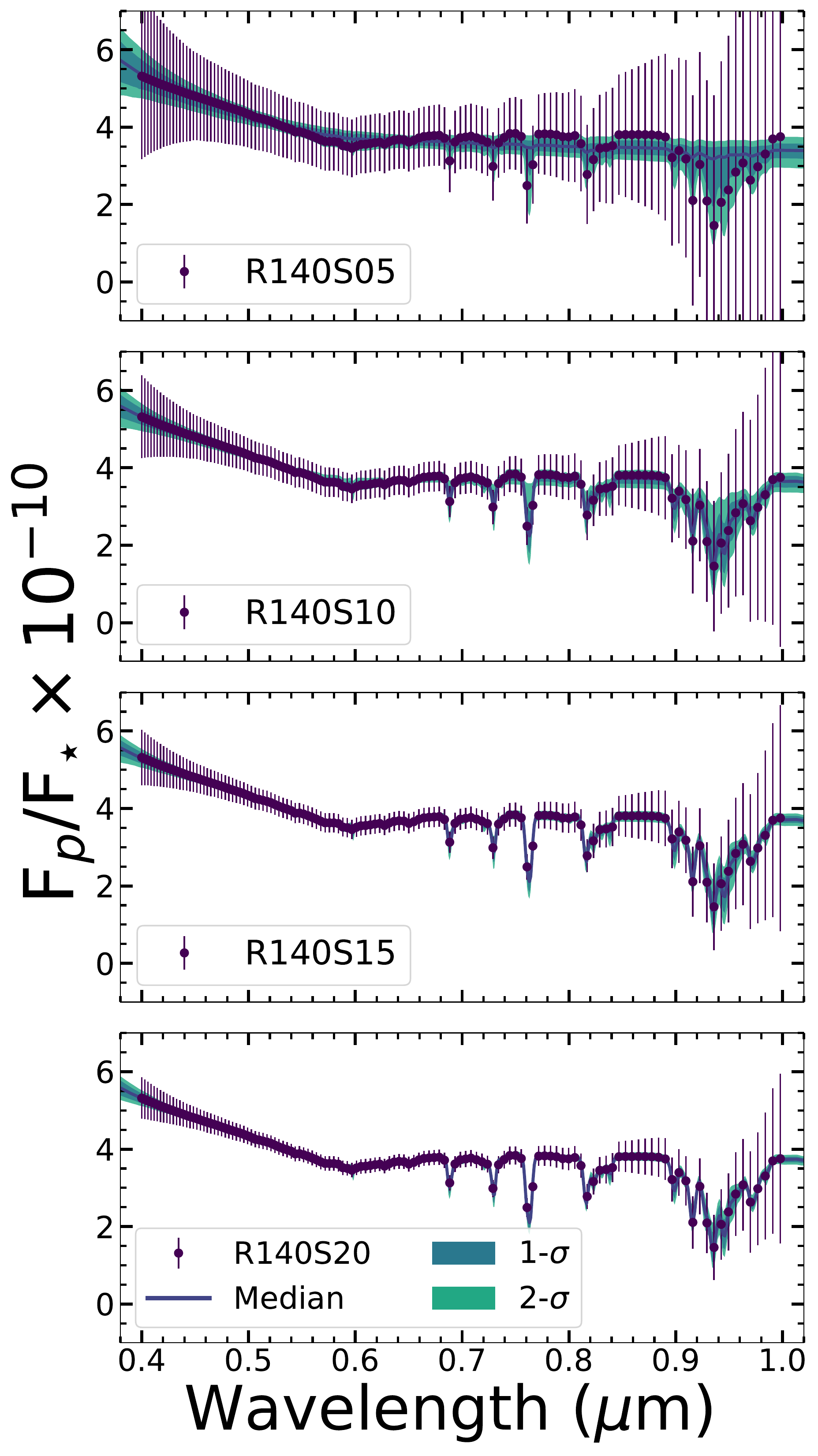}
    \includegraphics[scale = 0.25]{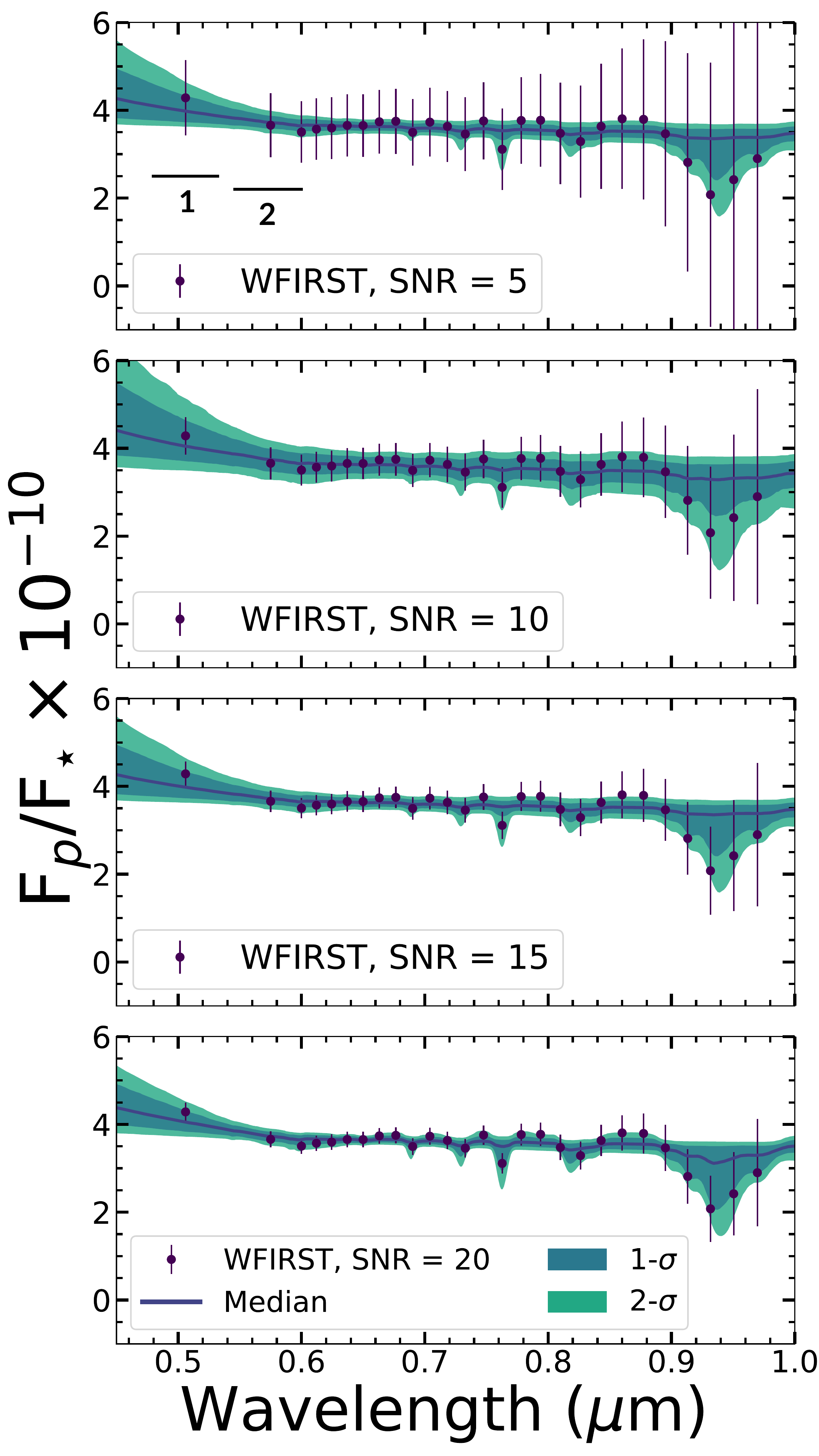}

\caption{Spectra generated with 1000 randomly drawn sets of parameters sampled with the retrievals plotted with \textbf{left:} $R = 70$ data for SNR$=5,\, 10,\,15,\, 20$; \textbf{middle:} $R = 140$ data for SNR$=5,\,10,\,15,\, 20$; and \textbf{right:} \textit{WFIRST} rendezvous data at SNR$\ =5,\, 10,\, 15,\, 20$. ``1'' and ``2'' mark the span of the \textit{WFIRST} Design Cycle 7 filters (see Table \ref{tab:runs}). Lighter contour (light green) represents 2-$\sigma$ fits while darker contour (blue-green) represents 1-$\sigma$ fits. Solid line represents the median fit. }
\label{fig:yarr}
\end{figure*}

For the \textit{WFIRST} rendezvous scenario, we utilized the Design Cycle 7 instrument parameters to set the locations of the two photometric points and the range and resolution of the spectrometer ($R=50$; see Table \ref{tab:runs}). Because of this particular set-up, we reference the SNRs in our grid (5, 10, 15, 20) at 600~nm, and assign the photometric points the same SNR as at 600~nm.  Table \ref{tab:wfirstresults} lists the median and 1-$\sigma$ values of all retrieved parameters for each SNR variant. Figure \ref{fig:compareWfirst} presents the posterior distributions for the four \textit{WFIRST} rendezvous variants with respect to the input values. Figure \ref{fig:yarr} shows the spread in fits and median fit to the data for each variant.

\begin{deluxetable*}{lrrrrr}
\tablecaption{$R = 70$ retrieval results, with median value and 1-$\sigma$ uncertainties of the parameters. \label{tab:r70results}}
\tablewidth{0pt}
\tabletypesize{\scriptsize}
\tablehead{Parameter	&  Input & SNR$=5$	& SNR$=10$ & SNR$=15$ & SNR$=20$}
\startdata
$\log \, \rm H_2 O$ & $-2.52$ & ${-5.07}_{-1.92}^{+2.34}$ & ${-3.85}_{-2.60}^{+1.77}$ & ${-3.12}_{-1.71}^{+0.97}$ & ${-2.76}_{-0.88}^{+0.62}$ \\
$\log \, \rm O_3$ & $-6.15$ & ${-7.55}_{-1.46}^{+1.49}$ & ${-6.79}_{-1.81}^{+0.93}$ & ${-6.37}_{-0.84}^{+0.55}$ & ${-6.24}_{-0.60}^{+0.47}$ \\
$\log \, \rm O_2$ & $-0.68$ & ${-5.12}_{-3.23}^{+3.25}$ & ${-4.51}_{-3.61}^{+3.24}$ & ${-1.86}_{-3.99}^{+1.29}$ & ${-1.00}_{-1.01}^{+0.66}$ \\
$\log \, \rm P_0$ & $0.0$ & ${0.02}_{-0.84}^{+1.35}$ & ${-0.03}_{-0.70}^{+0.87}$ & ${0.28}_{-0.56}^{+0.85}$ & ${0.25}_{-0.49}^{+0.56}$ \\
R$_{\rm p}$ & $1.0$ & ${1.23}_{-0.58}^{+1.54}$ & ${1.33}_{-0.52}^{+1.23}$ & ${0.97}_{-0.27}^{+0.68}$ & ${0.98}_{-0.25}^{+0.44}$ \\
$\log \, \rm g$ & $0.99$ & ${1.33}_{-0.77}^{+0.48}$ & ${1.48}_{-0.68}^{+0.38}$ & ${1.28}_{-0.66}^{+0.51}$ & ${1.24}_{-0.69}^{+0.55}$ \\
$\log \, \rm A_s$ & $-1.3$ & ${-0.96}_{-0.74}^{+0.58}$ & ${-1.05}_{-0.59}^{+0.55}$ & ${-0.70}_{-0.62}^{+0.37}$ & ${-0.63}_{-0.46}^{+0.29}$ \\
$\log \, \rm p_t$ & $-0.22$ & ${-1.14}_{-0.61}^{+0.97}$ & ${-1.19}_{-0.56}^{+0.93}$ & ${-0.92}_{-0.71}^{+0.86}$ & ${-0.94}_{-0.73}^{+0.84}$ \\
$\log \, \rm dp$ & $-1.0$ & ${-1.67}_{-0.92}^{+1.24}$ & ${-1.71}_{-0.91}^{+1.18}$ & ${-1.35}_{-1.14}^{+1.17}$ & ${-1.43}_{-1.06}^{+1.11}$ \\
$\log \, \tau$ & $1.0$ & ${0.10}_{-1.43}^{+1.30}$ & ${0.21}_{-1.48}^{+1.23}$ & ${0.49}_{-1.66}^{+1.03}$ & ${0.61}_{-1.66}^{+0.93}$ \\
$\log \, \rm f_c$ & $-0.3$ & ${-1.43}_{-1.07}^{+0.99}$ & ${-1.33}_{-1.12}^{+0.94}$ & ${-0.93}_{-1.32}^{+0.71}$ & ${-1.05}_{-1.27}^{+0.80}$ \\
\enddata
\end{deluxetable*}

\begin{deluxetable*}{lrrrrr}
\tablecaption{$R = 140$ retrieval results, with median value and 1-$\sigma$ uncertainties of the parameters. \label{tab:r140results}}
\tablewidth{0pt}
\tabletypesize{\scriptsize}
\tablehead{Parameter	& Input & SNR$=5$	& SNR$=10$ & SNR$=15$ & SNR$=20$}
\startdata
$\log \, \rm H_2 O$ & $-2.52$ & ${-4.56}_{-2.35}^{+2.14}$ & ${-2.74}_{-1.07}^{+0.69}$ & ${-2.61}_{-0.65}^{+0.47}$ & ${-2.43}_{-0.56}^{+0.39}$ \\
$\log \, \rm O_3$ & $-6.15$ & ${-7.36}_{-1.65}^{+1.26}$ & ${-6.26}_{-0.68}^{+0.53}$ & ${-6.18}_{-0.48}^{+0.42}$ & ${-6.03}_{-0.48}^{+0.34}$ \\
$\log \, \rm O_2$ & $-0.68$ & ${-4.45}_{-3.69}^{+3.08}$ & ${-1.06}_{-1.43}^{+0.76}$ & ${-0.76}_{-0.79}^{+0.51}$ & ${-0.60}_{-0.59}^{+0.43}$ \\
$\log \, \rm P_0$ & $0.0$ & ${0.07}_{-0.84}^{+1.01}$ & ${0.20}_{-0.49}^{+0.72}$ & ${0.12}_{-0.36}^{+0.49}$ & ${0.07}_{-0.31}^{+0.39}$ \\
R$_{\rm p}$ & $1.0$ & ${1.25}_{-0.52}^{+1.16}$ & ${1.01}_{-0.28}^{+0.60}$ & ${0.99}_{-0.23}^{+0.42}$ & ${1.05}_{-0.27}^{+0.42}$ \\
$\log \, \rm g$ & $0.99$ & ${1.36}_{-0.74}^{+0.46}$ & ${1.31}_{-0.77}^{+0.49}$ & ${1.14}_{-0.65}^{+0.56}$ & ${1.20}_{-0.64}^{+0.50}$ \\
$\log \, \rm A_s$ & $-1.3$ & ${-0.98}_{-0.60}^{+0.54}$ & ${-0.67}_{-0.50}^{+0.32}$ & ${-0.68}_{-0.44}^{+0.29}$ & ${-0.79}_{-0.69}^{+0.34}$ \\
$\log \, \rm p_t$ & $-0.22$ & ${-1.23}_{-0.55}^{+1.03}$ & ${-0.96}_{-0.71}^{+0.80}$ & ${-0.79}_{-0.82}^{+0.70}$ & ${-0.66}_{-0.85}^{+0.53}$ \\
$\log \, \rm dp$ & $-1.0$ & ${-1.72}_{-0.91}^{+1.25}$ & ${-1.43}_{-1.09}^{+1.13}$ & ${-1.55}_{-1.00}^{+1.10}$ & ${-1.49}_{-0.98}^{+1.00}$ \\
$\log \, \tau$ & $1.0$ & ${0.18}_{-1.49}^{+1.31}$ & ${0.50}_{-1.66}^{+1.09}$ & ${0.61}_{-1.61}^{+0.98}$ & ${0.79}_{-1.40}^{+0.87}$ \\
$\log \, \rm f_c$ & $-0.3$ & ${-1.30}_{-1.13}^{+0.93}$ & ${-1.31}_{-1.21}^{+0.94}$ & ${-0.99}_{-1.27}^{+0.76}$ & ${-0.76}_{-1.26}^{+0.54}$ \\
\enddata
\end{deluxetable*}

\begin{deluxetable*}{lrrrrr}
\tablecaption{\textit{WFIRST} rendezvous retrieval results, with median value and 1-$\sigma$ uncertainties of the parameters. \label{tab:wfirstresults}}
\tablewidth{0pt}
\tabletypesize{\scriptsize}
\tablehead{Parameter & Input	&  SNR$=5$	& SNR$=10$ & SNR$=15$ & SNR$=20$}
\startdata
$\log \, \rm H_2 O$ & $-2.52$ & ${-4.94}_{-2.05}^{+2.35}$ & ${-4.89}_{-2.11}^{+2.48}$ & ${-4.03}_{-2.52}^{+1.87}$ & ${-3.11}_{-1.71}^{+1.17}$ \\
$\log \, \rm P_0$ & $0.0$ & ${-0.16}_{-0.80}^{+1.32}$ & ${-0.19}_{-0.71}^{+1.03}$ & ${0.03}_{-0.74}^{+1.16}$ & ${0.45}_{-0.85}^{+1.01}$ \\
$\log \, \rm O_3$ & $-6.15$ & ${-7.66}_{-1.59}^{+1.65}$ & ${-7.53}_{-1.66}^{+1.54}$ & ${-7.16}_{-1.66}^{+1.19}$ & ${-6.80}_{-1.30}^{+0.94}$ \\
$\log \, \rm O_2$ & $-0.68$ & ${-5.05}_{-3.39}^{+3.26}$ & ${-4.89}_{-3.54}^{+3.43}$ & ${-3.43}_{-4.41}^{+2.50}$ & ${-2.26}_{-3.88}^{+1.71}$ \\
$\log \, \rm P_0$ & $0.0$ & ${-0.16}_{-0.80}^{+1.32}$ & ${-0.19}_{-0.71}^{+1.03}$ & ${0.03}_{-0.74}^{+1.16}$ & ${0.45}_{-0.85}^{+1.01}$ \\
R$_{\rm p}$ & $1.0$ & ${1.13}_{-0.50}^{+1.60}$ & ${1.13}_{-0.48}^{+1.27}$ & ${1.02}_{-0.38}^{+1.10}$ & ${0.80}_{-0.19}^{+0.81}$ \\
$\log \, \rm g$ & $0.99$ & ${1.42}_{-0.82}^{+0.42}$ & ${1.45}_{-0.75}^{+0.41}$ & ${1.41}_{-0.83}^{+0.43}$ & ${1.26}_{-0.83}^{+0.52}$ \\
$\log \, \rm A_s$ & $-1.3$ & ${-0.89}_{-0.74}^{+0.63}$ & ${-0.84}_{-0.70}^{+0.56}$ & ${-0.95}_{-0.68}^{+0.64}$ & ${-0.76}_{-0.79}^{+0.53}$ \\
$\log \, \rm p_t$ & $-0.22$ & ${-1.26}_{-0.52}^{+0.98}$ & ${-1.24}_{-0.55}^{+0.83}$ & ${-1.23}_{-0.57}^{+0.84}$ & ${-0.84}_{-0.73}^{+0.89}$ \\
$\log \, \rm dp$ & $-1.0$ & ${-1.75}_{-0.87}^{+1.16}$ & ${-1.70}_{-0.87}^{+1.12}$ & ${-1.46}_{-1.06}^{+1.14}$ & ${-1.49}_{-1.03}^{+1.49}$ \\
$\log \, \tau$ & $1.0$ & ${0.03}_{-1.39}^{+1.33}$ & ${0.05}_{-1.40}^{+1.42}$ & ${0.61}_{-1.55}^{+0.94}$ & ${0.99}_{-1.44}^{+0.73}$ \\
$\log \, \rm f_c$ & $-0.3$ & ${-1.41}_{-1.07}^{+0.96}$ & ${-1.42}_{-1.07}^{+1.00}$ & ${-0.82}_{-1.28}^{+0.60}$ & ${-0.58}_{-0.97}^{+0.41}$ \\
\enddata
\end{deluxetable*}

\section{Discussion} \label{discussion}
The results from our retrieval analyses enable us to identify the SNR required, at a given spectral resolution, to constrain key planetary and atmospheric quantities.  These findings have important implications for the development of future space-based direct imaging missions.  We discuss these ideas below, and also touch on impacts of certain model assumptions and ideas for future research directions.

In what follows, we define a ``weak detection'' for a given parameter as having a posterior distribution that has a marked peak but which also has a substantial tail towards extreme values (indicating that, e.g., for a gas we could not definitively state that the gas is present in the atmosphere).  A ``detection'' implies a peaked posterior without tails towards extreme values but whose 1-$\sigma$ width is larger than an order of magnitude.  We use the term ``constraint'' to indicate a detection whose posterior distribution has 1-$\sigma$ width smaller than an order of magnitude.  A non-detection would be indicated by a flat posterior distribution across the entire (or near-entire) prior range.  For planetary radius, which is not retrieved in logarithmic space, we distinguish between a ``detection'' and a ``constraint'' when the 1-$\sigma$ uncertainties are small enough to firmly place the planet in the Earth/super-Earth regime \citep[i.e., with a radius below $1.5R_{\oplus}$,][]{rogers2016,chen2017}.  A visual summary of weak detections, detections, and constraints as a function of SNR for our different observing scenarios and for a selection of key parameters are given in Tables \ref{tab:r70detect}, \ref{tab:r140detect}, and \ref{tab:wfirstdetect}.

\subsection{Influence of SNR on Inferred Properties}
For $R = 70$ at SNR = 5, Figure \ref{fig:compareR70} shows there is only a weak detection of $P_0$ and a detection of $R_{\rm p}$, which merely suggests the planet has an atmosphere and is not a giant planet. As SNR increases to 10, the \ozone~posterior distribution has a weak peak near the fiducial value, and the gas is only weakly detected. Once the SNR is equal to 15, we weakly detect \water, \ozone, and \oxygen. At a SNR of 20, it is possible to detect each of \water, \ozone, and \oxygen.  At this SNR, the oxygen mixing ratio is estimated to be above roughly $10^{-3}$, indicating that we are unable to determine if \oxygen\ is a major atmospheric constituent (i.e., present at the 1\% level or more). Gravity (and, thus, planetary mass) remain undetected at all SNRs, similar to the findings of \citet{lupu2016}. The surface albedo is unconstrained (or worse) at all SNRs, but shows a weak bias toward a higher value of $A_s \approx 0.3$ ($\log{A_{\rm s}} \approx -0.5$) at the highest SNRs, which is likely due to the relatively large error bars at red wavelengths (driven primarily by low detector quantum efficiency) where we have the most sensitivity to the surface. We are able to get weak detections of $\tau$ and $f_{\rm c}$, which are shown in Figure \ref{fig:b4stairs} to be correlated. Yet, with these posteriors, we cannot rule out scenarios without cloud cover. We note the drop-off in the posteriors of $p_{\rm t}$ and $dp$ at higher pressure values likely result from the limiting conditions that the cloud base cannot extend below the surface pressure and the upper limit of the $P_0$ prior.  The improved signal-to-noise ratio leads to a posterior more concentrated around the true value for $p_{\rm t}$, $dp$, and $R_{\rm p}$.  For improved constraints on cloud properties, it may be beneficial to observe time variability with photometry \citep[e.g.,][]{ford2001} or use polarimetry \citep[e.g.,][]{rossi2017}.

At a higher spectral resolution ($R = 140$), the improvement in detections and constraints begin at a lower SNR, as illustrated by Figure \ref{fig:compareR140}. Gravity remains undetected for all SNRs. At a SNR equal to 5, $P_0$ and $R_{\rm p}$ have a weak detection and a detection, respectively. At SNR = 10, it is possible to detect \water, \ozone, and \oxygen. As with the $R=70$ case, surface albedo is unconstrained (or worse) at all SNRs, and, at the highest SNRs, the model is biased towards $A_{\rm s} \approx 0.3$ (as with $R = 70$). Moving to SNR = 15 adds a constraint to $R_{\rm p}$, $P_0$, and \ozone, as well as weak detections of cloud parameters. Increasing the SNR to 20 does not dramatically change the posterior distributions, although the posteriors for \water\ and \oxygen\ become narrow enough to offer constraints.  Here, the constraint on \oxygen\ suggests it is a major constituent in the atmosphere. In spite of the generous SNR, though, the 1-$\sigma$ uncertainties on the gas mixing ratios are not more precise than roughly an order of magnitude (see Table~\ref{tab:r140results}).

Considering both $R = 140$ and $R = 70$, we see that SNR = 5 data offer very little information about the planetary atmosphere. In the case of $R = 140$, SNR = 10 data offer detections but no constraints, and SNR = 20 data are required to constrain all included gas species.  In other words, the conclusions we would draw about the planet (e.g., the amount of gases, the bulk and cloud properties) improve significantly between SNR = 10  and SNR = 20. With $R = 70$, the boost from SNR = 10 to SNR = 15 provides weak detections of key atmospheric and surface parameters, and SNR = 20 data offer detections but few constraints (i.e., except on planetary radius).

For the \textit{WFIRST} rendezvous data sets, we are able to infer very little information at a SNR of 5 or 10 except for weak detection of surface pressure and a detection of the planetary radius. All gases remain undetected at these SNRs. The posterior distributions for most parameters do not vary much as SNR improves, although there are weak detections of cloud optical depth and fractional coverage at the highest SNRs. Like all previous cases, we do not detect the surface gravity.  At SNR$\ = 15,\ 20$, the detection of $f_c$ is unable to rule out scenarios with little cloud cover. To obtain weak detections of the atmospheric gases we require a SNR of 20, but, even here, the posteriors have tails that extend to near-zero mixing ratios.

\begin{figure*}[ht!]
\centering
\subfigure[Mixed runs SNR$=10$ bulk parameters]{
    \includegraphics[width = 0.45\textwidth]{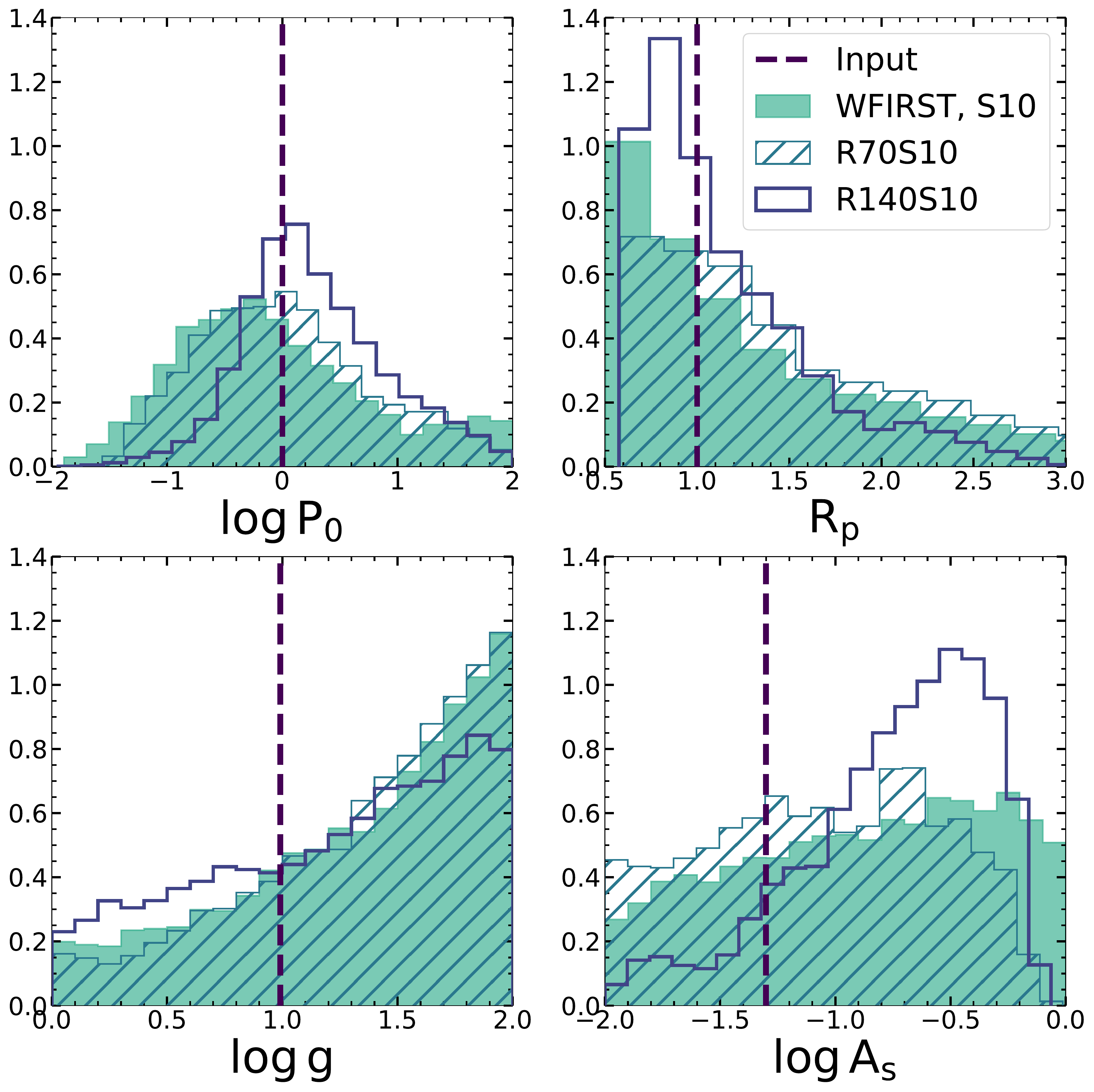}
    \label{mixedbulk}}\qquad
\subfigure[Mixed runs SNR$=10$ cloud parameters]{
    \includegraphics[width = 0.45\textwidth]{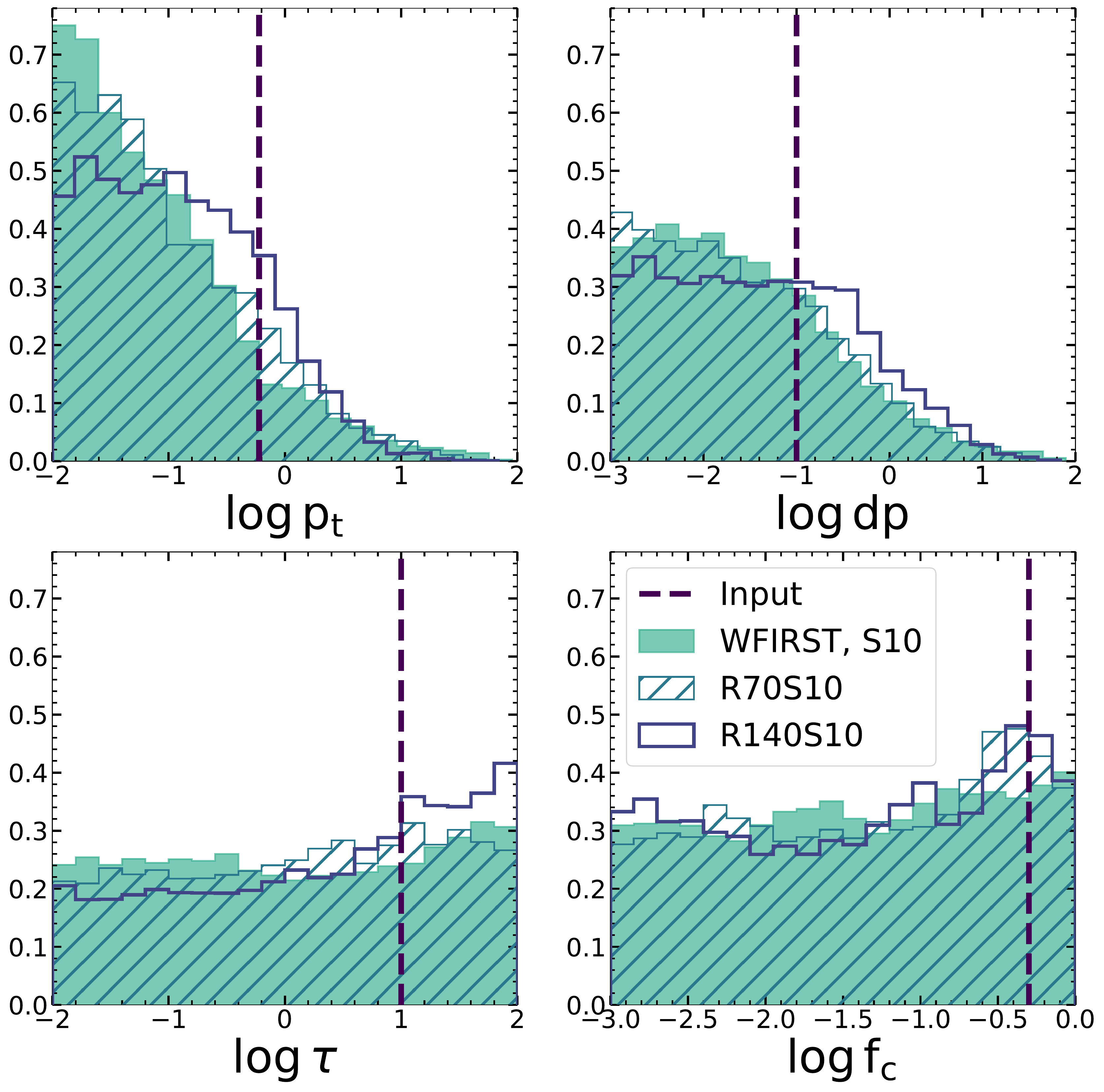}
    \label{mixedclouds}}
\subfigure[Mixed runs SNR$=10$ gas mixing ratios]{
    \includegraphics[width = 0.8\textwidth]{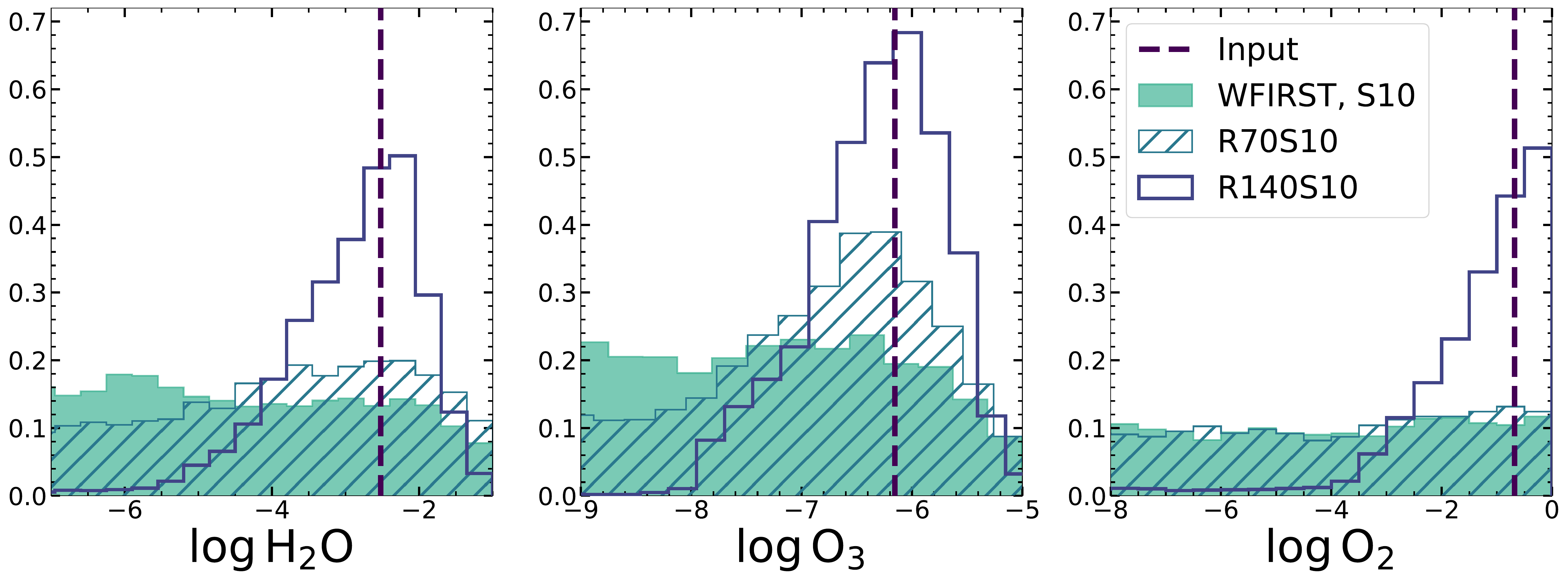}
    \label{mixedgases} \qquad
    }
\caption{Comparing the posteriors for all parameters for SNR$\ =10$ cases of \textit{WFIRST} rendezvous, $R = 70$, and $R = 140$. Overplotted dashed line represents the fiducial values from Table \ref{tab:params}.}
\label{fig:compareMixed}
\end{figure*}

To compare the performance of a \textit{WFIRST} rendezvous scenario against HabEx/LUVOIR scenarios at $R = 70$ and $R = 140$, we plot together the posterior distributions of the parameters for the SNR$\ = 10$ results from \textit{WFIRST} rendezvous, $R = 70$, and $R = 140$ in Figure \ref{fig:compareMixed}. While this comparison sheds light on the corresponding trade-off in terms of parameter estimation for the same SNR, these cases do not represent equal integration times, which scales with resolution and SNR.  If the dominant noise source does not depend on resolution (e.g., detector noise), the cases of $R = 140$ at SNR = 10, $R = 70$ at SNR = 20, and $R = 50$ at SNR = 28 would be roughly equal in integration times. However, if the dominant noise source does depend on resolution (e.g., exozodiacal dust), the cases of $R = 140$ at SNR = 10, $R = 70$ at SNR = 14, and $R = 50$ at SNR = 17 would roughly have equivalent integration times. Tables \ref{tab:r70detect} through \ref{tab:wfirstdetect} allow approximate comparisons of these different scenarios, excluding a \textit{WFIRST} rendezvous scenario at high SNR = 28 that we have not considered. 

\begin{deluxetable}{lcccc}
\tablecaption{$R = 70$: Strength of detection for a set of key parameters as a function of SNR. \label{tab:r70detect}}
\tablewidth{0pt}
\tabletypesize{\scriptsize}
\tablehead{Parameter	& SNR$=5$	& SNR$=10$ & SNR$=15$ & SNR$=20$}
\startdata
\water & $-$   & $-$   & W     & D \\
\ozone & $-$ &W & W&D \\
\oxygen &$-$ & $-$ & W &D \\
$P_0$ &W &W &W &D \\
$R_p$ &D &D &D & C\\
\enddata
\tablecomments{Weak detection (``W'') corresponds to a posterior distribution with a marked peak but also a substantial tail towards extreme values. Detection (``D'') refers to a peaked posterior without tails towards extreme values but a 1-$\sigma$ width larger than an order of magnitude. Constraint (``C'') is defined as a peaked posterior distribution with a 1-$\sigma$ width less than an order of magnitude. Non-detection, or flat posteriors across the entire (or near-entire) prior range, are marked with ``$-$''.}
\end{deluxetable}

\begin{deluxetable}{lcccc}
\tablecaption{$R = 140$: Strength of detection for a set of key parameters as a function of SNR. \label{tab:r140detect}}
\tablewidth{0pt}
\tabletypesize{\scriptsize}
\tablehead{Parameter	& SNR$=5$	& SNR$=10$ & SNR$=15$ & SNR$=20$}
\startdata
\water & $-$ &D &D &C \\
\ozone & $-$ &D & C&C \\
\oxygen &$-$ &D &D &C \\
$P_0$ &W    &D & C &C \\
$R_p$ &D    &D & C& C\\
\enddata
\tablecomments{Weak detection (``W'') corresponds to a posterior distribution with a marked peak but also a substantial tail towards extreme values. Detection (``D'') refers to a peaked posterior without tails towards extreme values but a 1-$\sigma$ width larger than an order of magnitude. Constraint (``C'') is defined as a peaked posterior distribution with a 1-$\sigma$ width less than an order of magnitude. Non-detection, or flat posteriors across the entire (or near-entire) prior range, are marked with ``$-$''.}
\end{deluxetable}

\begin{deluxetable}{lcccc}
\tablecaption{\textit{WFIRST}: Strength of detection for a set of key parameters as a function of SNR. \label{tab:wfirstdetect}}
\tablewidth{0pt}
\tabletypesize{\scriptsize}
\tablehead{Parameter	& SNR$=5$	& SNR$=10$ & SNR$=15$ & SNR$=20$}
\startdata
\water & $-$ &$-$ &$-$ &W \\
\ozone & $-$ &$-$ & $-$ &W \\
\oxygen &$-$ &$-$ &W &W \\
$P_0$ & W& W& W&W \\
$R_p$ &D &D &D & D\\
\enddata
\tablecomments{Weak detection (``W'') corresponds to a posterior distribution with a marked peak but also a substantial tail towards extreme values. Detection (``D'') refers to a peaked posterior without tails towards extreme values but a 1-$\sigma$ width larger than an order of magnitude. Constraint (``C'') is defined as a peaked posterior distribution with a 1-$\sigma$ width less than an order of magnitude. Non-detection, or flat posteriors across the entire (or near-entire) prior range, are marked with ``$-$''.}
\end{deluxetable}

From Figure \ref{fig:compareMixed}, we see that the performance of the \textit{WFIRST} rendezvous retrieval is similar to that of $R = 70$ at SNR = 10. The noticeable difference is a weak detection of \ozone\ with $R = 70$. Because we adopt the photometric setup from \textit{WFIRST} Design Cycle 7 through the shorter wavelengths, the data do not provide complete spectroscopic coverage across the significant \ozone\ feature from $0.5 - 0.7\ \mu$m, as in the case of HabEx/LUVOIR simulated data. Figure \ref{fig:labeledSpec} shows the sampling of the forward model spectrum for the three types of data sets we considered. We compare the spectral fits in Figure \ref{fig:yarr} and note the much wider spread in the possible fits for wavelengths shorter than $0.6\ \mu$m for \textit{WFIRST} rendezvous versus $R = 70$ or $R = 140$, which have continuous coverage in the full range.  The $R = 140$, SNR = 10 data set was able to offer {\it detections} of all atmospheric gases, setting it apart from the other two.  We stress, however, that {\it constraints} were only found at SNR = 20 and $R = 140$. 

\subsection{Considering Multiple Noise Instances} \label{subsec:multiple}

Our parameter estimations are likely to be optimistic as a consequence of our adoption of non-randomized spectral data points in our faux observations.  Thus, the requisite SNRs for detection detailed above should be seen as lower limits. Ultimately, our decision to use non-randomized data points stemmed from computational limitations (preventing us from running large numbers of randomized faux observations for each of our $R$/SNR pairs) and from a desire to avoid the biases that can occur from attempting to make inferences from retrievals performed on a single, randomized faux observation \citep{lupu2016}.

However, we deemed it necessary to investigate the consistency of our findings with respect to different noise instances. To work within our computational restrictions, we realized that cases such as $R=70$ with SNR $=5$ yielded little detection information for any parameter even in the ideal scenario of non-randomized data. We then decided to focus on two ``threshold'' cases based on the results from the non-randomized data: $R=140$ with SNR $=10$ and $R=70$ with SNR $=15$. We ran 10 noise instances of these two cases where it is likely the optimistic non-randomized data makes the difference between detection and constraint for several parameters (see Tables \ref{tab:r70detect} and \ref{tab:r140detect}).

\begin{figure*}[h]
\centering
    \includegraphics[width=0.8\linewidth]{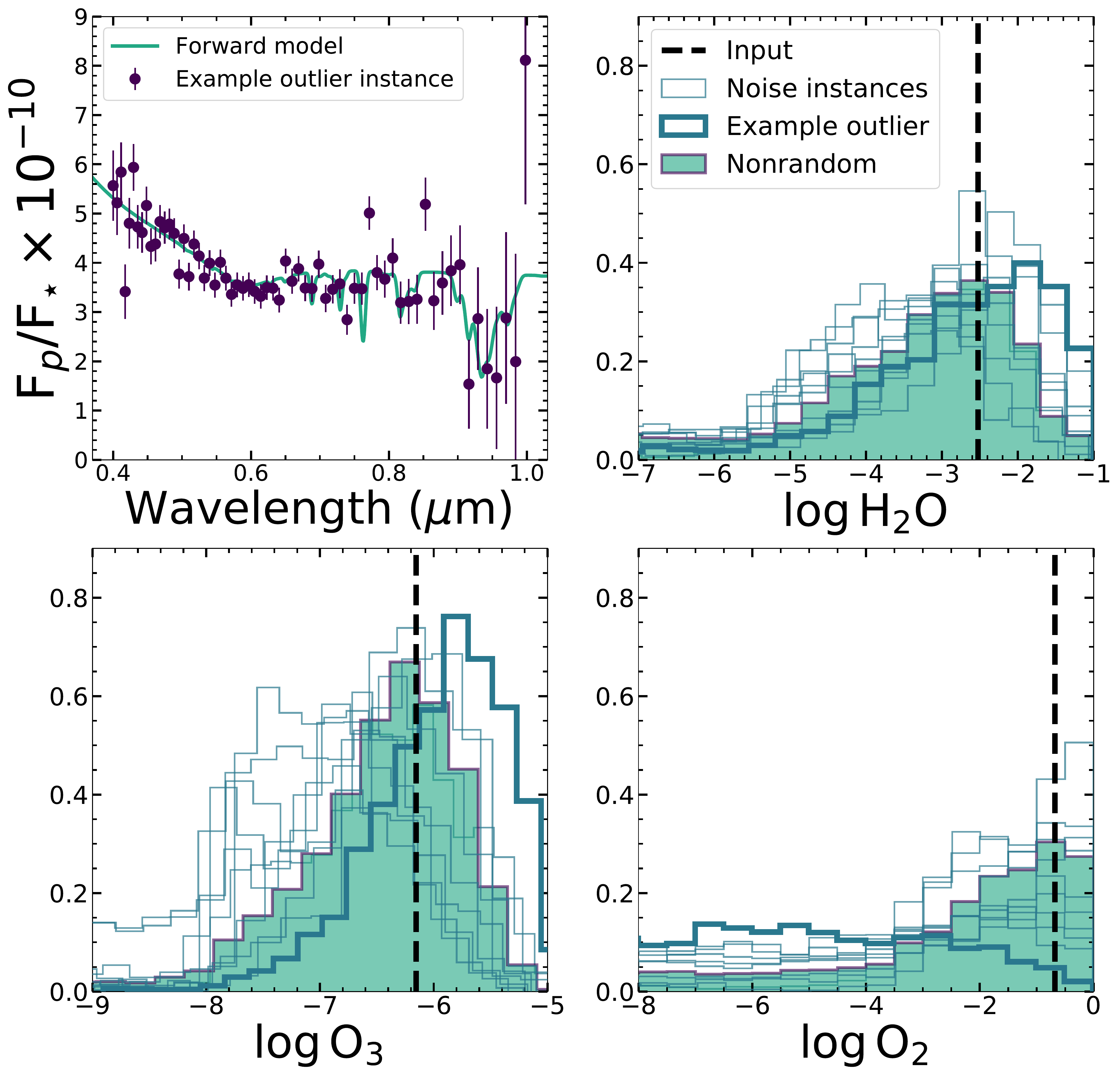}\hfil
\caption{The top left panel shows one of the 10 noise instances we retrieved on for $R=70$, SNR $=15$ data, plotted along with the forward model spectrum at $R\sim70$. The remaining three panels show the gas mixing ratio posteriors (\water, \ozone, \oxygen) of all the 10 noise instances of $R=70$, SNR $=15$. In addition, we are showing the corresponding posterior distributions from the non-randomized data set (seen originally in Figure \ref{fig:compareR70}) for comparison. The set of posteriors that correspond to the noise instance in the top left panel is the set of bolded distributions. The vertical dashed lines represent the input values of the parameters.}
\label{fig:demoOutlier}
\end{figure*}

Each noise instance is run for at least 10000 steps in {\tt emcee}. Figure \ref{fig:demoOutlier} shows all the individual posteriors for the gas mixing ratios from each noise instance for $R=70$, SNR $=15$. We highlight the posteriors from one ``outlier'' case where there is no oxygen detection. The corresponding set of data points are shown as well. This highlights the fact that single noise instances can mislead our interpretation and the benefit of having many noise instances run to obtain a more comprehensive understanding of the state of an atmosphere.

\begin{figure*}[ht!]
\centering
\subfigure[$R=70$, SNR$=15$: Combined bulk parameter posteriors from 10 noise instances]{
    \includegraphics[width = 0.45\textwidth]{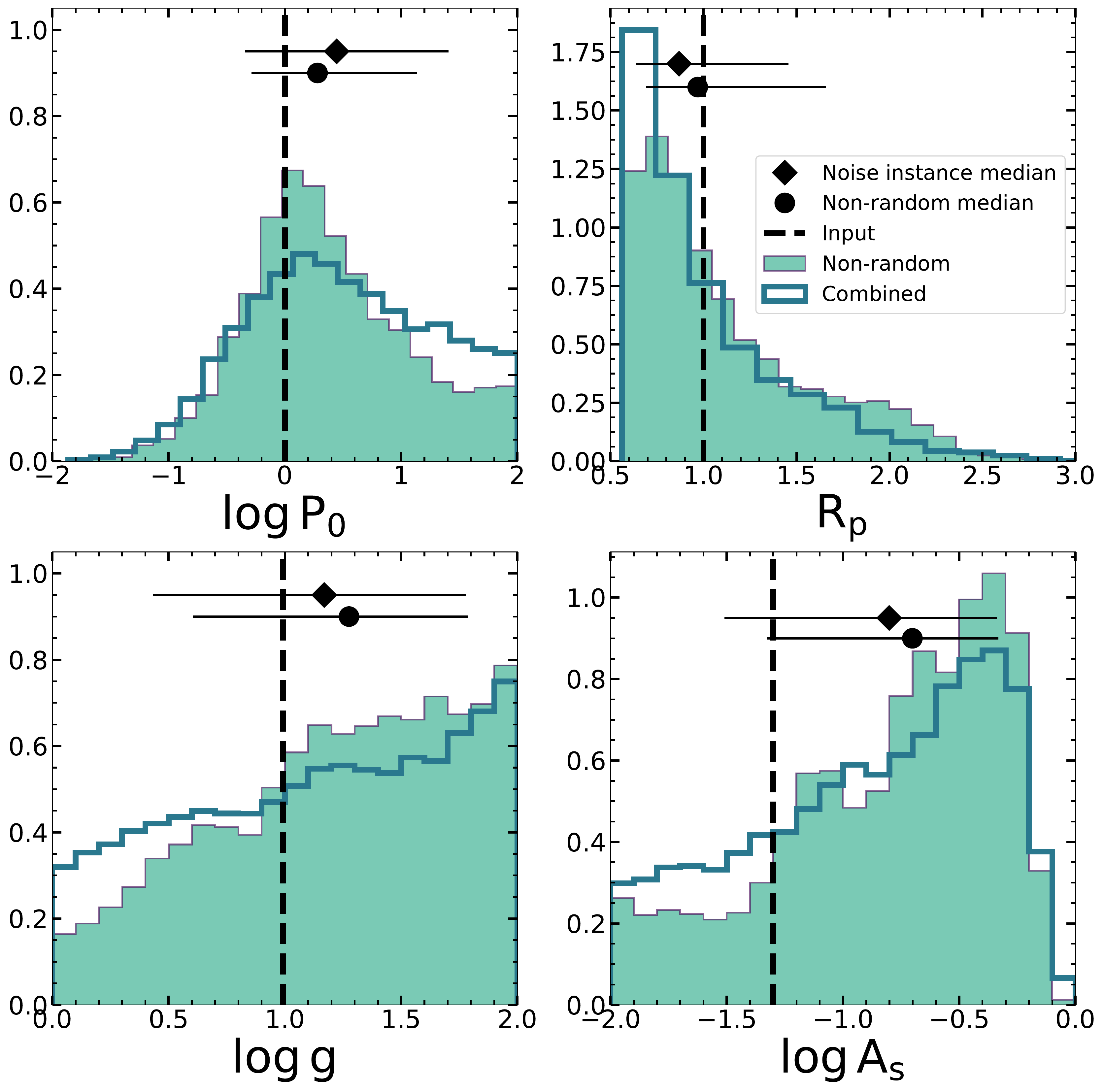}
    \label{mixedbulk70}}\qquad
\subfigure[$R=70$, SNR$=15$: Combined cloud parameter posteriors from 10 noise instances]{
    \includegraphics[width = 0.45\textwidth]{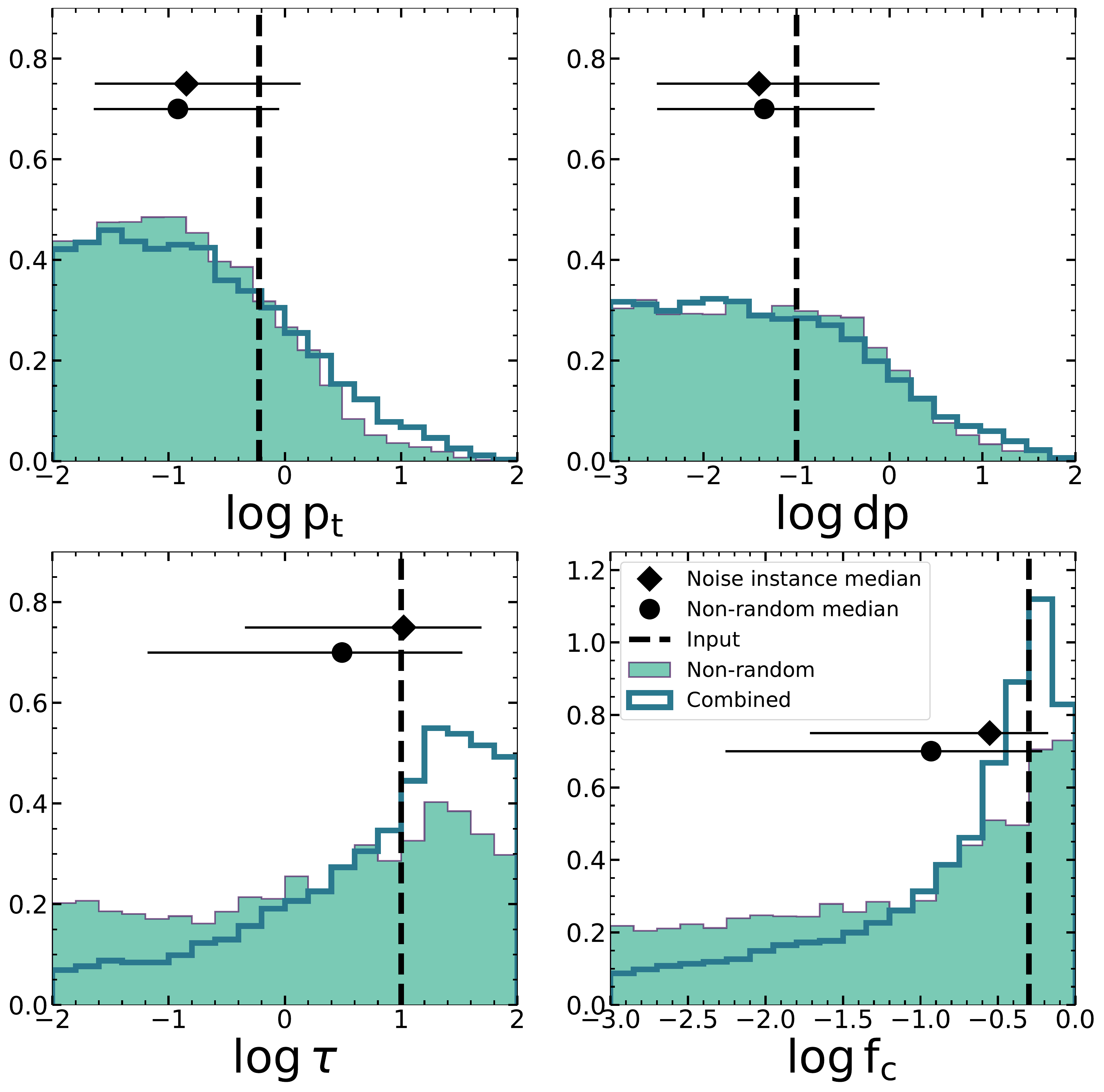}
    \label{mixedclouds70}}
\subfigure[$R=70$, SNR$=15$: Combined gas mixing ratio posteriors from 10 noise instances]{
    \includegraphics[width = 0.8\textwidth]{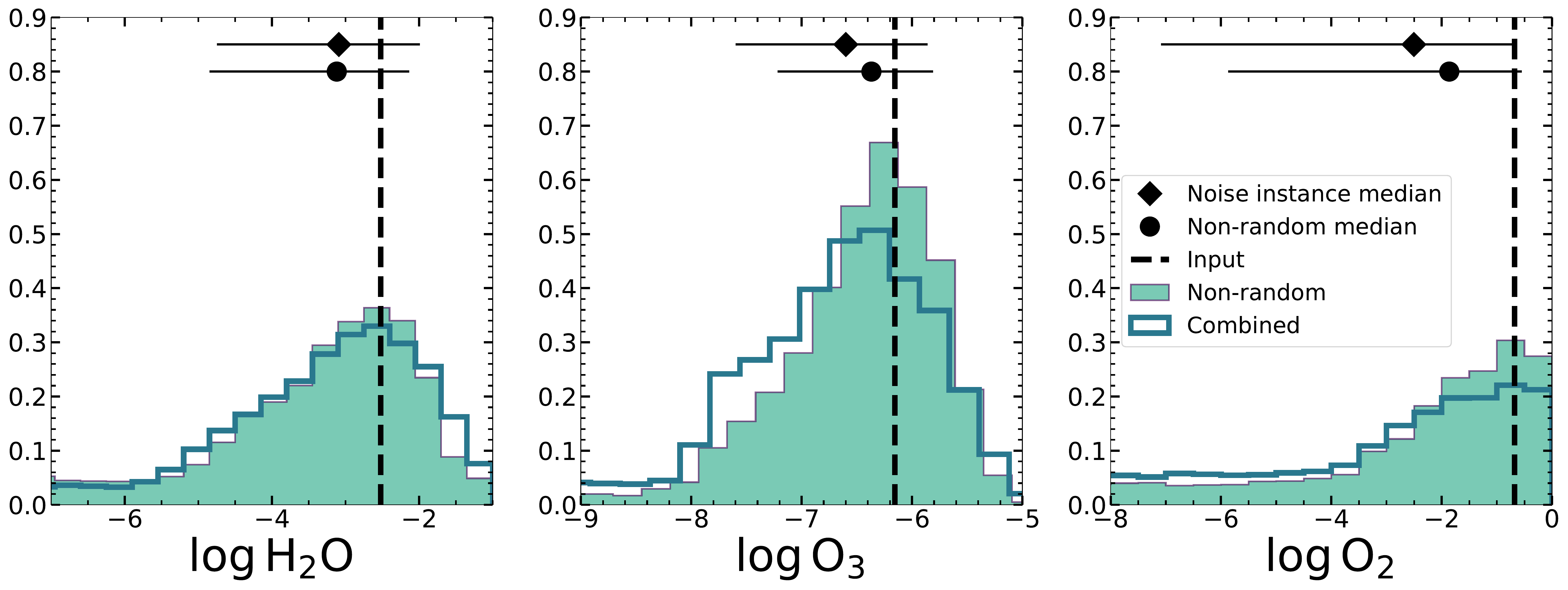}
    \label{mixedgases70}} \qquad
\caption{The combined posteriors distributions from 10 noise instances of $R=70$, SNR$=15$ compared to the posteriors from the non-randomized data set (see also Figure \ref{fig:compareR70}). The diamond represents the median value of each combined posterior, while the circle is the median of the non-randomized data set posterior. Each median is plotted along with the 68\% confidence interval from the same distribution. The vertical dashed lines represent the input values of the parameters.}
\label{fig:noiseR70}
\end{figure*}

\begin{figure*}[ht!]
\centering
\subfigure[$R=140$, SNR$=10$: Combined bulk parameter posteriors from 10 noise instances]{
    \includegraphics[width = 0.45\textwidth]{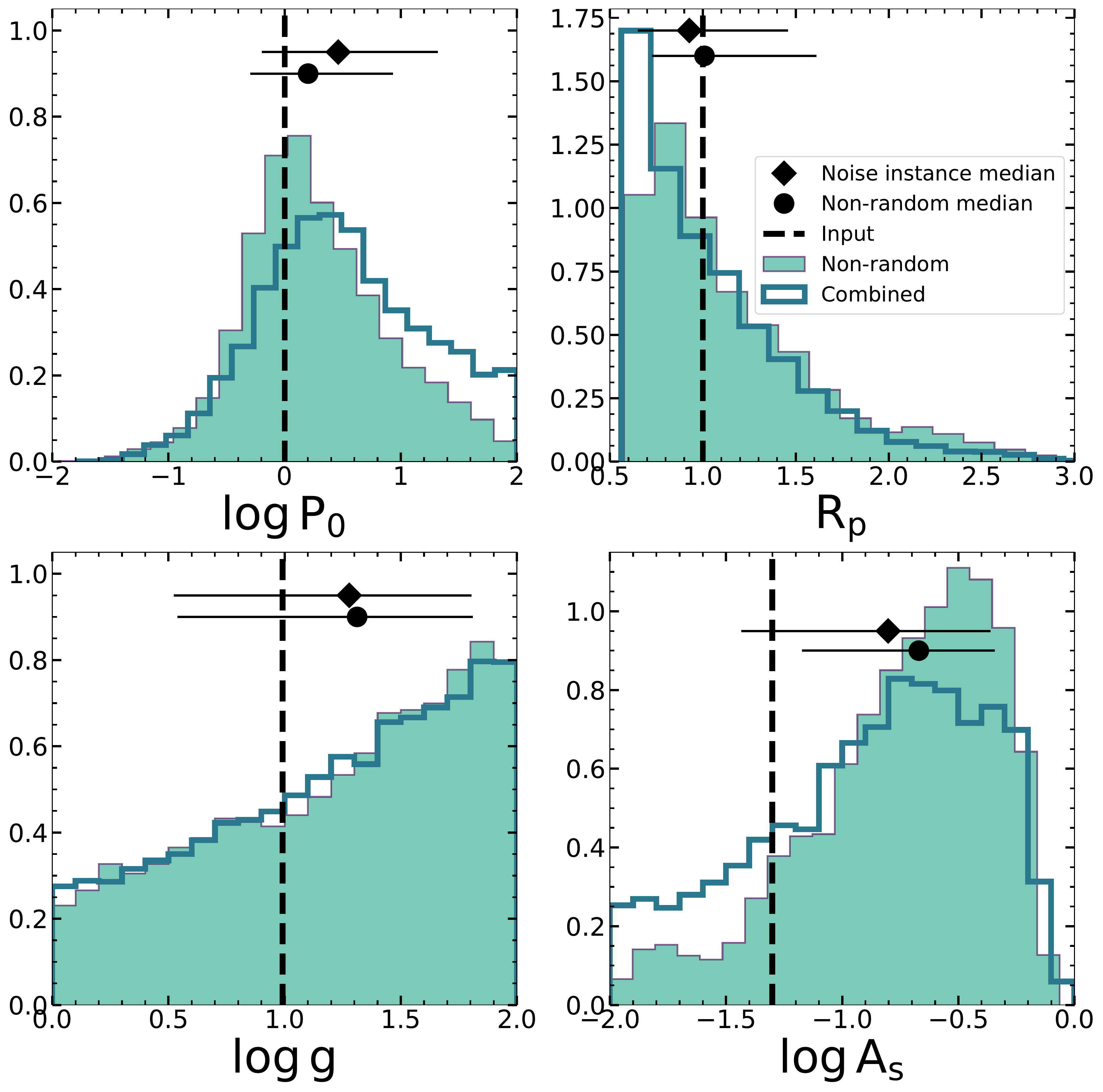}
    \label{mixedbulk140}}\qquad
\subfigure[$R=140$, SNR$=10$: Combined cloud parameter posteriors from 10 noise instances]{
    \includegraphics[width = 0.45\textwidth]{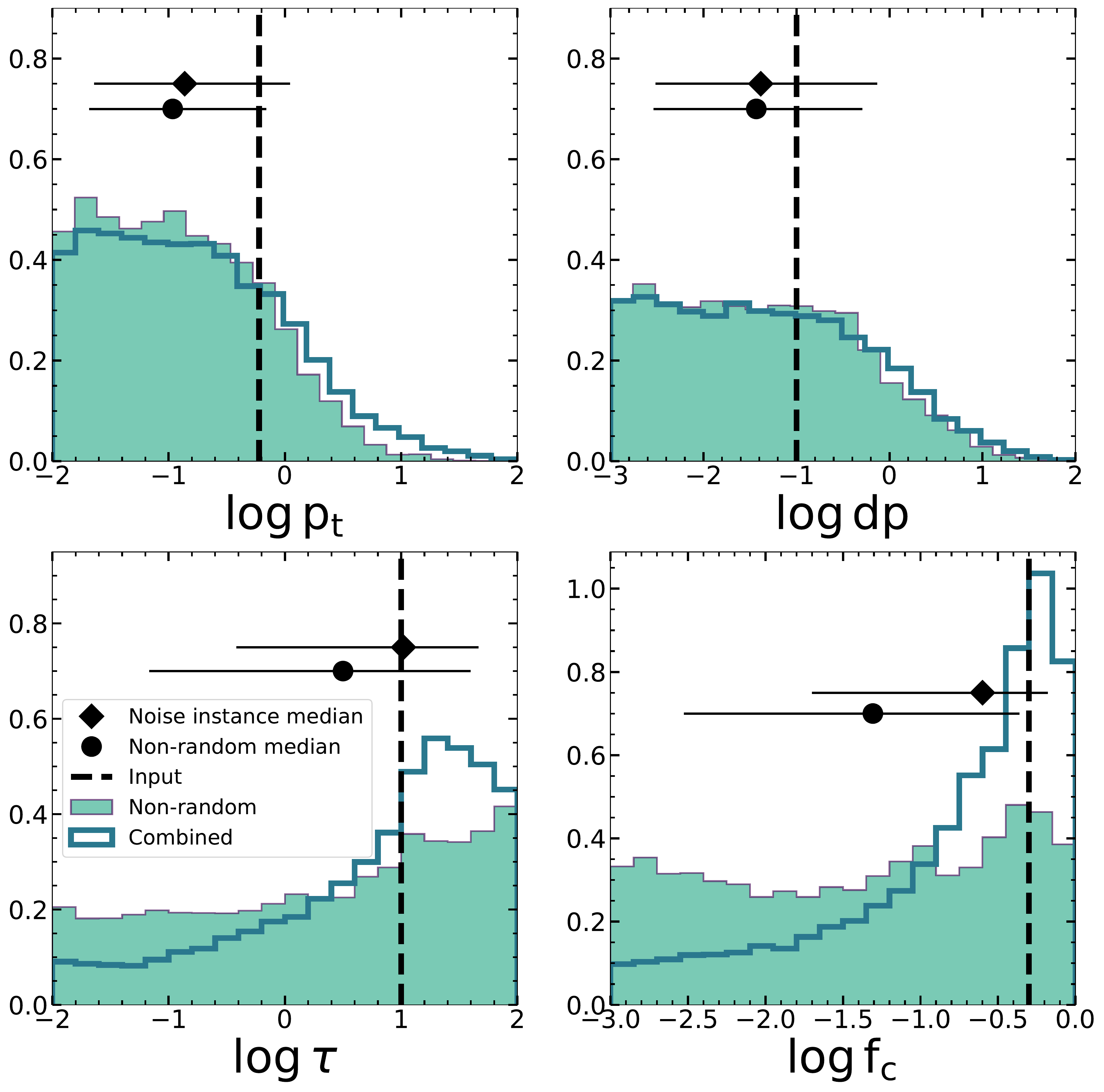}
    \label{mixedclouds140}}
\subfigure[$R=140$, SNR$=10$: Combined gas mixing ratio posteriors from 10 noise instances]{
    \includegraphics[width = 0.8\textwidth]{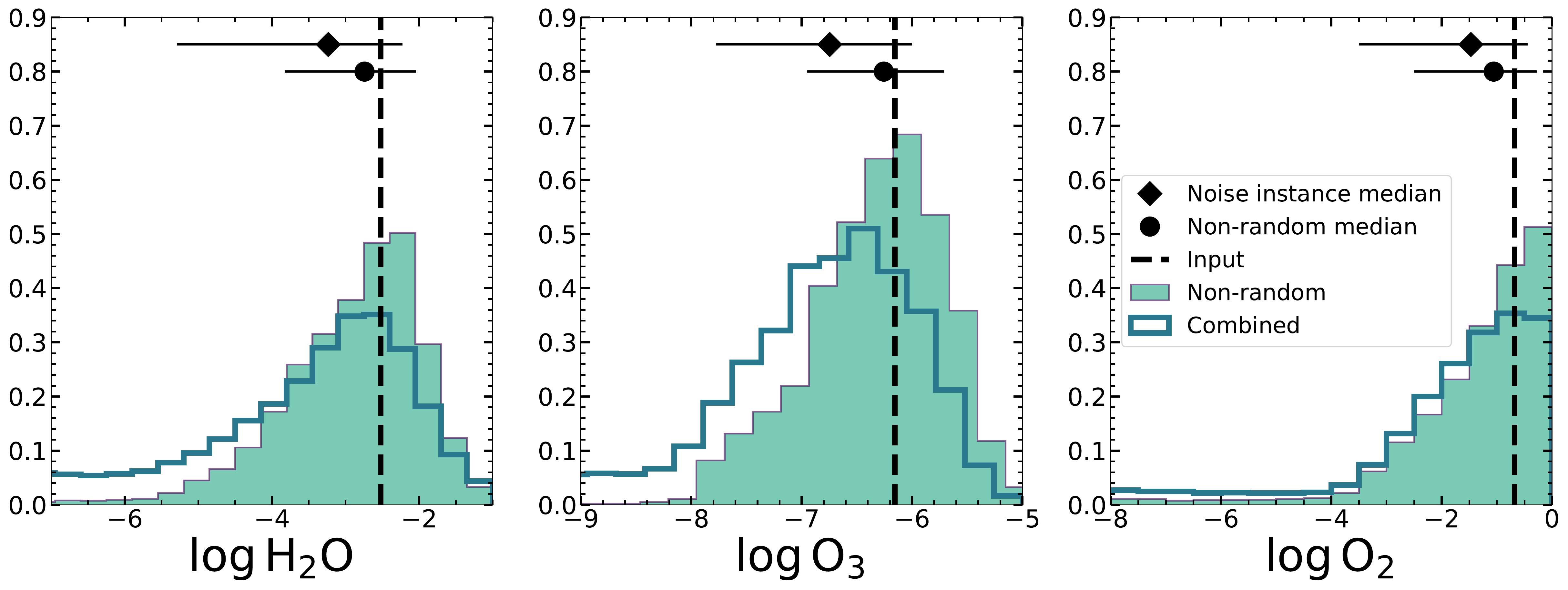}
    \label{mixedgases140} \qquad
    }
\caption{The combined posteriors distributions from 10 noise instances of $R=140$, SNR$=10$ compared to the posteriors from the non-randomized data set (see also Figure \ref{fig:compareR140}). The diamond represents the median value of each combined posterior, while the circle is the median of the non-randomized data set posterior. Each median is plotted along with the 68\% confidence interval from the same distribution. The vertical dashed lines represent the input values of the parameters.}
\label{fig:noiseR140}
\end{figure*}

To summarize the noise instance results, we concatenate samples from the last 1000 steps in each noise instance and construct an averaged set of posteriors. We are able to do this because the noise instances are equally likely, having been drawn in the same manner from a Gaussian with set parameters (i.e., the same SNR as the standard deviation). In Figure \ref{fig:noiseR70}, we plot up the combined posteriors of the 10 noise instances of $R=70$, SNR $=15$ and compared them to the posterior from the last 5000 steps of the non-randomized data case. We illustrate the same comparison for $R=140$, SNR $=10$ in Figure \ref{fig:noiseR140}. We overplot the truth values as well as the 68\% confidence interval and median value for each parameter from the combined noise-instances posterior and the non-randomized data posterior.

For all parameters in both the $R=70$ and $R=140$ cases, we find that the average posterior from the 10 noise instances agree with the posterior from the non-randomized data set qualitatively. Their medians and 68\% confidence interval ranges are also similar with significant overlap. The overall conclusions we can draw from the average posteriors do not appear to differ much from those using the non-randomized data set posteriors.

\subsection{Implications for Future Direct Imaging Missions}

Future space-based direct imaging missions will have a diversity of goals for exoplanet studies, and will likely emphasize the detection and characterization of Earth-like exoplanets.  For the detection of oxygen and ozone---which are key biosignature gases---in the atmospheres of Earth twins, our results indicate that spectra at a minimum characteristic SNR of 10 will suffice if at $R=140$, while data at SNR of at least 15--20 would be needed at $R=70$.  For a {\it WFIRST} rendezvous-like observing setup, these gases would only be weakly detected even at a SNR of 20.  Methane, which is another important biosignature gas, has no strong signatures in the visible wavelength range for the modern Earth, so we did not consider detection of this gas.  Thus, we could not use our simulated data and retrievals to argue for detections of atmospheric chemical disequilibrium \citep{sagan1993,krissansen2016}.

Key habitability indicators include atmospheric water vapor and surface pressure.  Detecting the former requires a SNR of 15--20 at $R=70$, but only a SNR of 10 at $R=140$.  Surface pressure can be constrained to within an order of magnitude for SNR~$\gtrsim 15$ at $R=140$, although the overall lack of temperature information in these reflected-light spectra would make it impossible to use pressure/temperature data to argue for habitability \citep{robinson2017}.  Surface temperature information may then need to come from climate modeling investigations that are constrained by retrieved gas mixing ratios.

For all of our observing setups, the data yield detections of, and in some cases constraints on, the planetary radius. Except at SNR of 20 for $R=70$ or SNR~$>15$ for $R=140$, the posterior distributions are not well-enough constrained to distinguish a Earth/super-Earth ($R_p < 1.5\,R_{\oplus}$) from a mini-Neptune based on size alone, although the data do rule out planetary sizes larger than Neptune.  Additional atmospheric information (e.g., composition) could potentially be used to help distinguish between terrestrial planets and mini-Neptunes.  These findings are consistent with the gas giant-focused work of \citet{nayak2017}, who note that observations at multiple phase angles can also help to better constrain planetary size.  Our overall lack of surface gravity constraints, paired with the weak constraints on planet size, implies that we do not have a constraint on the planetary mass.  Follow-up (or precursor) radial velocity observations (or, potentially, astrometric observations) could offer additional constraints on planet mass.

We can make rough comparisons of our $R$/SNR results to those of \citet{brandt2014}, who used minimally parametric models to investigate detections of \oxygen\ and \water\ for Earth twins.  These comparisons are not direct, however, as \citet{brandt2014} were fitting for fewer parameters (8 versus our 11) and also only assumed that SNR was proportional to planetary reflectance (versus our more complicated scaling, as shown in Figure~\ref{fig:snrscaling}).  For \oxygen, \citet{brandt2014} find $R=150$ and SNR~$=6$ for a 90\% detection probability, which is consistent with our $R=140$ posteriors moving from a non-detection at SNR~$=5$ to a detection at SNR~$=10$.  When investigating \water, \citet{brandt2014} find $R=40$ and SNR~$=7.5$ or $R=150$ and SNR~$=3.3$ for a 90\% detection probability.  Using Figure~\ref{fig:snrscaling} to scale our SNRs to 890~nm (i.e., to the continuum just shortward of 950~nm water vapor band), at $R=50$ we only find a weak detection of \water\ for SNR$_{\rm 890~nm}$~$=10$, and at $R=140$ we transition from a water vapor non-detection to detection between a SNR$_{\rm 890~nm}$ of 2.5--5.  Taken altogether, these comparisons indicate that we agree with \citet{brandt2014} at higher spectral resolution ($R=140$--150), but that detection of \water\ at lower spectral resolution ($R=50$) will likely require higher SNRs than originally indicated.

The discussion above emphasizes mere detections, not constraints (which, again, we define as having peaked posterior distributions with 1-$\sigma$ widths less than an order of magnitude).  While uncertain, we anticipate that characterization of climate, habitability, and life likely require constraints, not simple detections.  Here, as is shown in Table \ref{tab:r140detect}, only $R=140$ and SNR = 20 observations offer the appropriate constraints.  Thus, future space-based high contrast imaging missions with goals of characterizing Earth-like planetary environments are likely to need to achieve $R=140$ and SNR = 20 observations (or better).  Of course, combining near-infrared capabilities, which would provide access to additional gas absorption bands, may help loosen these requirements.

\subsection{Impacts of Model Assumptions}

Several key assumptions adopted in this study warrant further comment.  First, as noted earlier, we do not retrieve on planetary phase angle and planet-star distance, both of which influence the planet-to-star flux ratio.  Thus, in effect, we are assuming that the planetary system has been revisited multiple times for photometric and astrometric measurements, such that the planetary orbit is reasonably well-constrained (i.e., that the orbital distance and phase angle are not the dominant sources of uncertainty when interpreting the observed planet-to-star flux ratio spectrum).  If the orbit is not well-constrained, \citet{nayak2017} showed that strong correlations can exist between the retrieved phase angle and the planet radius.

Second, we have assumed detector-dominated noise and a quantum efficiency appropriate for the {\it WFIRST}/CGI for all of our observational setups.  While this is likely a fair assumption for our {\it WFIRST} rendezvous studies, it is likely that detector development will lead to major improvements in instrumentation for a HabEx/LUVOIR-like mission.  Here, the rapid decrease into the red due to detector quantum efficiency may not be as dramatic, implying that spectra would have relatively more information content at red wavelengths as compared to the present study.  Furthermore, a HabEx/LUVOIR-like mission may no longer be in the detector-dominated noise regime.  In the limit of noise dominated by astrophysical sources (e.g., exo-zodiacal light or stellar leakage), the SNR only varies as $\sqrt{q\mathcal{T}B_{\lambda}}$.

Finally, we adopt a relatively simple parameterization of cloud three-dimensional structure.  Specifically, we allow for only a single cloud deck in the atmosphere, and we then permit these clouds to have some fractional coverage over the entire planet. This parameterization of fractional cloudiness implies uniform latitudinal and longitudinal distribution of patchy clouds.  In reality, clouds on Earth have a complex distribution in altitude, latitude, and longitude \citep{stubenrauch13}, and variations in time also have an observational impact \citep{cowan2009,cowanfujii2017}.  However, given the overall inability of our retrievals to constrain cloud parameters \citep[at least at the SNRs investigated here; see also ][]{lupu2016,nayak2017}, it seems challenging for future space-based exoplanet characterization missions to detect (or constrain) more complex cloud distributions with the types of observations studied here and data of similar quality.

\subsection{Future Work}

Our current forward model is able to include both CO$_2$ and CH$_4$, although we did not retrieve on these gases in the current study due to their overall lack of strong features in the visible wavelength range for modern Earth.  However, these species do have stronger features in the near-infrared wavelength range.  As both of the HabEx and LUVOIR concepts are considering near-infrared capabilities, it will be essential to extend our current studies to longer wavelengths and to investigate whether or not constraints on additional gases (i.e., beyond water, oxygen, and ozone) can be achieved at these wavelengths.

Additionally, given the likely huge diversity of exoplanets that will be discovered by future missions (and that have already been identified and studied by {\it Kepler}, {\it Hubble}, and {\it Spitzer}), it will be necessary to extend our parameter estimation studies to include a wider range of worlds.  Both super-Earths and mini-Neptunes are more-favorable targets for a \textit{WFIRST} rendezvous mission, and may also be easier targets for HabEx/LUVOIR-like missions.  Our forward model is already capable of simulating these types of worlds, and we anticipate emphasizing a variety of exoplanet types in future studies.  Such future studies may also include retrievals on planetary phase angle, which would be relevant to observing scenarios where the planetary orbit is poorly constrained.

\section{Summary} \label{conclusion}

We have developed a retrieval framework for constraining atmospheric properties of an Earth-like exoplanet observed with reflected light spectroscopy spanning the visible range ($0.4 - 1.0\mu$m). We have upgraded an existing, well-tested albedo model to generate high-resolution geometric albedo spectra used to simulate data at resolutions and quality relevant to future telescopes, such as the HabEx and LUVOIR mission concepts. We combined our albedo model with Bayesian inference techniques and applied MCMC sampling to perform parameter estimation. The data we considered were for \textit{WFIRST} paired with a starshade (i.e., the rendezvous scenario), $R = 70$, and $R = 140$ at SNR$ = 5, 10, 15, 20$. We validated our forward model, and we demonstrated the successful application of our retrieval approach by gradually adding complexity to our inverse analyses.

Following work by \citet{lupu2016} and \citet{nayak2017}, who have constructed a retrieval framework for gas giants in refleted light, we made several modifications to the albedo model featured in these previous studies. Our model has a reflective surface, absorption due to water vapor, oxygen, and ozone, Rayleigh scattering from nitrogen and other key gases, pressure-dependent opacities, an adaptive pressure grid, and a single-layer water vapor cloud layer with fractional cloudiness. We performed our retrievals with the goal of estimating our ability to detect and constrain the atmosphere of an Earth twin. We found that $R = 70$, SNR $=15$ data allowed us to weakly detect surface pressure as well as water vapor, ozone, and oxygen. At $R = 140$, we found that SNR $=10$ was needed to more firmly detect these parameters. At $R = 140$, a SNR of 20 was needed to constrain key planetary parameters, and $R = 70$ data at this SNR offered extremely few constraints.  A \textit{WFIRST} rendezvous scenario, with its photometric points and lower resolution spectrum ($R = 50$), is only able to offer limited diagnostic information. For example, at SNR $=10$, we only weakly detect and detect surface pressure and planetary radius, respectively. To weakly detect the gases, \textit{WFIRST} rendezvous data needed to be at least SNR $= 20$. Throughout our runs, we found that we are unable to accurately constrain surface albedo or place estimates on the surface gravity, although we can straightforwardly rule out planetary sizes above roughly the radius of Uranus or Neptune.

Our findings demonstrate that direct imaging of Earth-like exoplanets in reflected light offers a promising path forward for detecting and constraining atmospheric biosignature gases.  Instrument spectral resolution for future missions strongly impacts requisite SNRs for detection and characterization, and this must be taken into consideration during mission design.  Thus, the scientific yield of future space-based exoplanet direct imaging missions can only be maximized by simultaneously considering mission characterization goals, integration time constraints, and instrument spectral performance.

\acknowledgments YKF is supported by the National Science Foundation Graduate Research Fellowship under Grant DGE1339067.  TR gratefully acknowledges support from NASA through the Sagan Fellowship Program executed by the NASA Exoplanet Science Institute.  The authors thank the 2016 Kavli Summer Program in Astrophysics, its Scientific and Local Organizing Committees, the program founder, Pascale Garaud, and the Kavli Foundation for supporting the genesis of this work.  This work was made possible by support from the UCSC Other Worlds Laboratory and the {\it WFIRST} Science Investigation Team program.  The results reported herein benefited from collaborations and/or information exchange within NASA's Nexus for Exoplanet System Science (NExSS) research coordination network sponsored by NASA's Science Mission Directorate.  Certain essential tools used in this work were developed by the NASA Astrobiology Institute's Virtual Planetary Laboratory, supported by NASA under Cooperative Agreement No. NNA13AA93A.  Computation for this research was performed by the UCSC Hyades supercomputer, which is supported by National Science Foundation (award number AST-1229745) and UCSC. We thank Cecilia Leung, Asher Wasserman, Daniel Thorngren, Eric Gentry, and Chris Stark for stimulating discussions and essential guidance.


\clearpage


\end{document}